\documentclass[article]{aastex62}
   
\shortauthors{Kuraszkiewicz et al.}
\shorttitle{Beyond Simple AGN Unification with Chandra-observed 3CRR Sources}

\usepackage{graphics}
\usepackage{bm}
\usepackage{rotating}
\usepackage{times,graphicx}
\usepackage{natbib}

\def\q24 {q$_{24}$}

\def\deg{\arcdeg}

\def\nh {$N_{\rm H}$}
\def\lax    {${_<\atop^{\sim}}$}
\def\gax    {${_>\atop^{\sim}}$}
\def\lax{$\la$}
\def\gax{$\ga$}
\def\chandra {{\it Chandra}}

\def\tauSi {$\tau_{\rm 9.7}$}
\def\rcd {$R_{\rm CD}$}
\def\5to8 {L$_{5\mu \rm m}$/L$_{8\mu \rm m}$}
\def\lxlr {$L_{\rm X}/L_{\rm R}$}

\def\lx {$L_{\rm X}$}

\def\spitz {{\it Spitzer}}
\def\fe {{Fe K$\alpha$}}
\newcommand{\oiii}{\hbox{\sc [O~iii]}}
\newcommand{\oii}{\hbox{\sc [O~ii]}}
\newcommand{\civ}{\hbox{\sc C~iv}}
\newcommand{\mgii}{\hbox{\sc M}g~{\sc ii}}
\newcommand{\NH}{ {N_{\rm{H}}} }

\begin{document}

\title{Beyond Simple AGN Unification with Chandra-observed 3CRR Sources at $0.5 < z < 1$} 
  
\author[0000-0001-5513-029X]{Joanna Kuraszkiewicz}
\affil{Harvard-Smithsonian Center for Astrophysics, Cambridge, MA 02138}

\author[0000-0003-1809-2364]{Belinda J.~Wilkes}
\affil{Harvard-Smithsonian Center for Astrophysics, Cambridge, MA 02138}

\author[0000-0002-3357-5228]{Adam Atanas}
\affil{Harvard University, Cambridge MA}

\author[0000-0003-0426-6634]{Johannes Buchner}
\affil{Pontificia Universidad Cat\'{o}lica de Chile, Instituto de Astrof\'{i}sica, Casilla 306, Santiago 22, Chile}
\affil{Millenium Institute of Astrophysics, Vicu\~{n}a MacKenna 4860, 7820436 Macul, Santiago, Chile}
\affil{Max Planck Institute for Extraterrestrial Physics, Giessenbachstrasse,
85741 Garching, Germany}

\author[0000-0002-7093-295X]{Jonathan C.~McDowell}
\affil{Harvard-Smithsonian Center for Astrophysics, Cambridge, MA 02138}

\author[0000-0002-9895-5758]{S.~P.~Willner}
\affil{Harvard-Smithsonian Center for Astrophysics, Cambridge, MA 02138}

\author[0000-0002-3993-0745]{Matthew L.~N.~Ashby}
\affil{Harvard-Smithsonian Center for Astrophysics, Cambridge, MA 02138}

\author[0000-0001-6004-9728]{Mojegan Azadi}
\affil{Harvard-Smithsonian Center for Astrophysics, Cambridge, MA 02138}

\author[0000-0002-0106-5776]{Peter Barthel}
\affil{Kapteyn Institute, University of Groningen, The Netherlands}

\author[0000-0002-7284-0477]{Martin Haas}
\affil{Astronomisches Institut, Ruhr-University, Bochum, Germany}

\author[0000-0002-1516-0336]{Diana M.~Worrall}
\affil{H.H. Wills Physics Laboratory, University of Bristol, UK}

\author[0000-0002-1858-277X]{Mark Birkinshaw}
\affil{H.H. Wills Physics Laboratory, University of Bristol, UK}

\author{Robert Antonucci}
\affil{Department of Physics, University of California, Santa  Barbara, CA 93106}

\author{Rolf Chini}
\affil{Astronomisches Institut, Ruhr-University, Bochum, Germany}
\affil{Instituto de Astronom\'{i}a, Universidad Cat\'{o}lica del Norte, Antofagasta, Chile}

\author{Giovanni G.~Fazio}
\affil{Harvard-Smithsonian Center for Astrophysics, Cambridge, MA 02138}

\author{Charles Lawrence}
\affil{JPL, Pasadena, CA 91109}

\author[0000-0002-3471-981X]{Patrick Ogle}
\affil{Space Telescope Science Institute, Baltimore, MD 21218}

\begin{abstract}

Low-frequency radio selection finds radio-bright galaxies
regardless of the amount of obscuration by gas and dust. We
report \chandra\ observations of a complete 178~MHz-selected, and so
orientation unbiased, sample of 44 $0.5<z<1$ 3CRR sources.  The sample
is comprised of quasars and narrow-line radio galaxies (NLRGs) with
similar radio luminosities, and the radio structure serves as
both an age and an orientation indicator. Consistent with Unification,
intrinsic obscuration (measured by \nh, X-ray hardness ratio, and
X-ray luminosity) generally increases with inclination. However, the
sample includes a population not seen in high-$z$ 3CRR sources: NLRGs
viewed at intermediate inclination angles with
\nh~$<10^{22}$~cm$^{-2}$. Multiwavelength analysis suggests these
objects have lower $L/L_{\rm Edd}$ than typical NLRGs at similar
orientation. Thus both orientation and $L/L_{\rm Edd}$ are important,
and a ``radiation-regulated Unification'' provides a better
explanation of the sample's observed properties. In comparison with
the 3CRR sample at $1<z<2$, our lower-redshift sample shows a higher
fraction of Compton-thin NLRGs (45\% vs.\ 29\%) but similar
Compton-thick fraction (20\%), implying a larger covering factor of
Compton-thin material at intermediate viewing angles and so a more
``puffed-up'' torus atmosphere.  We posit that this is due to a range
of $L/L_{\rm Edd}$ extending to lower values in this sample.  In
contrast, at high redshifts the narrower range and high $L/L_{\rm Edd}$
values allowed orientation (and so simple Unification) to dominate the
sample's observed properties.

\end{abstract}

\keywords{Active galactic nuclei (16)
  Quasars (1319)
  Radio loud quasars (1349)
  X-ray quasars (1821)}

\section{Introduction}
\label{sec:intro}

Active Galactic Nuclei (AGN) are among the most luminous non-transient
objects in the Universe and are responsible for the majority of
accretion (as opposed to stellar) power output. Their activity is
centered in a small nuclear region (the central engine), where the
standard model invokes a supermassive black hole surrounded by
accreting gas forming an accretion disk (emitting in the
visible--UV--soft-X-rays) and a hot corona (emitting hard-X-rays).  Much
of this radiation is then absorbed and reprocessed by gas and dust
(emitting in the infrared) in a disk/torus-like structure
surrounding the accretion disk, as described by the Standard Unification
model \citep{1989ApJ...336..606B,
  1993ARA&A..31..473A,1995PASP..107..803U,2015ARA&A..53..365N}. In
the Standard model, observationally different AGN and radio galaxies are
related to each other via the viewing angle. The broad-line
(``Type~1'') AGN (Seyfert~1s, quasars, broad-line radio galaxies) are
viewed along the poles of the dusty disk/torus, where the
(``face-on'') view of the central engine and the broad emission line
region (BLR) is unobscured.  The narrow-line (``Type~2'') AGN
(Seyfert~2s, narrow-line radio galaxies) are viewed edge-on to the
torus, so the central engine and the BLR are blocked from view, and
only the narrow emission lines, formed farther out, are visible. In
some Type~2s, the emission from the central engine reveals itself in
scattered polarized light \citep{2005AJ....129.1212Z}.

In its most basic version
\citep{1993ARA&A..31..473A}, Unification assumes a compact, smooth
torus \citep{1992ApJ...401...99P,1997ApJ...486..147G} with the same
opening angle for all AGN independent of their intrinsic
luminosity. Simple Unification is an oversimplification (already pointed
out by Antonucci in his 1993 review), and a ``receding torus model''  where the
inner sublimation radius increases with AGN luminosity, was introduced
\citep{1995A&A...298..395F,1991MNRAS.252..586L} to explain the
observed decrease of the fraction of Type~2 AGN with
increasing luminosity. 
Further refinement of Unification and the introduction of clumpy torus
models
\citep{2008ApJ...685..147N,2008ApJ...685..160N,2010A&A...515A..23H,2012MNRAS.420.2756S,2015A&A...583A.120S}
introduced the covering factor as an additional, independent variable
(\citealp{2012ApJ...747L..33E}; ``realistic'' Unification). In this
scenario, AGN at a given intrinsic luminosity have a distribution of
covering factors. The ratio of Type~2 to Type~1 AGN depends on the
mean covering factor of the sample, and the Type~2s will
preferentially be drawn from a population of AGN that have covering
factors higher than the mean, while the Type~1s are drawn from a
population with covering factors below the mean.  It was recently
shown \citep{2017Natur.549..488R,2017MNRAS.472.3492E} that the
covering factor of the obscuring dusty gas is strongly dependent on a
fundamental parameter of the central engine -- the Eddington ratio
$L/L_{\rm Edd}$ -- and is lowest in AGN with the highest $L/L_{\rm
  Edd}$. This dependence is explained as due to clearing out of the
(Compton-thin) gas and dust clouds within the opening angle of the torus via
radiation pressure, creating larger torus opening angles in sources
with higher $L/L_{\rm Edd}$.  Labelled 
``radiation-regulated Unification'', the effect results in the probability of
finding an obscured AGN increasing with decreasing $L/L_{\rm Edd}$
ratio.

Obscuration in AGN is not only highly anisotropic and likely $L/L_{\rm
  Edd}$ dependent, it is also strongly wavelength-dependent, which
will cause complex selection effects
and result in strong biases against specific subsets of AGN depending
on the wavelength of a sample's selection. A significant fraction of
the AGN population is largely unobserved as was demonstrated by the
Cosmic X-ray Background (CXRB, \citealp{2007A&A...463...79G}), which
requires equal numbers of unobscured and moderately (Compton-thin)
obscured ($10^{21} < $ \nh/cm$^{-2}$~$< 10^{23}$) sources, and a
comparable number of highly-obscured, Compton-thick (\nh\ $\geq 1.5
\times 10^{24}$~cm$^{-2}$) AGN. This last, Compton-thick population
has not yet been found.  The Two Micron All-Sky Survey (2MASS)
revealed a significant population of red, moderately obscured
($10^{21} < $ \nh/cm$^{-2}$~$< 10^{23}$,
\citealp{2002ApJ...564L..65W,2005ApJ...634..183W,2009ApJ...692.1180K,2009ApJ...692.1143K})
Type~1 and Type~2 AGN with a number density comparable to that of blue
optically-selected (Type~1) AGN at low redshifts
\citep{2002ASPC..284..127C}. The Sloan Digital Sky Survey (SDSS),
using optical color selection techniques \citep{2003AJ....126.1131R},
and the Hamburg Quasar Spectral Survey \citep{1995A&AS..111..195H}
revealed many Type~1 AGN with much redder colors than those found in
AGN samples typically selected based on blue optical colors.  {\it
  Chandra\/} and {\it Spitzer\/} facilitated many deeper,
multi-wavelength surveys such as GOODS \citep{2004ApJ...600L..93G},
SWIRE \citep{2003PASP..115..897L}, Bo\"{o}tes
\citep{2007ApJ...671.1365H}, ChaMP \citep{2007ApJS..169..401K}, COSMOS
\citep{2007ApJS..172....1S}, AEGIS
\citep{2007ApJ...660L...1D,2004ApJS..154...48E}, CANDELS
\citep{2011ApJS..197...35G}, and HERMES \citep{2012MNRAS.424.1614O}
which through hard-X-ray and/or infrared (IR) selection probed deeply
into the AGN population revealing larger numbers of obscured AGN than
the traditional optical surveys \citep{2003AJ....126..539A,
  2006ApJ...642..673D}.  However, even as more are being found, bias
against finding Compton-thick AGN remains. They are difficult to find
as their direct light is obscured even at {\it Chandra\/} and {\it
  XMM-Newton\/} energies ($<$10~keV). Harder X-ray surveys carried out using
{\it Swift}/BAT, {\it NuSTAR} and {\it INTEGRAL},
\citep{2011ApJ...728...58B,2015ApJ...815...66A,2012ApJ...757..181S}
also miss the most Compton-thick AGN, which is not surprising as
direct X-ray light from NGC 1068, a canonical nearby Type 2, is
undetected to energies \gax 100~keV
\citep{1997A&A...325L..13M}. Selection at IR wavelengths
\citep{2004ApJS..154..166L, 2005ApJ...631..163S} provides a way to
search for highly obscured AGN, but these are difficult to identify
among a much larger population of IR galaxies
\citep{2006ApJ...642..126B, 2010ApJ...717.1181P}.

Low-frequency radio selection (although limited to bright radio-loud
sources) is based on the optically thin and nearly isotropic
emission from the extended radio lobes. It is largely independent of orientation
and provides a reliable way to assemble radio-loud AGN
samples that are complete  and free of orientation-related bias.    
Accordingly, the 3CRR catalog of \cite{1983MNRAS.204..151L} 
delivers a complete, randomly oriented sample out to redshift $z=2.5$ down to a
limiting flux density of 10~Jy at 178~MHz and includes 173 radio
galaxies and quasi-stellar radio sources (quasars).  At these low
frequencies 3CRR sources are dominated by emission from the
extended radio lobes resulting in a sample free of orientation bias.

In the present work, we focus on the complete (orientation unbiased) subset of 
$0.5<z<1$ 3CRR sources and analyze X-ray, IR, optical, and radio
properties in relation to orientation and obscuration effects, thus
constraining the properties and geometry of the obscuring material. This
paper extends our studies of the $1<z<2$ 3CRR sample
\citep{2013ApJ...773...15W}, allowing investigations of redshift and
luminosity-dependent effects on obscuration relative to orientation
and testing Unification schemes. The medium-$z$ 3CRR sample is
described in Section~2. The supporting, non-X-ray, data are presented
in Section~3. The analysis of new and existing \chandra\ X-ray data is given in
Section~4, and the relation of the X-ray, radio, and infrared
properties to obscuration and orientation in Section~5. The discussion
of the results in the context of Unification models is presented in
Sections~6 and 7, and a summary is given in Section~8. 
Throughout the paper we assume a $\Lambda$CDM cosmology with
$H_{0}=69.6$~km~s$^{-1}$~Mpc$^{-1}$, $\Omega_M=0.286$, and
$\Omega_{\Lambda}=0.714$ \citep{2014ApJ...794..135B}

\section{The Sample}
\label{sec:sample}

The 3CRR catalog \citep{1983MNRAS.204..151L} contains a complete,
178~MHz radio-flux limited sample of 173 quasars and radio galaxies
brighter than 10~Jy extending to $z = 2.5$.  At these low frequencies,
the emission, whether for radio galaxies or quasars, is dominated by
extended radio lobes, which are optically thin and emit nearly
isotropically, resulting in a sample that is unbiased by the effects
of orientation and obscuration.  The radio morphologies, radio sizes,
and lobe separations are well known for all 3CRR sources. The
higher-frequency 5~GHz radio data (where the radio core emission is
more pronounced than in low frequency radio) provide an independent
estimate of orientation via the radio core fraction
\citep{1982MNRAS.200.1067O} \rcd~$\equiv F_{\rm core}$(5~GHz)$/F_{\rm
    lobe}$(5~GHz), which is defined as the ratio of the beamed radio
  core (unresolved on arcsecond scales) to the extended, nearly
  isotropic emission from the radio lobes. Additionally the lengths of
  the radio jets provide an estimate of the AGN ages
  (e.g. \citealp{2015A&A...575A..80P}).

\cite{2013ApJ...773...15W} studied the $1<z<2$ subset
of the 3CRR sources (hereafter the high-$z$ sample). In this work we
focus on the $0.5<z<1$ 3CRR sample (hereafter the medium-$z$ sample;
Table~\ref{tb:obs}), 
which includes 44 sources. All 3CRR sources at $z>0.5$ are of
Fanaroff-Riley type II  (FR~II; \citealp{1974MNRAS.167P..31F}) 
characterized by powerful double radio lobes (often extending far
beyond the host galaxy) that are edge-brightened (i.e., having bright
hotspots at the ends of their lobes) and showing high radio powers
$P_{\rm 178\,MHz} > 10^{26.5}$~W\,Hz$^{-1}$\,sr$^{-1}$. At these
redshifts, the radio luminosities are comparable to those of the most
powerful radio sources found at earlier epochs ($2.5 < z < 6$) when
the quasar activity peaked. This ensures that the objects in our
sample are powerful AGN.  Studies of redshift and size distributions
\citep{1993MNRAS.263..139S} and the detection of X-ray emission
(Section~\ref{sec:xdat_analys}) confirm the presence of an AGN in all
sources. All 3CRR sources in the medium-$z$ sample have now been
observed with \chandra.

The medium-$z$ sample can be divided into two types:
\begin{enumerate}
\item{broad-line radio galaxies and quasars, hereafter collectively 
  referred to as quasars (14 objects),}
\item{narrow-line radio galaxies (NLRGs; 29 sources) and 1
  low-excitation radio galaxy (LERG),  hereafter collectively referred to as
  radio galaxies.}

\end{enumerate}

Most of the 3CRR quasars and radio galaxies have steep radio spectra
($\alpha > 0.5$; $F_{\nu} \propto \nu^{-\alpha}$) and extended,
lobe-dominated radio emission at 178~MHz. However, six quasars and
three NLRGs with steep radio spectra have compact ($<$10~kpc)
structure. These are compact steep spectrum (CSS) sources
\citep{1998PASP..110..493O, 1985A&A...143..292F, 2012ApJ...760...77A},
thought to be either evolutionarily young or to have their jets
frustrated due to interaction with large amounts of material. 
There are no strongly beamed,
radio core-dominated quasars in this sample, with only two marginally
core-dominated radio sources (3C~380 with log~\rcd~$=0.18$ and 3C~216 with
log~\rcd~$=0.15$ ), so beamed emission is not dominant across the sample.

The one LERG (Hine \& Longair 1979) in the sample is 3C~427.1. LERGs
have inherently weak (unobscured) X-ray \citep{2009MNRAS.396.1929H}
and mid-IR emission \citep{2006ApJ...647..161O} and are possibly
powered by a radiatively inefficient accretion flow
\citep{2009MNRAS.396.1929H, 2006ApJ...642...96E, 2006ApJ...647..161O,
  2001A&A...379L...1G}. They reside mostly in FRI-type radio sources
\citep{1974MNRAS.167P..31F} or lower-radio power ($P_{\rm
  178\,MHz}$\,$\sim$\,10$^{26.5}$\,W~Hz$^{-1}$\,sr$^{-1}$) FRII-type
sources \citep{2002A&A...394..791C, 2004MNRAS.349..503G}. 3C~427.1 is
one of the latter.

The medium-$z$ 3CRR quasars and NLRGs occupy the same $\sim$1.5~dex
range in 178~MHz radio luminosity density(Figure~\ref{fg:LrLx}~{\it left}),
where $10^{35.2} < L_{\nu}(178~{\rm MHz})$/erg~s$^{-1}$\,Hz$^{-1} <
10^{36.6}$. The K-S test reveals no difference in $L_{\nu}(178~{\rm MHz})$
distributions of quasars and NLRGs. In comparison, the distribution of
radio luminosities in the high-$z$ 3CRR sample
\citep{2013ApJ...773...15W} is narrower (1~dex) and covers higher
radio luminosities $10^{35.9} < L_{\nu}(178~{\rm
  MHz})$/erg~s$^{-1}$\,Hz$^{-1} < 10^{36.8}$ (Figure~\ref{fg:LrLx}~{\it left: inset}).

Because of their high flux densities ($F_{\nu}(178\,{\rm MHz}) >$\,10\,Jy), 
their high luminosities, the complete nature of the survey, and the
availability of comprehensive multi-wavelength data, the 3CRR sources
constitute an excellent AGN sample with which to study
orientation-based effects and test Unification schemes.  One caveat is
that only 10\% of the AGN population is radio loud, and caution is
required when generalizing results to the whole AGN
population. Additionally, the radio-emitting plasma may affect the
opening angle of the torus \citep{1995A&A...298..395F} and contribute
to the X-ray emission (especially in strongly beamed  sources).

%
%
%

\section{Supporting data}
\label{sec:supp_dat}

\subsection{Radio Data}
\label{sec:radio_dat}

The 5~GHz radio core and extended radio lobe flux densities have been
compiled from the literature. The radio core, the total (core+lobe)
flux densities, and the total luminosity densities at 5\,GHz
($L_{\nu}(5\,{\rm GHz})$) are presented in
Table~\ref{tb:obs}.  The radio core fraction \rcd\ 
is also given, which 
is often used as an orientation indicator in radio-loud 
AGN \citep{1982MNRAS.200.1067O, 1993ApJ...407...65G} and gives, in
general, an estimate of the inclination angle accurate to within
$\pm$20\deg\ \citep{1995ApJ...448L..81W}
and in the case of the $z \ge 1$ 3CRR sources to $\pm$10\deg\ or less
\citep{2016ApJ...830...82M}. When available, we used the same
reference for the radio core and extended radio lobe luminosities when
calculating \rcd . Other references were checked for flux
consistency.  For sources with no 5\,GHz data, the 8\,GHz flux density
was used to estimate the 5\,GHz flux density, assuming a radio
spectral index of $\alpha = 0.7$ (typical of extended emission from
radio galaxies e.g., \citealp{1999MNRAS.304..271D}) for the radio
lobes and $\alpha = 0.3$ (a compromise between a flat spectrum and
steep spectrum core) for the radio core (where $F_{\nu} \propto
\nu^{-\alpha}$). 
In the medium-$z$ 3CRR sample, log~\rcd\  spans values
from 0.15 to less than $-$3.5, which according to
\cite{2016ApJ...830...82M} correspond to a range of viewing angles
measured in respect to the radio jets that range between 8\deg\ (close to pole-on) 
and 90\deg\ (perpendicular to the jet or edge-on to the torus). 

\vspace{-0.6mm}
\subsection{IR Data}
\label{sec:ir_dat}

\spitz~\citep{2004ApJS..154....1W} IRAC and MIPS photometry has been
obtained and analyzed for the full 3CRR sample
(\citealp{2008ApJ...688..122H} for $z > 1$ and
\citealp{2006ApJ...647..161O} for $z<1$ sources). IRS spectroscopy is
also available for sources in the redshift range $0.4<z<1.4$
(\citealp{2007ApJ...660..117C} for $0.4<z<1.2$ and
\citealp{2010ApJ...717..766L} for $1<z<1.4$). 
All sources were observed in the far-IR during {\it Herschel}
guaranteed time (PI Barthel) with PACS and SPIRE, and their IR SEDs
(including 2MASS, WISE, \spitz, and {\it Herschel} data) were analyzed
by \citet{2015A&A...575A..80P} for $z>1$ and
\citet{2016AJ....151..120W} for $z<1$.
The near-to-mid-IR (3--40$\mu$m) emission, dominated by the AGN, was
found to be stronger in quasars than in radio galaxies, while the
far-IR component, dominated by dust heated by star formation, is
comparable in strength for the two classes. The difference in the mid-IR
emission is consistent with 
the Unification scenario where the hot dust from the inner regions
is directly visible in face-on quasars but obscured in NLRGs,
which are viewed edge-on to the dusty torus. At $z<1$, an additional
population of weak  mid-IR AGN was found (LERGs and weak-MIR sources),
possibly representing a different class of objects (nonthermal,
jet-dominated with low accretion power) or different evolutionary
stage from the mid-IR-bright sources \citep{2006ApJ...647..161O} .

\section{X-ray Data}
\label{sec:xdat_analys}

Of the 44 sources in the present sample, fourteen (7 quasars 3C~207,
254, 263, 275.1, 309.1, 334, 380 and 7 NLRGs 3C~6.1, 184, 228, 280,
289, 330, 427.1) had archival \chandra\ observations. One of these
(3C~184) was also observed with {\it XMM}.  For the remaining 30
sources \chandra\ ACIS-S observations of 23 sources were obtained (PI
Kuraszkiewicz, proposal number 14700660) between 2013 Jan~21 and
Oct~20 followed by observations of 7 sources (PI Massaro proposal
number 15700111; between 2014 Jun~15 and 2015 May~20
\citep{2018ApJS..234....7M}.  The exposure times were set to ensure
detection at flux levels expected for NLRGs and quasars as a function
of redshift.  Sub-arrays were used for the brightest quasars to avoid
pileup.  The nuclei of all but two sources (3C~220.3, 441) were
detected.  There is a wide range of signal-to-noise ($S/N$) ratios
extending from a few net counts for the faintest NLRGs to $\sim$10000
net counts for the brightest quasars found in the archive (3C~207,
334).  All \chandra\ observations are listed in Table~\ref{tb:obs}
together with references to the existing \chandra\ and XMM data and
spectral analysis.

The X-ray emission from radio-quiet AGN includes multiple components
\citep{1993ARA&A..31..717M}: 1) an accretion-related power-law
dominating the X-ray emission of luminous broad-lined AGN, absorbed in
narrow-lined AGN, 2) a soft-X-ray excess, linked to the accretion
disk, 3) reflected emission from hot and/or cold material surrounding
the nucleus, 4) emission lines \citep{2003A&A...402..849O}, and 5)
scattered nuclear light. Components 3,~4, and 5 become  more significant
in  AGN  with higher inclination angles, where 
the direct nuclear light is obscured \citep{1993ARA&A..31..717M}.

The X-ray emission of radio-loud AGN additionally includes
non-thermal, synchrotron and/or inverse-Compton components associated
with radio structures: jets, lobes, and hot spots (resolved with the
high spatial resolution of \chandra\, \citealp{2012ApJ...745...84W,
  2009A&ARv..17....1W,2006ARA&A..44..463H}) and jets dominating the
emission of beamed, core-dominated (face-on), broad-lined, radio-loud
AGN, which have on average $\sim 3 \times$ higher soft X-ray
luminosity and harder spectra in comparison with the radio-quiet AGN
(\citealp{1981ApJ...245..357Z,1987ApJ...323..243W, 1987ApJ...313..596W, 1990ApJ...360..396W, 2011ApJ...726...20M},  but see \citealp{2020MNRAS.496..245Z}
who suggest a corona-jet interpretation).  The amount of X-ray excess jet
emission, above that expected from radio-quiet AGN, depends on the radio
spectral slope and radio loudness  
and is a factor 0.7--2.8$\times$\ higher for radio-intermediate quasars,
$\sim$3$\times$\ higher for radio-loud quasars
and 3.4--10.7$\times$\ higher for extremely radio-loud (strongly beamed
sources). 
The X-ray jet-linked emission is less beamed (has a lower bulk Lorentz
factor) than the radio jet emission \citep{2011ApJ...726...20M}.  
At $z<1$, it is possible to distinguish or place limits on the
relative contributions from nuclear jet- and accretion-related X-ray
components
\citep{2009MNRAS.396.1929H,2006ApJ...642...96E,2006MNRAS.366..339B} in
the higher signal-to-noise X-ray data. However, none of the sources in
our sample are strongly beamed in our line of sight, and therefore the
X-ray jet component is not expected to be strong
\citep{1999MNRAS.309..969H}.

\subsection{Data Processing and Analysis}
\label{sec:dat_proc}

The \chandra~data, both new and archival, were reprocessed using the
standard pipeline to apply the latest calibration products appropriate
for their observation dates and assure that processing was uniform
across the sample.
The counts for each source were extracted from a 2\farcs2 radius
circle (to enclose the full point-spread function) centered on the
radio core coordinates or when not available on the AGN X-ray position
(Table~1). The background counts were extracted from an annulus
with inner and outer radii of 15\arcsec\ and 35\arcsec, respectively centered on
the AGN, then scaled for area and subtracted to determine the net counts
for each source. In a few sources, the background annulus was adjusted 
to exclude bright incidental X-ray sources.  For nine sources 
(3C~172, 175, 228, 263, 265, 268.1, 330, 334, 340, 337, 441) for which the
radio lobes showed substantial and extended X-ray emission, 
two circular regions with a 15\arcsec\ radius lying outside the extended
emission were used for background count estimation.

We use the following X-ray energy bands: broad ($B=0.5-8.0$~keV),
soft ($S=0.5-2.0$~keV), and hard ($H=2.0-8.0$~keV).  The 
broadband net source and background counts
for each source are given in Table~\ref{tb:flux} (columns 3 and
4). The soft and hard band source and background counts were used to
calculate hardness ratios (column 14). 

\subsection{Initial Flux Estimate from Srcflux: low count sources}
\label{sec:srcflux}

To provide uniformly derived X-ray fluxes, the X-ray data for
\chandra-observed sources were initially processed with {\it Srcflux},
a program in CIAO (Chandra Interactive Analysis of Observations;
Fruscione et al. 2006), which is particularly useful in calculating
the net count rates and fluxes in low count sources, where spectral
fits are poorly constrained.  {\it Srcflux} performs no spectral fits
but instead fits the normalization based on the observed count rate
for an assumed source spectrum and source and background regions. 
This results in 
fluxes  estimated in a consistent manner, particularly for sources with
highly absorbed or complex spectra and low $S/N$ data. 
We assumed a power-law spectrum with a canonical photon index
$\Gamma=1.9$ \citep{2007ApJ...665.1004J,1993ARA&A..31..717M} and
Galactic absorption characterized by the equivalent hydrogen column
density from \citet{1990ARA&A..28..215D} and quoted in
Table~\ref{tb:obs}.  The same source and background regions as
described in Section~\ref{sec:dat_proc} were used.  The ``srcflux
fluxes'' and ``srcflux luminosities'' (K-corrected assuming a
power-law with $\Gamma$=1.9) in the 0.5--8~keV range are given in
Table~\ref{tb:flux} (columns~5 and 6).  We will use these values as
X-ray fluxes and luminosities throughout the paper for sources with
$<$10 counts.

\subsection{Spectral Fits}
\label{sec:spec_fit}

We performed X-ray spectral modeling of all sources in the sample with
{\it Sherpa} \citep{2001SPIE.4477...76F}, a modeling and fitting
package in CIAO.  We used the Levenberg-–Marquardt optimization method
with the $\chi^2$ statistic including the Gehrels variance function,
which allows for a Poisson distribution for low-count sources.  First
a power-law with a canonical photon index $\Gamma=1.9$ and Galactic
absorption was fit to binned spectra. For sources with $\ge$30 net
counts, a second step including intrinsic absorption (\nh ) at the redshift of
the source was added to the fit.  For sources with \gax700 net
counts (mostly quasars), the power-law photon index was then freed in
the final spectral fit.  The results of the analysis are presented in
Table~\ref{tb:flux}.  Significantly detected \nh , indicating
absorption in excess of the Galactic column density, is most likely
absorption intrinsic to the quasar associated with the nucleus and/or
the host galaxy. Although unlikely, a contribution from absorption by
intervening material/sources along the line-of-sight cannot be ruled
out.

For eight archival sources with more than a few thousand counts resulting in
 $\delta >$5\% pileup, the CIAO pileup model ({\it jdpileup}) was included in
the spectral fits.
The pileup fraction is reported in Table~\ref{tb:complex}, and
the pileup corrected fluxes are presented in Table~\ref{tb:flux}.

\subsection{Complex Spectra}
\label{sec:complex}

Several NLRGs displayed complex X-ray spectra.
In particular 3C~265, 280, 330 showed excess 
soft X-ray emission above the absorbed primary power-law.
This soft excess may be due to thermal emission from a surrounding
cluster, emission from the accretion disk or inner region of the jets, intrinsic
AGN emission visible due to partial covering of the AGN, or scattered
emission from material close to the nucleus. For example 3C~265, a
NLRG, shows a Sy1 spectrum in visible, polarized light
\citep{2006A&A...455..773V}, implying scattered intrinsic AGN emission
which may extend to the X-rays.  Four galaxies, 3C~184, 265, 330, 427.1
show a strong 6.4~keV fluorescent \fe\ line arising from the
reflection of the hard X-ray power-law on the (relatively) cold matter
in an accretion disk or torus (Fabian et al. 2000 and references
therein). Higher $S/N$ {\it XMM-Newton} data of 3C~184 require a soft
excess and \nh\ $= 4.9^{+2.2}_{-1.2}\times
10^{23}$~cm$^{-2}$ \citep{2006MNRAS.366..339B}. 3C~265 and 3C~330 display
both a soft excess and a \fe\ line.  The soft excess and the \fe\ line
become pronounced in the heavily obscured sources, when the
contribution of the intrinsic power-law is significantly reduced.

The fits of complex spectra were built up using an iterative
approach. In the initial stage, a model consisting of an absorbed
power-law was fitted as described in Section~\ref{sec:spec_fit}. If an
\fe\ line was visible in the fit residuals, the power-law was then
fitted over the energy range excluding the line.  Next the fitted
parameters were frozen, and an additional component, the soft excess
or the \fe\ line, was added to the model. 
For two sources that required both the \fe\ line and the soft excess, the
soft excess component was added and fitted first. 
The soft excess was modeled as an unabsorbed power-law with a fixed
$\Gamma=1.9$. Then the slope was freed and fitted,
after which the primary, intrinsic power-law normalization and
\nh\ were freed and fitted. 
The \fe\ line was modeled with a Gaussian and fitted iteratively.  
First the \fe\ line amplitude was fitted assuming an approximate peak
position at 6.4~keV (restframe), appropriate for neutral \fe, and an
arbitrary full width at half maximum (FWHM) of 0.2~keV. Then the line
amplitude and FWHM were freed and fitted simultaneously. For 3C~265,
where the iron line is particularly strong, the position of \fe\ peak
was also fitted.  As a next step, the \fe\ line parameters were frozen, 
and all other non-iron parameters (i.e., intrinsic power-law normalization,
\nh , soft excess power-law slope and normalization) were refitted
followed by another \fe\ line-only fit.
The resulting best-fit parameters for the soft excess and the \fe\
line (in the complex spectra) are given in Table~\ref{tb:complex}.  For
pileup sources, the pileup fraction is also shown in this table.  Spectral
fits for all complex sources are plotted in Figure~\ref{fg:complex}.

\subsection{Intrinsic \nh\ and \lx\  estimation from Hierarchical Bayesian Model}
\label{sec:HBM}

Here we explore the Hierarchical Bayesian Modeling
(hereafter HBM), to constrain individual and whole-sample {\it
  intrinsic} luminosities and column densities and  the
obscured and Compton-thick AGN fractions in the sample.  HBM is a
statistical method that facilitates inferences about a population
based on individual objects and their observations (and vice versa). 
Our hierarchical model has three layers: The bottom layer is formed by
the observed data (X-ray spectra) and is fixed. The middle layer
contains the parameters for each object, namely their {\it intrinsic}
X-ray luminosity $L$(0.5--10~keV) and column density \nh .  The top
layer describes the $L$(0.5--10~keV)  and \nh\ distributions of the
whole population. The HBM 
simultaneously finds posteriors on individual and population
parameters. 
It ``shrinks'' individual parameter estimates toward the population
mean, which lowers RMS errors and naturally deals with large
uncertainties and upper limits. The uncertainty is determined via
nested sampling. The Appendix presents a detailed explanation of the
method.

To apply HBM to our sample, we first used Bayesian inference in
analyzing the X-ray spectra assuming flat, uninformative priors for
$L$(0.5--10~keV) and \nh , which were then updated using Bayes'
theorem to posterior priors taking into account parameter
distributions of the whole population. The Bayesian X-ray Analysis
module was used (BXA; \citealp{Buchner2014}) for {\it Sherpa}
\citep{2006SPIE.6270E..1VF}, assuming an AGN with intrinsic
obscuration and taking into account Compton scattering and iron
fluorescence (\texttt{BNTORUS} model; \citealp{2011MNRAS.413.1206B})
with an added warm-mirror power-law (same as the scatterd light
component in Section~\ref{sec:complex}).  All normalizations had wide
log-uniform priors, and the intrinsic photon index was assigned a
Gaussian prior centered at $1.95$ with standard deviation of
$0.15$. The warm-mirror normalization can reach up to $10\%$ of the
intrinsic AGN power-law component.  The above setup is described
e.g., by \citet{Buchner2014}. The analysis gives preliminary posterior
probability distributions for the parameters in the middle layer,
i.e., the individual posterior HBM $L$(0.5--10~keV) and \nh , which
are shown in Figures~\ref{fg:HBM_NH},~\ref{fg:HBM_Lx}, and
\ref{fg:HBM_NH_L}.
The effect of the HBM is that weak observations are informed by
well-constrained observations, which indicate probable parameter
values. For example, extremely high luminosities are suppressed.  The
HBM median values of intrinsic $L$(0.5--10~keV) and \nh\ for each
source are given in Table~\ref{tb:BXApost}.

\section{Comparison of X-ray Properties of quasars and NLRGs}
\label{sec:results_comparison}

\subsection{Observed X-ray luminosity and Hardness Ratio}
\label{sec:result_Lx}

\begin{figure}
\epsscale{1.0}
\plottwo{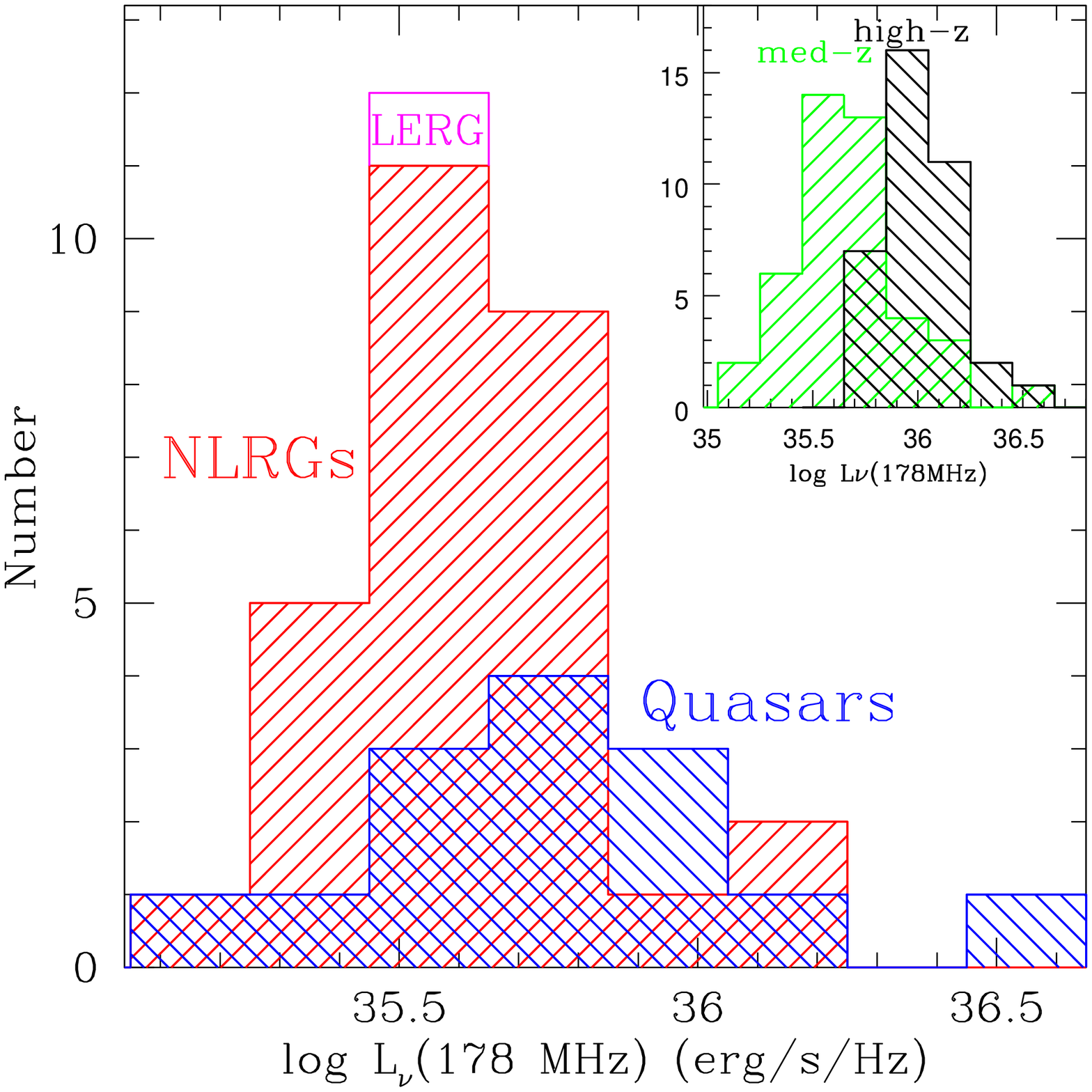}{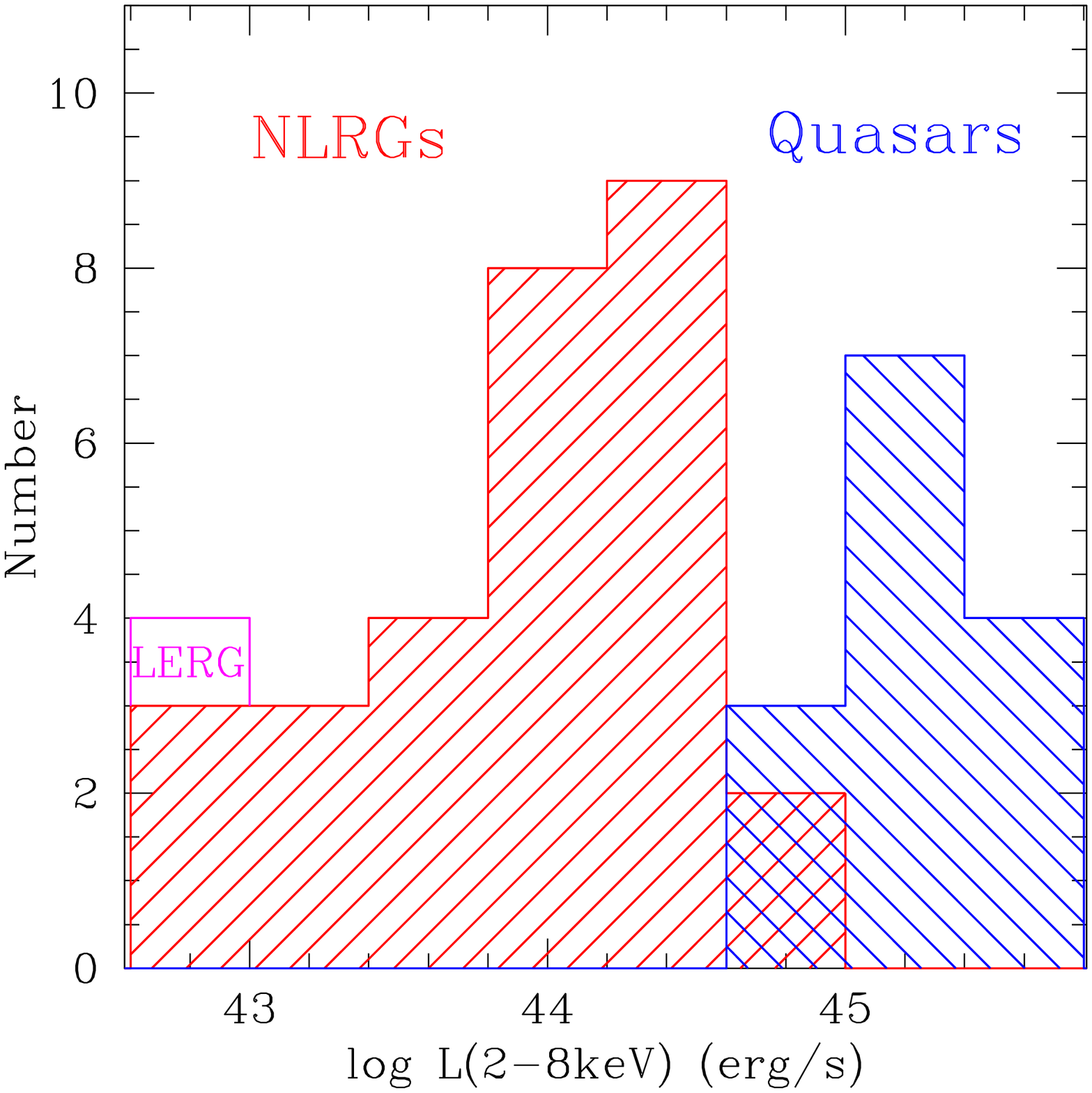}
\caption{{\it Left:} The distribution of the total, rest-frame,
  178~MHz radio luminosity density $L_{\nu}(178\,{\rm MHz})$ for the
  medium-$z$ ($0.5 < z < 1$) 3CRR sample. The blue histogram shows
  quasars, the red histogram NLRGs, and the LERG is plotted in
  magenta. The inset in the upper right corner, shows the distribution
  of $L_{\nu}(178\,{\rm MHz})$ for all sources in the medium-$z$ 3CRR
  sample in green and the high-$z$ 3CRR sample
  \citep{2013ApJ...773...15W} in black. {\it Right:} The distribution
  of the 2--8~keV hard-X-ray luminosity uncorrected for intrinsic
  absorption for the medium-$z$ sample. Quasars are plotted in
  blue, NLRGs in red, and the LERG in magenta. The range of radio
  luminosities is narrow ($\sim$1.5~dex) with quasars and NLRGs having
  similar 178~MHz luminosities (a proxy for intrinsic AGN luminosity).
  The hard-X-ray luminosity distribution covers a wider range
  ($\sim$3~dex) with the NLRGs being 10-1000 times fainter in X-rays
  than quasars, which can be explained by higher intrinsic
  obscuration in NLRGs (Section~\ref{sec:result_Lx}).}
\label{fg:LrLx}
\end{figure}

The quasars and NLRGs in the medium-$z$ 3CRR sample have comparable
(to within $\sim$1.5~dex) extended 178~MHz radio luminosities
(Section~\ref{sec:sample}, Figure~\ref{fg:LrLx} {\it left}) which
implies similar intrinsic AGN luminosities.  In contrast, the
2$-$8~keV luminosities, uncorrected for intrinsic
absorption, hardly overlap (Figure~\ref{fg:LrLx}\,{\it right}), where
the NLRGs show 10--1000~times lower hard-X-ray luminosities than
quasars, suggesting higher obscuration in NLRGs.  The widely different
apparent luminosities are consistent with the Unification model, where
the nuclei of NLRGs are thought to be viewed edge-on through a dusty,
torus-like structure and so are observed through higher amounts of
obscuration than the quasars.

The X-ray hardness ratio, defined as ${\rm HR} \equiv
\frac{(H-S)}{(H+S)}$, where $H$ and $S$ are the (2$-$8~keV) and
(0.5$-$2~keV) counts respectively,
is often used as a measure of intrinsic \nh\ and is particularly
useful in lower-count sources, where spectral fitting is not
possible. Higher (harder) hardness ratio indicates higher obscuration,
and lower (softer) hardness ratio lower obscuration.  A few sources in
our sample have low counts, so we determined the hardness ratios using
the Bayesian Estimation of Hardness Ratios (BEHR) method
\citep{2006ApJ...652..610P}, which accounts for the Poissonian nature
of the data and correctly deals with non-Gaussian error propagation,
appropriate for both the low- and high-count regimes.  These hardness
ratios are provided in Table~\ref{tb:flux} (column 14), and their
distribution is presented in Figure~\ref{fg:HRdist}.  All quasars
(plotted in blue) have soft HR$<0$, with the mean HR$=-0.36\pm0.15$,
consistent with an AGN power-law with $\Gamma=1.5^{+0.32}_{-0.33}$ and
low obscuration.  3C~196, the quasar with the hardest HR ($=-0.07$) in
the sample has intermediate obscuration of \nh $= 3\times
10^{22}$~cm$^{-2}$ and is classified as a Type~1.8 based on its
optical spectrum. 
In contrast, the NLRGs (plotted in red) span a wide range of hardness ratios
$-0.6<HR<0.9$,  implying a large range of intrinsic obscuration. 

\subsection{Hardness Ratio vs.  X-ray Absorption} 
\label{sec:result_Xabs}

Figure~\ref{fg:NHvsHR} shows the dependence of the observed hardness
ratio on \nh\  compared to trends
expected from modeling. The intrinsic \nh\ was obtained from X-ray
spectral fitting (Sec.~\ref{sec:spec_fit},~\ref{sec:complex}) for
sources with at least 30~cts.  Most of the sources lie on the track of
the pure absorbed power-law models with photon index $1.5 < \Gamma < 2.2$
and $10^{20} < $~\nh /cm$^{-2}$~$< 10^{25}$.  The exceptions are
3C~172, 184, 265, 280, 330, 427.1, for which the hardness ratios are softer
than predicted from an absorbed power-law with the measured
\nh. Apart from 3C~172, for which low $S/N$ (32~cts) does not allow
for a complex fit, these are the sources with complex spectra
discussed in Section~\ref{sec:complex}. These sources' spectra include
an additional soft excess component (besides the heavily obscured
power-law and the \fe\ line) which is possibly due to scattered
nuclear light or extended X-ray emission from gas surrounding the
nucleus, galaxy cluster or the radio/X-ray jet.

The \chandra\ data of 3C~184 had too few counts ($\sim$48) to justify
a complex fit, but the higher $S/N$ {\it XMM-Newton} data require a
soft excess, high column density (\nh~$=4.9^{+2.2}_{1.2}\times
10^{23}$~cm$^{-2}$), and an \fe\ line \citep{2006MNRAS.366..339B}.

\subsection{Hardness Ratio vs. \lx\  Dependence}
\label{sec:result_L_HR}

The observed (uncorrected for \nh ) broad-band 0.5--8~keV X-ray
luminosities are plotted against hardness ratios in
Figure~\ref{fg:LxvsHR}. These are compared with a pure absorbed
power-law model ($\Gamma=1.9$; red dotted curve), and other absorbed
power-law models ($\Gamma=1.5, 2.2$) with an added soft excess
componenf of varying strength (0.1\%, 1\%, 5\% of intrinsic light;
blue and green curves). The quasars have high observed \lx\ and soft
hardness ratios indicating low obscuration. The NLRGs show a broad
range of hardness ratios and lower observed \lx , indicating a varying
degree of intrinsic \nh\ and varying amount of scattered/extended
light emission. The majority of medium-$z$ NLRGs lie on 
models that include an absorbed power-law and a soft excess of varying
strength, which makes their HR softer than the ones expected from
a pure absorbed power-law model. 
Figure~\ref{fg:HRvsLxLr} is a modified version of
Figure~\ref{fg:LxvsHR} where the observed, 0.5--8~keV X-ray luminosity
is normalized to the total radio luminosity at 178~MHz (a surrogate
for intrinsic AGN luminosity).  Quasars show
$L$(0.5$-$8~keV)/$L$(178~MHz)$>1$ and soft HR.  NLRGs have
$L$(0.5$-$8~keV)/$L$(178~MHz)$<1$ and a range of HR.  A group of five
soft NLRGs (3C~6.1, 175.1, 228, 263.1, 455) has almost quasar-like
$L$(0.5$-$8~keV)/$L$(178~MHz) $\sim 1$, indicating low obscuration. These
will be discussed further  in 
Sections~\ref{sec:lowNH_NLRGS}~and~\ref{sec:expl_lowNH_NLRGS}.

\subsection{Comparison with the high-$z$ 3CRR sample}
\label{sec:medz-highz_comp}

The mean quasar hardness ratio of the medium-$z$ 3CRR sample
($-$0.36$\pm$0.15) is comparable to that of the high-$z$ 3CRR sample
($-$0.44$\pm$0.20). However, the median is harder ($-$0.34
vs. $-$0.51) implying flatter primary power-law slopes ($\Gamma$=1.5
vs. 1.9) and/or higher \nh\ in the medium-$z$ quasars, which may
reflect the fact that low \nh\ is easier to measure at lower redshifts
as the softer X-rays move into the \chandra\ observed band. Piled-up
quasars, present at medium-$z$, will also contribute to the harder
mean and median hardness ratios.  For NLRGs, the mean hardness ratio
(0.14$\pm$0.43) is comparable, within uncertainties, to the high-$z$
NLRG mean (0.10$\pm$0.45), while the median is softer (0.10 vs. 0.26)
implying a higher fraction of NLRGs with low \nh\ in the medium-$z$
sample (discussed in Section~\ref{sec:lowNH_NLRGS}).

The median 2--8~keV luminosity, uncorrected for intrinsic column
density, is 6$\times$ lower for NLRGs than quasars in the medium-$z$
sample (10$^{44.4}$~erg\,s$^{-1}$ vs. 10$^{45.2}$~erg\,s$^{-1}$
respectively), while it was $\sim100\times$ lower in the high-$z$ 3CRR
sample \citep{2013ApJ...773...15W}, suggesting a higher number of
NLRGs with low obscuration in the medium-$z$ sample.

\section{Discussion}
\label{sec:results}

\subsection{Orientation-Dependent Obscuration}
\label{sec:nh_rcd}

The ratio of the observed broad-band 0.5--8~keV X-ray luminosity
(uncorrected for \nh) to the total radio luminosity at 178~MHz (\lxlr,
where $L_{\rm R}=\nu L_{\nu}(178\,{\rm MHz})$ is calculated from
the 178~MHz flux densities in \citealp{1983MNRAS.204..151L}), which is a
measure of gas obscuration, is plotted in Figure~\ref{fg:LxLr_HRvsR}a
as a function of the radio core fraction \rcd\ (an orientation
indicator). Sources with lower obscuration have higher \lxlr\ ratios
and show larger values of \rcd , i.e., are preferentially seen at
lower viewing angles in respect to the radio jet (i.e., face-on to the
torus). Sources with higher obscuration (lower \lxlr ) have lower
\rcd\ and so are preferentially viewed perpendicular to the radio jet
(i.e., edge-on to the torus).
To show this explicitly, the intrinsic column
density \nh\ (estimated from X-ray spectral fits in
Sec.~\ref{sec:spec_fit} ~and~\ref{sec:complex}), is plotted as a function
of the radio core fraction \rcd\ in Figure~\ref{fg:LxLr_HRvsR}\,b.  The
strong relation between \nh\ (and \lxlr) and \rcd\ implies that
obscuration is strongly dependent on orientation and increases with
increasing viewing angle. This relation is consistent with the
orientation-dependent obscuration invoked by the Unification model and
agrees with our results for the high-$z$ 3CRR sample
\citep{2013ApJ...773...15W}.
However, 
at intermediate viewing angles $-3<\log$~\rcd\ $<-2$  NLRGs
with a broad range of \nh\ exist. These include typical, obscured
NLRGs with \nh~$>10^{22}$~cm$^{-2}$ and a peculiar class of NLRGs, not
present in the high-$z$ 3CRR sample, with low intrinsic column densities
\nh~\lax $10^{22}$~cm$^{-2}$.
These low-\nh\ NLRGs cannot be explained by a simple Unification model
dependent solely on orientation, and suggest that a second parameter
(clumpy torus, different obscurer, or different $L/L_{\rm Edd}$ ratio)
is needed.  
We will focus on the low-\nh\ NLRGs next. 

\subsection{Observational properties of low-\nh\ NLRGs}
\label{sec:lowNH_NLRGS}

One quarter of NLRGs (3C~6.1, 175.1, 228, 263.1, 455),  
or 14\% of the medium-$z$ 3CRR sample, have low 
 \nh\ ($10^{21}-10^{22}$~cm$^{-2}$), similar to the
unobscured BLRGs and quasars. As a result of low obscuration, these
NLRGs have soft, quasar-like hardness ratios ($HR<0$) and the highest
\lxlr\ amongst the NLRGs
(Figure~\ref{fg:LxLr_HRvsR}\,a). These low-\nh\ NLRGs have
intermediate core fractions ($-2.7 < $~log~\rcd~$ < -2$) and so are
likely viewed at angles skimming the edge of the accretion disk or
torus. No such sources were present in the high-redshift 3CRR sample,
where all NLRGs had higher intrinsic column densities of
log~\nh/cm$^{-2}$~$> 22.7$ and log \rcd~$< -2$.  Although it is
easier to measure low \nh\ values in sources at medium-$z$ than at
high-$z$ (as the softer-energy X-rays move into the \chandra-observed
band), the spectra also become more complex, often including an
additional, soft excess component.
The low-\nh\ NLRGs have enough counts (90--1700) to model the soft excess,
but
none of them required one. We hence conclude that the low intrinsic 
column densities in these NLRGs are measured correctly and are not underestimated
due to the lack of soft excess modeling in low-$S/N$ spectra.

The low-\nh\ NLRGs show relatively low mid-IR~(30$\mu$m) emission when 
compared to their radio emission.
The $L$(30~$\mu$m)/$L$(178~MHz) ratios are the lowest in the sample
(Figure~\ref{fg:MIRfigs}\,a), $\sim$10 times lower than in
quasars. Because the X-ray emission is also weaker by a factor of 10 
relative to radio emission (see Figure~\ref{fg:LxLr_HRvsR}\,a), the
$L$(30~$\mu$m)/$L$(2$-$8~keV) ratios are comparable to those of
quasars (see Figure~\ref{fg:MIRfigs}\,b).
The spectral energy distributions (SEDs) of low-\nh\ NLRGs show no
infrared or big blue bump (see Figures~4,~5,~7 of
\citealp{2016AJ....151..120W}), and the specific star formation rates
are close to those of normal galaxies \citep{2016AJ....151..120W}.

Three of the low-\nh\ NLRGs were observed with {\it HST}
(3C~6.1, 228, 263.1). The optical images show compact host galaxies with
no visible dust lanes \citep{1997ApJS..112..415M}.
The optical SDSS spectra are red (3C~175.1, 228, 263.1, 455). 3C~6.1
shows a weak optical continuum dominated by the host galaxy (visible
$\lambda$4000\AA\ absorption feature) with an 8~Gyr old stellar
population \citep{1979ApJ...231..307S}.  3C~455 has conflicting
optical types (Type~1 or 2) in the literature, but we classify this
source as a Type~2 based on the spectrum presented by
\cite{1994ApJS...91..491G}, which shows a weak continuum and no broad
H$\beta$ emission line.  H$\alpha$, however, was not covered to check
for intermediate Type~1.8 or 1.9.

\subsection{Understanding the low-\nh\ NLRGs}
\label{sec:expl_lowNH_NLRGS}

Possible scenarios that can explain low column densities,
lack of broad emission lines, and weak IR emission in the
low-\nh\ NLRGs are the following:

\begin{itemize}

\item These are {\bf ``true'' type 2} objects
  \citep{2002A&A...394..435P,2003ApJ...583..632T,2010ApJ...714..115S,2014MNRAS.437.3550M}, 
  which show no detectable broad lines and have low X-ray
  absorption. In such sources the broad line region (BLR) has faded
  due to recent weakening of the continuum or has not formed due to
  very low $L/L_{\rm Edd}$~$\ll 10^{-2}$
  \citep{2000ApJ...530L..65N}.  In the latter scenario, such low
  $L/L_{\rm Edd}$ ratios would result in more than $100-1000\times$
  weaker 0.5--8~keV luminosities (as accretion disk SEDs strongly depend
  on $L/L_{\rm Edd}$ -- see e.g., \citealp{1996astro.ph..9180C}, Fig.~1),
  but the  values of \lxlr\ only few-to-$10\times$
  lower than in quasars (Figure~\ref{fg:LxLr_HRvsR}\,a) rule out this
  scenario.

\item The {\bf obscuration is non-standard}, caused not by a torus but
  by a dust lane or a host galaxy disk mis-aligned with the dusty
  torus (as in the red 2MASS AGN; \citealp{2009ApJ...692.1180K,
    2009ApJ...692.1143K}) which would result in \nh\ $\leq
  10^{22}$~cm$^{-2}$. Such column density is low enough not to obscure
  significantly the intrinsic X-ray emission or the IR emission from the
  dusty torus, but is sufficient to hide the AGN's optical+UV
  continuum and the broad emission line region. In this scenario, the
  weak IR emission in low-\nh\ NLRGs cannot be easily explained 
  unless the dusty torus is absent.
  
\item {\bf Low $L/L_{\rm Edd}$ ratio}.  The low-\nh\ NLRGs are found
  at intermediate viewing angles ($-3 <$~log \rcd\ $<-2$), 
  together with  NLRGs that have
  higher column densities of $10^{22.5} <$~\nh/cm$^{-2}$~$<10^{23.5}$
  (Figure~\ref{fg:LxLr_HRvsR}\,b). 
  Therefore, a scenario is needed in which clouds with a large range of
  column densities may exist at such viewing angles.
  \citet{2008MNRAS.385L..43F} showed that the distribution of column
  densities of the gas and dust clouds surrounding an AGN is a function
  of  $L/L_{\rm Edd}$,  and only clouds with 
  \nh /cm$^{-2}$~ $\ge 5\times 10^{23}\times$~$L/L_{\rm Edd}$ can withstand the
  AGN's radiation pressure, while the lower-\nh\ clouds are blown away.
  At low $L/L_{\rm Edd} \sim 0.01$, clouds with column
  densities ranging from $\sim 10^{22}$~cm$^{-2}$ to Compton-thick can
  exist, whereas at high $L/L_{\rm Edd} \sim 1$, only
  those with Compton-thick column densities will survive.
  Applying the scenario to our sample, the NLRGs with low \nh\ must
  have low $L/L_{\rm Edd}$, while NLRGs with high \nh\ (viewed at
  similar, intermediate angles) have high $L/L_{\rm Edd}$. The
  scenario is further confirmed by the finding that the X-ray
  luminosities, uncorrected for intrinsic absorption, are comparable
  for the low-\nh\ and high-\nh\ NLRGs having the same intermediate
  viewing angles ($0.2< L$(0.5$-$8~keV)/$L$(178~MHz)$< 1$;
  Figure~\ref{fg:LxLr_HRvsR}\,a), despite significantly different
  column densities. Thus we conclude that indeed the low-\nh\ NLRGs
  have lower intrinsic X-ray luminosities and hence lower $L/L_{\rm
    Edd}$ than the high-\nh\ NLRGs.

\item {\bf Weak IR emission due to low $L/L_{\rm Edd}$}. At
  high $L/L_{\rm Edd}$ (strong big blue bump), only a torus
  that is compact and Compton-thick can withstand the intense UV radiation and
  strong winds. Dust in such a compact geometry will strongly radiate
  in the near-to-mid-IR, producing an SED with a strong IR bump
  \citep{1992ApJ...401...99P}.  At lower $L/L_{\rm Edd}$,  where
  the big blue bump is weaker and provides less illuminating flux for
  the torus, the torus may become clumpy and extended, resulting in a
  weaker IR bump \citep{2015A&A...583A.120S, 2010A&A...515A..23H,
    2008ApJ...685..147N,2008ApJ...685..160N, 2003ApJ...590..128K}. 
  Figure~\ref{fg:LMIR_Ledd_fig} shows the dependence of the
  30~$\mu$m luminosity on the 2--8~keV {\it intrinsic} luminosity (estimated from
  the HBM model), which is related to the $L/L_{\rm Edd}$
  ratio (e.g., \citealp{1996astro.ph..9180C} Fig.~1). Both luminosities are normalized
  by the extended radio luminosity $L$(178~MHz) to remove any redshift
  dependence on the IR and X-ray luminosities. There is a strong
  correlation between $L$(30~$\mu$m)/$L$(178~MHz) and $L$(2--8~keV)/$L$(178~MHz)
  with a probability 0.01\% of occurring by chance in both the
  generalized Kendall rank and Spearman rank tests. The correlation indicates that
  higher $L/L_{\rm Edd}$ sources (=higher intrinsic $L$(2$-$8~keV)) have
  stronger mid-IR luminosities. The low-\nh\ NLRGs have relatively
  low $L/L_{\rm Edd}$  (i.e., $L$(2$-$8~keV)/$L$(178~MHz)~$< 1$) and
  so their weak mid-IR emission can be explained as due to low
  $L/L_{\rm Edd}$.

  Two sources, 3C~220.3 and 3C~343, do not lie on the overall
  correlation in Figure~\ref{fg:LMIR_Ledd_fig}. They have relatively
  low $L/L_{\rm Edd}$ but show strong mid-IR emission.  3C~220.3 is
  lensing a background submm galaxy \citep{2014ApJ...790...46H}, which
  results in amplification of its IR luminosity. We suggest that
  perhaps 3C~343 may also be lensing a background galaxy. 
  Another outlier is 3C~172,
  with high $L/L_{\rm Edd}$ and low mid-IR
  emission.  The low IR emission can be explained by either extreme
  Compton-thick obscuration of \nh~$> 10^{25}$~cm$^{-2}$ or low
  amounts of dust due to 1000$\times$ lower than Galactic
  dust-to-gas ratio.  The former explanation is not suported by our
  low $S/N$ X-ray spectral modeling, which gives \nh~$\sim
  10^{24}$~cm$^{-2}$. 
  The latter is in conflict with typical AGN dust-to-gas ratios which
  are 1--100 times lower than Galactic \citep{2001A&A...365...28M,
    2012A&A...539A..48M, 2016A&A...586A..28B} with a few exceptions
  having this ratio a few times higher \citep{2017MNRAS.469..693O,
    2010ApJ...725.1749T}.

\end{itemize}

In summary, 
a simple Unification model where obscuration changes only with
orientation cannot fully describe the observed multiwavelength
properties of the medium-$z$ 3CRR sample, and a range of $L/L_{\rm
  Edd}$ ratios, extending to low values, is required to explain the
existence and the properties of the low-\nh\ NLRGs. In contrast, 
the multiwavelength properties  of the high-$z$ 3CRR sample were explained by
pure Unification, suggesting that $L/L_{\rm Edd}$ 
had a narrower range and possibly higher values in comparison with the
medium-$z$ sample, allowing orientation effects to dominate the
observed properties of the sample.

\subsection{Heavily obscured NLRGs}
\label{sec:CT}

\subsubsection{Compton-thick (CT) candidates}
\label{sec:O3_Lx}

\begin{figure}
\epsscale{1.0}
\epsscale{1.0}
\plottwo{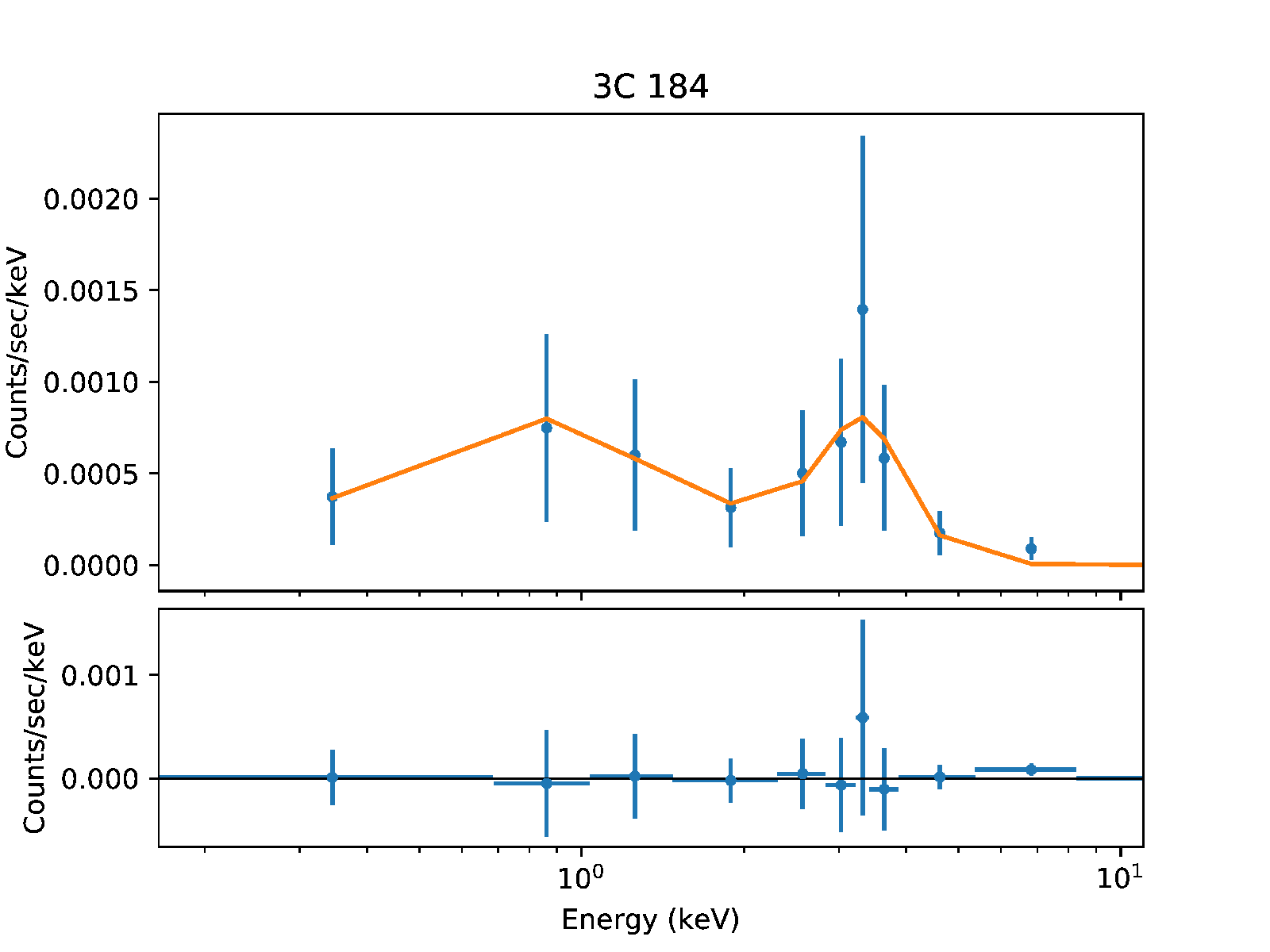}{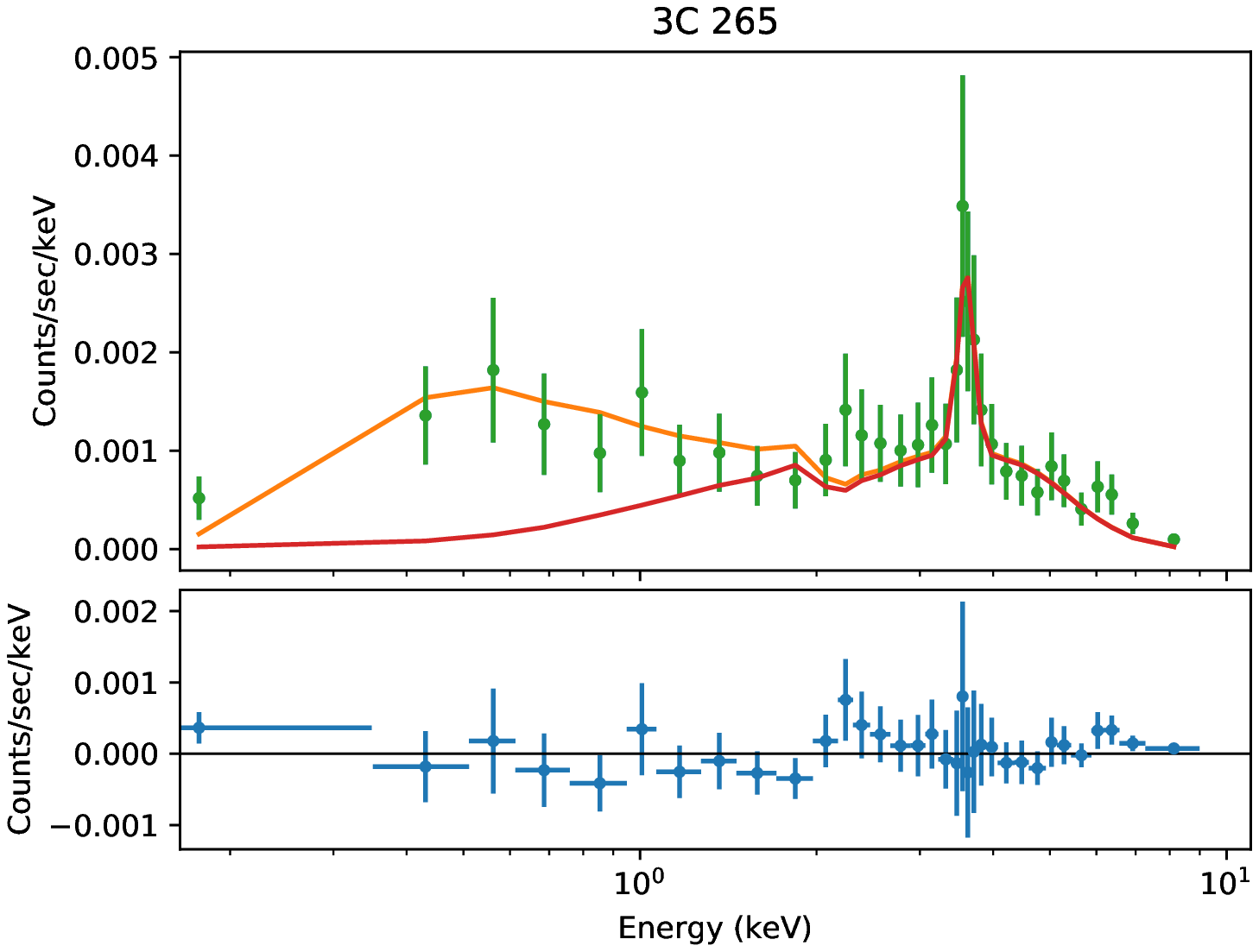}
\plottwo{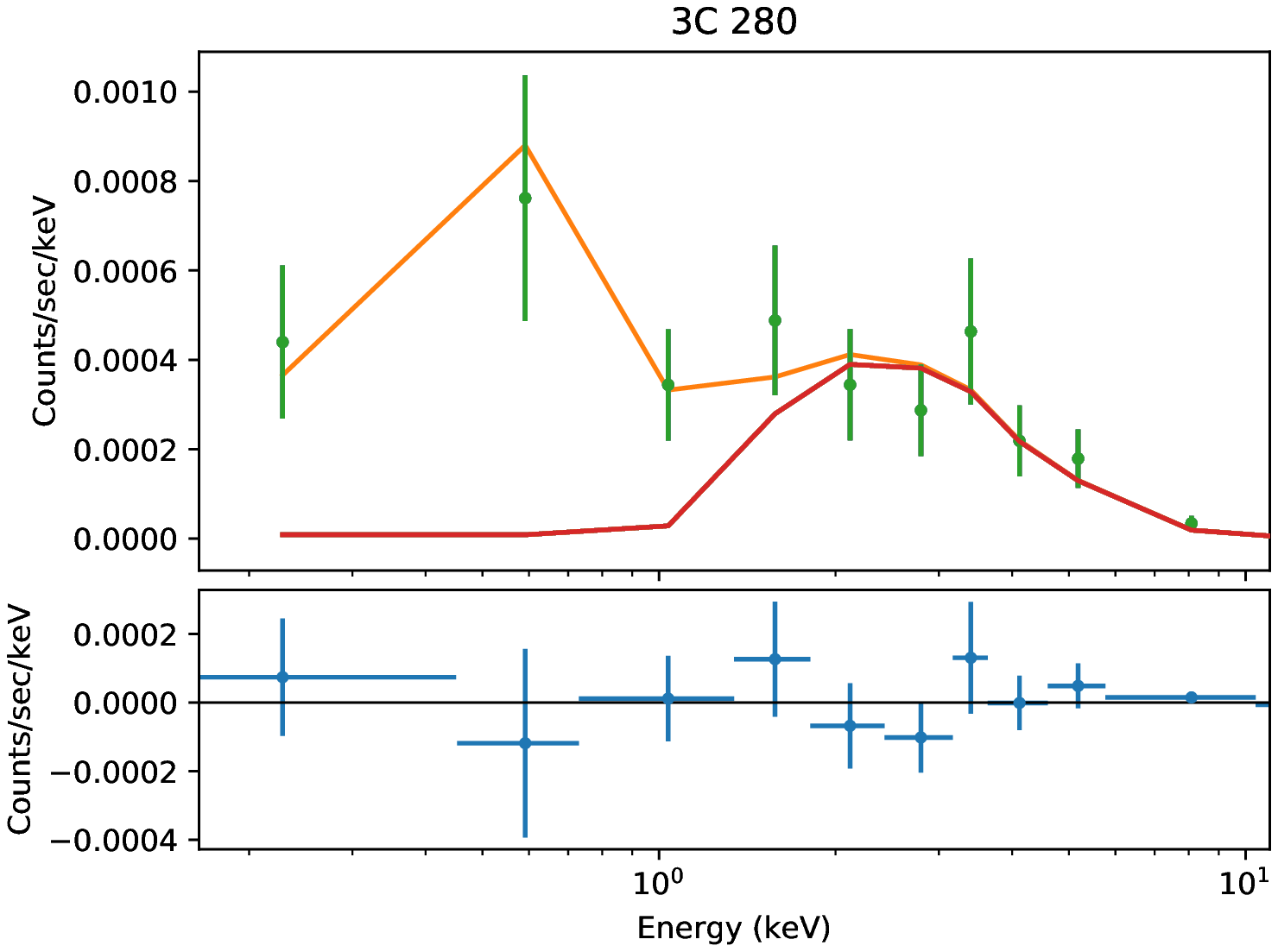}{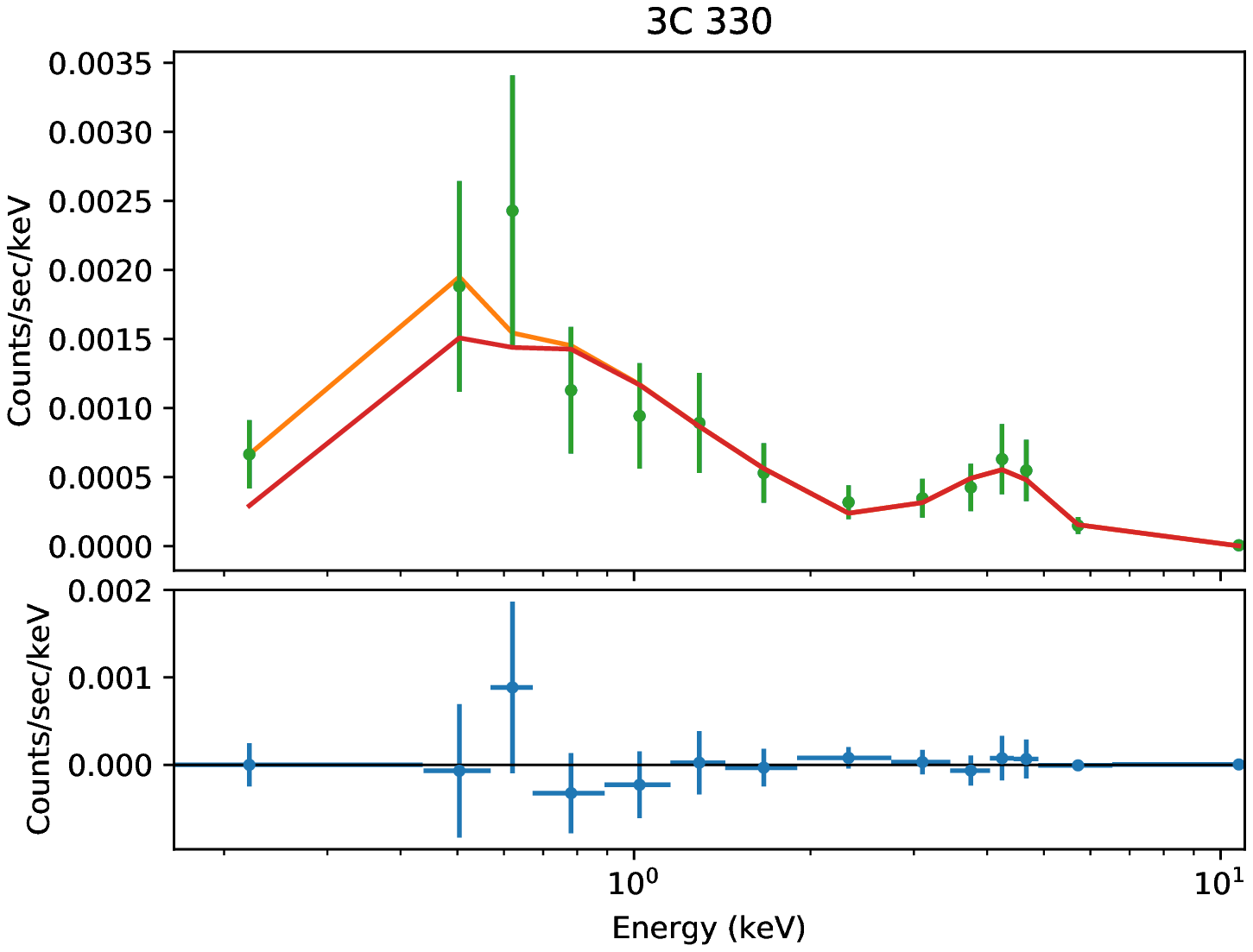}
\epsscale{0.5}
\plotone{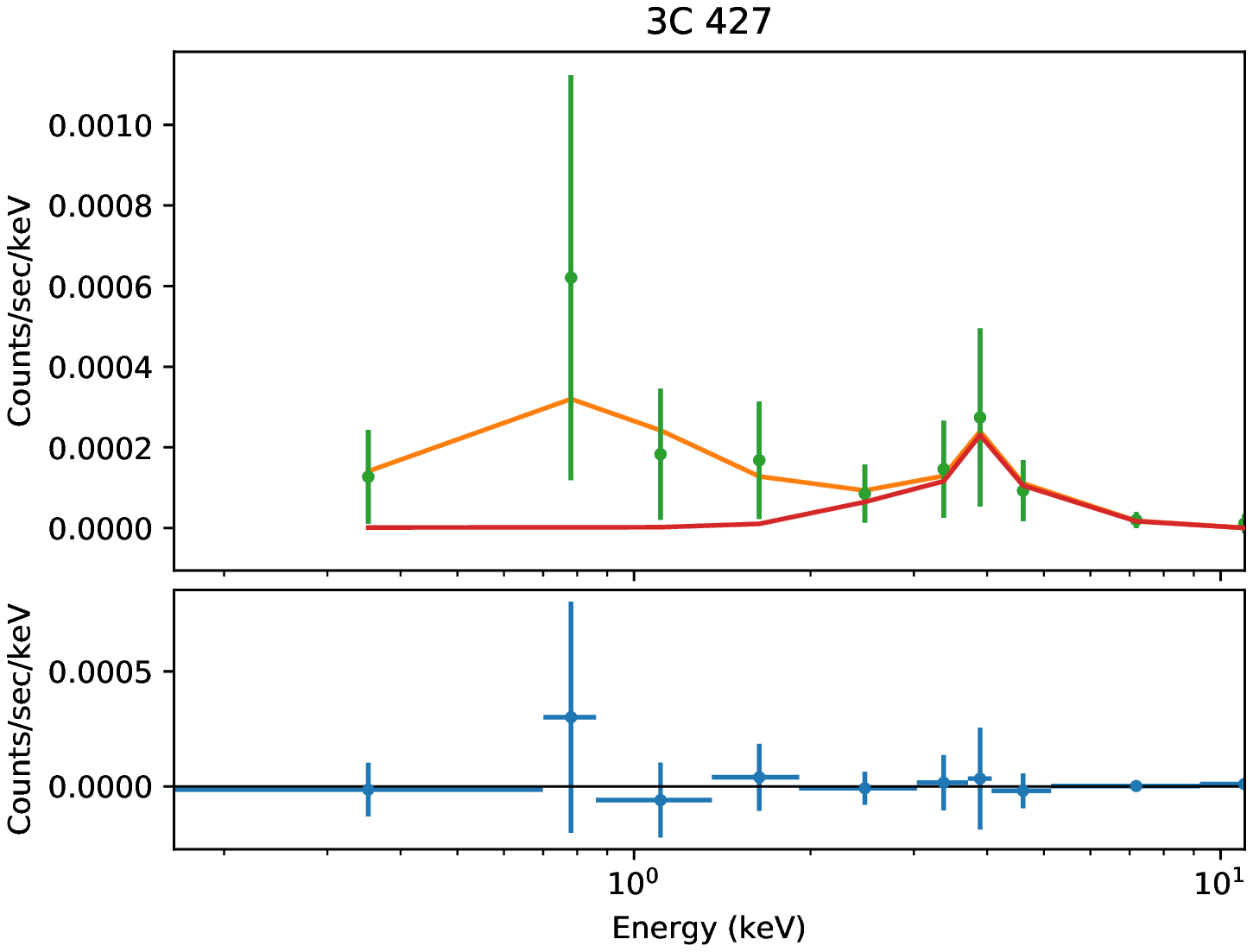}
\caption{{\it Sherpa} X--ray fits (yellow line) to sources with
  complex spectra (in observed frame) modeled with a power-law
  ($\Gamma=1.9$) absorbed by Galactic and intrinsic \nh , Fe~K$\alpha$
  line, and a soft excess modeled as an unobscured power-law
  ($\Gamma>1.9$ - see Table~\ref{tb:complex}) in 3C~265, 280, 330, 427.1 
  where also a fit without the soft excess is shown in red.  Fit
  parameters are given in Table~\ref{tb:flux} and
  \ref{tb:complex}. Residuals to the fits are shown in the bottom
  panel of each figure.}
\label{fg:complex}
\end{figure}

The luminosity of the \oiii$\lambda$5007 emission line (hereafter
$L$(\oiii)  was found to track the radio and intrinsic X-ray
luminosities for both the Type~1 and Type~2 AGN
\citep{1997MNRAS.286..241J, 1994ApJ...436..586M}. It is often used as
an indicator of intrinsic AGN luminosity \citep{1999ApJ...522..157R,
  2006A&A...455..173P} and has little or no inclination dependence at
high luminosities \citep{2004MNRAS.349..503G, 1997MNRAS.286..241J}.
The observed hard X-ray luminosity, on the other hand, is strongly
dependent on obscuration (especially at high \nh ), so the ratio of
$L$(\oiii)/$L$(2--8~keV) is often used to discriminate between
Compton-thin and Compton-thick (hereafter CT) sources
\citep{1999ApJ...522..157R, 2006A&A...455..173P}.
Figure~\ref{fg:O3LxvsR} shows the ratio $L$(\oiii)/$L$(2--8~keV) plotted against
the radio core fraction \rcd . The $L$(\oiii) values are from
\citet{2004MNRAS.349..503G} and are shown in
Table\ref{tb:obs}. Seventeen sources have actual \oiii\ measurements,
and for the remainder $L$(\oiii) was estimated from either the
\oii$\lambda$3727 emission line or the 151~MHz radio luminosity
(3C~292, 427.1). 
The dotted line in Figure~\ref{fg:O3LxvsR}, shows the dividing line
between Compton-thin and CT sources reported by
\cite{2011ApJ...736..104J} and seven sources: 3C~184, 220.3, 225B,
277.2, 280, 441 (all NLRGs, with L(\oiii ) estimated from L(\oii ))
and 3C~427.1 (a LERG) appear to be CT.  The HBM analysis
(Section~\ref{sec:HBM}, Table~\ref{tb:BXApost},
Figure~\ref{fg:HBM_NH}) gives CT probabilities ranging from
24\%--80\%.
3C~220.3, 225B, 277.2, 441 have too few counts ($<15$) to model the
X-ray spectrum to confirm the high \nh , but HBM implies CT
obscuration (see Figure~\ref{fg:HBM_NH}, Table~\ref{tb:BXApost}).  The
low-$S/N$ {\it Chandra } spectrum of 3C~184 (48~cts) shows a strong Fe
K$\alpha$ line (Figure~\ref{fg:complex}), implying heavy obscuration
(the reflection component becomes stronger as the intrinsic power-law
weakens with increasing obscuration), while the higher-$S/N$ {\it XMM}
data 
are fitted with high \nh , strong K$\alpha$ line, and a soft excess
\citep{2006MNRAS.366..339B}.
The {\chandra} X-ray spectrum of 3C~280 (117~counts) is modeled with a
strong soft excess and intermediate \nh\ (Figure~\ref{fg:complex},
Table~\ref{tb:flux}).

Five of the above CT candidates have measured $L$(30~$\mu$m)
\citep{2016AJ....151..120W} and all except 3C~427.1 have
log~$L$(30~$\mu$m)/$L$(2--8~keV)~$>1.8$ (Figure~\ref{fg:MIRfigs}b).
The {\it Spitzer}/IRS spectra of 3C~184 and 3C~441 show strong
9.7\,$\mu$m silicate absorption (an indicator of large amounts of
dust) with \tauSi~$ > 0.3$. 3C~280, despite being a CT candidate, has
no 9.7\,$\mu$m silicate absorption \citep{2011A&A...531A.116G}.  All
the above CT candidates, except for 3C~427.1, have log~\rcd~$<-3$
indicating inclination angles larger than 80\deg\ (i.e., orientation
edge-on to the torus).

For 3C~427.1, 
neither the \oiii\ nor \oii\ luminosity was measured directly, and
$L$(151~MHz) was used to estimate $L$(\oiii). To confirm this source's CT
nature we consider other CT  indicators.
3C~427.1 has the lowest $L$(0.5$-$8~keV)/$L$(178~MHz) in the sample
(Figure~\ref{fg:LxLr_HRvsR}a) suggesting low observed \lx, which may
be either due to CT obscuration, low $L/L_{\rm Edd}$, or X-rays being
recently turned off (the source is a LERG which harbors a low
luminosity AGN). The $L$(30~$\mu$m)/$L$(178~MHz) ratio is the lowest
in the sample (Figure~\ref{fg:MIRfigs}a). Low mid-IR emission is
typical for LERGs \citep{2016AJ....151..120W}, where low $L/L_{\rm
  Edd}$ results in weaker big blue bump emission, which provide less
illuminating flux for the circumnuclear dust emitting in the IR  
(Figure~\ref{fg:LMIR_Ledd_fig} shows the dependence between $L/L_{\rm
  Edd}$ and $L$(30~$\mu$m)/$L$(178~MHz)). Alternatively the mid-IR
emission could be suppressed by heavy obscuration, \nh~$\gtrsim
10^{25}$~cm$^{-2}$, resulting in a strong 9.7\,$\mu$m silicate
absorption which cannot be checked in the source for lack of a {\it
  Spitzer}/IRS spectrum. 
However, the presence of a strong Fe K$\alpha$ line (Figure~\ref{fg:complex}) 
implies that 3C~427.1 is indeed heavily obscured.

\subsubsection{CT and borderline CT candidates with low [OIII] emission}
\label{sec:otherCT}

There are five NLRGs that have low \rcd\ values implying extreme
(edge-on) inclination angles characteristic of the CT sources
described above but having low, Compton-thin $L$(\oiii)/$L$(2--8~keV)
ratios. Despite this, these sources are possibly CT or borderline CT
as explained below: 

3C~55: {\it Sherpa} modeling of the 15-count \chandra\ spectrum does
not give an estimate of intrinsic \nh , but HBM finds CT \nh\ 
and a 97\% probability of the source being CT
(Table~\ref{tb:BXApost}). The {\it Spitzer}/IRS spectrum shows strong
9.7~$\mu$m silicate absorption indicating heavy absorption.  Also the
$L$(30~$\mu$m)/$L$(2$-$8~keV) and $L$(0.5$-$8~keV)/$L$(151~MHz) ratios
have values consistent with other CT sources in the sample. The source
is definitely CT. 

3C~172: 
both {\it Sherpa} modeling of the 30-count X-ray spectrum and HBM imply
\nh\ consistent with CT (Table~\ref{tb:flux}
and~\ref{tb:BXApost}) with a 52\% probability of 
being CT.  This strong CT candidate is unusually weak in the IR
(Figure~\ref{fg:MIRfigs}a), having no {\it Herschel} detection, and
showing an SED with no IR bump \citep{2016AJ....151..120W}.  No {\it
  Spitzer}/IRS spectrum is available to estimate the strength of
the 9.7~$\mu$m silicate absorption.

3C~330:  the X-ray spectrum (143~counts) is modeled with a highly
absorbed (but not CT) power-law (Table~\ref{tb:flux}) and includes a
soft excess and medium strength iron K$\alpha$ line
(Figure~\ref{fg:complex}). HBM estimates a high but not CT \nh\ and
a  9\% CT probability (Table~\ref{tb:BXApost}).  The {\it
  Spitzer}/IRS spectrum shows a moderate 9.7~$\mu$m silicate
absorption \citep{2016AJ....151..120W}. The source is definitely
heavily obscured but not CT.

3C~337:  the low-$S/N$ spectrum (10~counts) does not allow for an
\nh\ estimate from {\it Sherpa} modeling. HBM gives an estimate of high but
not CT obscuration and a 13\% probability that this source is CT
(Table~\ref{tb:BXApost}).  No {\it Spitzer}/IRS spectrum is available
to estimate the strength of the silicate 9.7~$\mu$m absorption. The
$L$(30~$\mu$m)/$L$(2$-$8~keV) is in the range of highly obscured
sources (Figure~\ref{fg:MIRfigs}b). 3C~337 is weak in the mid-IR,
having one of the lowest $L$(30~$\mu$m)/$L$(178~MHz) ratios in the sample
(Figure~\ref{fg:MIRfigs}a) implying low $L/L_{\rm Edd}$
(Figure~\ref{fg:LMIR_Ledd_fig}).  
The intrinsic hard X-ray luminosity estimated from HBM
(Table~\ref{tb:BXApost}) is also one of the lowest in the sample, 
suggesting low $L/L_{\rm Edd}$. The source has low
$L$(0.5$-$8~keV)/$L$(178~MHz) and $L$(2$-$8~keV)/$L$(178~MHz) values
within the range of CT sources (Figure~\ref{fg:LxLr_HRvsR}a). 3C~337
is heavily obscured but not CT. 

3C~343 was classified in NED as a quasar   
\citep{1985PASP...97..932S, 1973ApJ...185..739B}, but
\citet{1994ApJS...93....1A} reclassified this source as a Type~2 based
on an optical spectrum that lacks a broad H$\beta$ emission line
(although H$\alpha$ was not covered). Also \citet{1996ApJS..107..541L} found
only narrow \mgii\ and \civ\ emission lines in their spectra.
The low $L$(2$-$8~keV)/$L$(178MHz) and
$L$(0.5--8~keV)/$L$(178~MHz) are in the range of other CT candidates
in the sample (Fig.~\ref{fg:LxLr_HRvsR}a).  The 
log~$L$(30~$\mu$m)/$L$(2$-$8~keV) and
log~$L$(30~$\mu$m)/$L$(178~MHz) are also consistent with other CT
candidates.  Strong 9.7~$\mu$m silicate absorption, visible in the
IRS/{\it Spitzer} spectrum \citep{2016AJ....151..120W} implies
heavy dust obscuration. Contrary to these CT indicators,  
$L$(\oiii)/$L$(2--8~keV) lies below the CT line  (Figure~\ref{fg:O3LxvsR}). 
The low $L(\oiii )$ was measured directly  \citep{2004MNRAS.349..503G}.
3C~343 is a CSS source, where the radio jets are thought to be young
or frustrated by large amounts of material. In the latter case, the
ionizing photons could be trapped by the dense material that is
frustrating the jets, resulting in low [OIII] emission and
a Compton-thin $L$(\oiii)/$L$(2$-$8~keV) ratio. 
The X-ray spectrum has too few counts (18-cts) to estimate
 \nh , but HBM gives a non-CT \nh , one of the lowest
intrinsic X-ray luminosities in the sample (possibly implying low $L/L_{\rm Edd}$),
and a 14\% probability that this source is CT
(Table~\ref{tb:BXApost}).
We conclude 3C~343 is heavily obscured but likely not CT.

Based on our multiwavelength analysis we find nine CT AGN (3C~55, 172,
184, 220.3, 225B, 277.2, 280, 427.1, 441) and three (3C~330, 337, 343)
heavily obscured but not CT objects in the medium-redshift 3CRR
sample. We conclude that 20\% of the sources in this sample are CT,
consistent with the 21\% found for the high-$z$ 3CRR sample
\citep{2013ApJ...773...15W}.

\subsection{Reliability of Compton-thick indicators}
\label{sec:compCTind}

Table~\ref{tb:CT_ind} summarizes the various CT indicators for each of
the CT candidates discussed above and shows that these indicators do
not always agree.  We analyze the reasons and give recommendations for
their use.

The distribution of the most widely used CT indicator
$L$(\oiii)/$L$(2$-$8~keV), where $L$(2$-$8~keV) is X-ray luminosity
not corrected for intrinsic \nh , is plotted in
Figure~\ref{fg:o3_Lx_MIR_Lx_fig}a.  Most (7/9=78\%) of the CT
candidates in our medium-$z$ 3CRR sample lie at
log~$L$(\oiii)/$L$(2$-$8~keV)$\ge-0.25$ the dividing line between the
Compton-thin and CT sources from \citep{2011ApJ...736..104J}.
Exceptions are 3C~55, 172, which together with the three borderline CT
sources (3C~330, 337, 343) show Compton-thin
$L$(\oiii)/$L$(2$-$8~keV). Interestingly, sources that make the CT cut
cover a full range of sample's intrinsic \lx\ (log~\lx~$=43-46$ - see
Table~\ref{tb:BXApost}) which means that they also cover the full
range of $L/L_{\rm Edd}$ in the sample, suggesting that
$L$(\oiii)/$L$(2$-$8~keV) is independent of $L/L_{\rm Edd}$.

The mid-IR (30\,$\mu$m) luminosity, similarly to the $L$(\oiii), is
used as a measure of intrinsic AGN luminosity, hence
$L$(30\,$\mu$m)/$L$(2$-$8~keV) can also be used as an indicator of CT
obscuration.  We plot this ratio as a function of \nh\ in
Figure~\ref{fg:MIR_NH_fig} and find that most of the sources with
\nh\ $> 10^{23}$~cm$^{-2}$ have log~$L$(30~$\mu$m)/$L$(2--8~keV)~$>1$.
The distribution of $L$(30\,$\mu$m)/$L$(2$-$8~keV) in
Figure~\ref{fg:o3_Lx_MIR_Lx_fig}b  shows that a value $>1.8$ finds
most CT sources in the sample: five out of seven (71\%) CT candidates
with measured $L$(30~$\mu$m) and one borderline CT source.  Relaxing
this criterion to $>1.2$ finds 6 out of those 7 CT sources (86\%), but
also picks three highly obscured, non-CT NLRGs.  3C~172 is the only
CT source with log~$L$(30~$\mu$m)/$L$(2$-$8~keV)~$<1$ due to the
unusually weak mid-IR emission (see Section~\ref{sec:otherCT}).
The $L$(30~$\mu$m)/$L$(2$-$8~keV)~$>1.8$  is therefore a robust CT
indicator, %
but it does not find CT sources exclusively. The ratio
may be enhanced by emission from lensed background galaxies (as in
3C~220.3; \citealp{2014ApJ...790...46H}). The fact that 3C427.1, a low
$L/L_{\rm Edd}$ source, does not make the cut suggests that Eddington
ratio also plays a role. 

The $L$(0.5$-$8~keV)/$L$(178~MHz) or $L$(2$-$8~keV)/$L$(178~MHz) ratios may
also be used to indicate heavy obscuration, where this time the
total radio luminosity at 178~MHz is a measure of intrinsic AGN
luminosity.  The log~$L$(0.5$-$8~keV)/$L$(178~MHz)$<0$ finds all the CT
and the heavily obscured (borderline CT) sources but also includes  one
Compton-thin NLRG (Figure~\ref{fg:LxLr_HRvsR}a).
Highly obscured, non-CT sources will make this cut if their 
$L/L_{\rm Edd}$ is low (resulting in low \lx).

Seven out of nine (78\%) CT sources have low radio core fractions
log~\rcd~$\le -3$, i.e., are highly inclined with viewing angles
$\theta > 80$\deg .  This low \rcd\ value may be used to find CT sources,
however other heavily absorbed sources with \nh~$>$~few$\times
10^{23}$~cm$^{-2}$ (Figure~\ref{fg:LxLr_HRvsR}b) 
also show similarly low \rcd. 

Out of the 9 CT thick candidates, four have IRS/{\it Spitzer} spectra
where three show strong 9.7\,$\mu$m silicate absorption (optical depth
\tauSi~$> 0.3$), while one (3C~280) despite being MIR bright does not.
Strong silicate absorption is a good indicator of heavy {\it dust}
obscuration, but lack thereof does not rule out that the source is
heavily obscured by gas. For example the nearby canonical CT galaxy
NGC~1068 lacks 9.7\,$\mu$m silicate absorption, and only half of the
nearby ($z<0.05$) CT AGN show \tauSi~$ > 0.5$
\citep{2012ApJ...755....5G}.
The strength of the 9.7$\mu$m silicate absorption is also affected
by dust lying farther out in the galaxy or  in a galaxy merger
environment, where the AGNs residing in mergers or post-mergers show the
strongest silicate absorption \citep{2012ApJ...755....5G}.

Summarizing:
\vspace{-0.3cm}

\begin{itemize}
\item log~$L$(\oiii)/$L$(2$-$8~keV)~$\ge -0.25$ is the most robust CT
  indicator of those studied here. 
  It is available for both the radio-quiet and radio-loud sources, finds exclusively (78\%) CT
  sources, and does not depend on $L/L_{\rm Edd}$.  
\item log~$L$(30~$\mu$m)/$L$(2$-$8~keV)~$>1.8$ identifies  71\%
  of CT sources  in the sample, but possibly only the ones
  with high $L/L_{\rm Edd}$ ratios. Lowering this criterion to
  $>1.2$ finds more CT sources
  (86\%), regardless of their $L/L_{\rm Edd}$ ratio. However, either criteria include
  heavily obscured sources that are not CT. This CT indicator is
  affected by $L/L_{\rm Edd}$ and any gravitational lensing.
\item  log~$L$(0.5--8~keV)/$L$(178\,MHz)~$<0$  is an indicator of
  heavy (both CT and borderline CT)  obscuration available for
  radio-loud sources. It is affected by  $L/L_{\rm Edd}$.
\item Low radio core fraction
  log~\rcd~$\leq -3$
  finds 78\% of the CT sources in our sample together with
  the highly obscured but non-CT objects.  It is a good
  indicator of high obscuration, both Compton thick and thin, 
  but only available for sources in which \rcd\ can be measured.
\item Strong 9.7$\mu$m silicate absorption (\tauSi~$ > 0.5$) is an
  indicator of heavy {\it dust} absorption, including by dust lying
  at larger, host-galaxy scales and dust related to mergers.  However,
  sources in which CT obscuration originates from dustless circumnuclear gas
  will not have strong silicate absorption (as 3C~280).
\end{itemize}

Out of all the CT indicators studied above
log~$L$(\oiii)/$L$(2$-$8~keV)~$\ge -0.25$ is the most reliable CT
indicator that finds exclusively CT sources, does not depend on
$L/L_{\rm Edd}$ ratio, and is available for both the radio-quiet and
radio-loud sources.  All other indicators pick up a small fraction of
highly obscured, but not CT sources and depend on $L/L_{\rm Edd}$,
lensing or the location of the obscurer.  None of the indicators find
all the CT sources in the sample, so we recommend examining all
that are available.

\section{The circumnuclear obscurer}

\subsection{Geometry}
\label{sec:geometry}

The strong dependence of \lxlr\ (where $L_{\rm X}$ is uncorrected for
\nh ) and  \nh\ on 
\rcd\ (Figure~\ref{fg:LxLr_HRvsR}, Section~\ref{sec:nh_rcd}) implies
that obscuration in the medium-$z$ 3CRR sample is
orientation-dependent, increases with viewing angle, and, to first
order, is consistent with the standard Unification model.  However, at
intermediate viewing angles, sources with a large range of
\nh\ between $10^{21.3}$ and $10^{23.5}$~cm$^{-2}$ are present, 
suggesting that another parameter independent of orientation (possibly
$L/L_{\rm Edd}$) contributes to the spread in \nh .

The number of sources as a function of \nh\ can provide constraints on
the covering factor of the obscuring material.  If we assume that the
3CRR sources have a geometry in which the obscuring material lies in
the plane perpendicular to the radio jet, and the sources lie randomly
oriented on the sky, the probability of finding a source lying in a
cone of angle $\phi$ is $P(\theta < \phi) = 1-cos \phi$
\citep{1989ApJ...336..606B}. Because 14 out of the 44 (32\%) sources  in the
sample are quasars with \nh\ $<10^{21.5}$~cm$^{-2}$, strong, broad
emission lines, and blue visible colors, this gives an estimate of the
half-opening angle of the obscuring material (torus) of
47\deg$\pm3$\deg . For comparison 60\deg$\pm8$\deg was found in the
high-$z$ sample.

Nine NLRGs are CT candidates characterized by the following
Compton-thick indicators: $L$(\oiii )/$L$(2--8~keV)~$\ge -0.25$,
$L$(30~$\mu$m)/$L$(2--8~keV)~$>1.8$,
$L$(0.5--8\,keV)/$L$(178MHz)~$<0$, low radio core fraction
(log~\rcd~$<-3$), and/or strong 9.7$\mu$m silicate absorption. In the
Unification model, these sources are viewed at the highest inclination
angles through optically thick material lying in the plane of the
torus/accretion disk. The CT candidates represent 20\% (9/44) of the
total sample which implies that CT material covers an angle of 12\deg
$\pm$3\deg\ above and below the equatorial plane of the obscuring
structure as shown in Figure~\ref{fg:geometry}. The remaining
Compton-thin NLRGs (with $10^{22.5} < $~\nh/cm$^{-2}$~$<1.5\times
10^{24}$) cover 21\deg $\pm$2\deg. Intermediate column density
($10^{21.5}<$~\nh/cm$^{-2}$~$<10^{22.5}$) sources including five
low-\nh\ NLRGs (Section~\ref{sec:lowNH_NLRGS}) and 3C~196, a red
broad-line radio galaxy with relatively high \nh~$=2.7\times
10^{22}$~cm$^{-2}$, cover 10\deg$\pm$4\deg. Figure~\ref{fg:geometry}
shows the geometry of the obscuring material found from these simple
estimates, together with the high-$z$ sample
\citep{2013ApJ...773...15W}, and a summary is given in
Table~\ref{tb:geom}. 
In both samples, the covering factor for CT material is similar (same
percentage of CT sources in both samples), but the opening angle of
the torus is smaller for the sample at medium-$z$ than at high-$z$
(47\deg\ vs. 60\deg ) implying that the Compton-thin
($10^{21.5}-10^{-24}$~cm$^{-2}$) part of the obscuring material (torus or
accretion disk wind) is more ``puffed-up'' in the medium-$z$ 3CRR
sample.

\citet{2008MNRAS.385L..43F} have shown that the long-lived gas and dust
clouds in the vicinity of an AGN have a range of column densities
that depend on  $L/L_{\rm Edd}$ where \nh /cm$^{-2}$  $> 5\times
10^{23}\times$~$L/L_{\rm Edd}$. 
\citet{2017Natur.549..488R} studied a sample of local AGN (both Type~1 and
2 with median $z=0.037$) from the all-sky hard-X-ray (14$-$195~keV) {\it
  Swift} Burst Alert Telescope (BAT) survey, for which reliable
estimates of BH mass, intrinsic column densities, X-ray luminosities,
and $L/L_{\rm Edd}$ were obtained. They found  the fraction of CT
sources in their hard-X-ray-selected sample to be $\sim$23$\pm$6\%,
independent of $L/L_{\rm Edd}$, and similar to the fraction in the medium-$z$ and high-$z$
3CRR samples. However, the fraction of Compton-thin but
obscured sources strongly decreases with $L/L_{\rm Edd}$ in their
sample from 0.8 for $L/L_{\rm Edd} < 0.01$ to 0.2 for 
$L/L_{\rm Edd} > 0.1$. \citet{2017Natur.549..488R} therefore
suggested a ``radiation-regulated Unification'' model, where the
covering factor of the Compton-thin gas 
($10^{22}<$~\nh/cm$^{-2}$~$<10^{24}$) increases with decreasing
$L/L_{\rm Edd}$ while the covering factor of the CT gas stays the
same.  In this model, for lower $L/L_{\rm Edd}$ the obscuring
structure (torus/accretion disk wind) is more puffed-up (see their
Fig.~4). Our results for the medium-$z$ sample imply the presence of a
puffed-up torus in the low-\nh\ NLRGs, suggesting that $L/L_{\rm Edd}$
extends to lower values than those in the high-$z$ sample.

\subsection{\nh\ Distribution} 
\label{sec:NH_distrib}

The distributions of the intrinsic \nh\  in the medium-$z$ and
the high-$z$ 3CRR samples are presented in Figure~\ref{fg:NHdist}.
The high-$z$ sample (on the right) shows a bimodal distribution, where
quasars have \nh~$<10^{22.5}$~cm$^{-2}$, consistent with low
obscuration at face-on inclination angles, while the NLRGs show
\nh~$>10^{22.5}$~cm$^{-2}$, implying higher obscuration at higher
inclination angles, consistent with Unification schemes.  There are
two quasars with moderate column densities ($10^{22.5} < $ \nh~/cm$^{-2}$~$ <
10^{23}$) and hard hardness ratios ($0 <$~HR~$<0.5$) in this
sample.
In the medium-$z$ sample, the distributions of quasars and NLRGs
overlap.  Although the quasars show
\nh~$<10^{22.5}$~cm$^{-2}$, similar to quasars at high redshifts, the
NLRGs have a much broader range of column densities that extend to
lower, quasar-like values in the low-\nh\ NLRGs. These NLRGs possibly
have low $L/L_{\rm Edd}$, which allows clouds with low column
density to form in the vicinity of the central engine
(Section~\ref{sec:expl_lowNH_NLRGS}). Such low $L/L_{\rm Edd}$ NLRGs
are missing from the high-$z$ sample.

Although a simple Unification model was sufficient to explain the
X-ray data and the bimodal \nh\ distribution in the high-$z$ sample,
this is not the case in the medium-$z$ sample.  An additional
parameter, a range of $L/L_{\rm Edd}$, is required to explain the
large range of \nh\ in NLRGs
seen at intermediate inclination angles, skimming the edge of the torus or
accretion disk atmosphere/wind.  
As a result, the broad range of $L/L_{\rm Edd}$ smears the
\nh\ distribution for NLRGs, removing the bimodality that was found in
the high-$z$ sample. Turning this argument around, because the
Unification model was sufficient for the high-$z$ 3CRR sample,
producing a bimodal and narrow \nh\ distribution, the $L/L_{\rm Edd}$
ratio must have a narrower range and higher values compared to the
medium-$z$ sample, allowing orientation effects to dominate the
properties of the high-$z$ sample. To test this hypothesis, we
compiled spectra of the high-$z$ 3CRR quasars (from the SDSS archive,
\citealp{1990A&AS...82..339B}, M.~Vestergaard, D.~Stern private
communication) and measured the black hole masses from the widths of
the \civ\ and \mgii\ emission lines.  The masses (measured
in 12 out of 20 high-$z$ quasars) are in the range of $M_{\rm BH} =
10^{7.7}--10^{9.0}$~M$_{\sun}$.  The radio-to-X-ray SEDs, compiled
using data from NED, provided estimates of bolometric luminosities.
The inferred $L/L_{\rm Edd}$ ratios are indeed high $>$0.3, implying
that orientation dominates the observed properties of the high-$z$
sample, and therefore simple Unification suffices.

\subsection{Distribution of intrinsic \lx} 
\label{sec:Lx_distrib}

The distribution of {\it intrinsic} 0.5--8~keV X-ray luminosity
(obtained from HBM modeling; Section~\ref{sec:HBM}) of the medium-$z$
and the high-$z$ 3CRR populations is presented in
Figure~\ref{fig:pop-lum-dist}.  The medium-$z$ sample peaks at lower
\lx\ (mean log~\lx /erg~s$^{-1} = 44.97\pm0.09$) and has a broader
intrinsic \lx\ distribution ($\sigma$=0.51), extending to $\sim$10 times
lower \lx\ values than the distribution for the high-$z$ sample (the
high-luminosity tail in the medium-$z$ sample is due to a
simplistic treatment of piled-up sources for which
\nh~$<10^{21}$~cm$^{-2}$ was assumed). The high-$z$ sample shows a
narrower distribution ($\sigma$=0.27), peaking at higher \lx\ values (mean
log~\lx /erg~s$^{-1} = 45.48\pm0.06$~erg~s$^{-1}$). Because the
intrinsic \lx\ depends on $L/L_{\rm Edd}$ (e.g.,
\citealp{1996astro.ph..9180C} Fig.~1), we interpret the difference as
due to a broad range of $L/L_{\rm Edd}$ in the medium-$z$ sample,
extending to lower values, while the high-$z$ sample has higher
$L/L_{\rm Edd}$ with a narrower range.  The different distributions of
intrinsic \nh\ in the two samples is also consistent with this
scenario (Section~\ref{sec:NH_distrib}). 

\subsection{Obscured fraction} 
\label{sec:CT_fractions}

Obscuration in AGN is highly anisotropic and strongly wavelength
dependent. 
Hence the ``obscured fraction'' defined as the ratio of the number of
obscured AGN (either optically classified Type~2s or those with
\nh~$>10^{22}$~cm$^{-2}$ in X-ray studies)
to all AGN, and its dependence on luminosity and/or redshift differ
for samples selected at different wavebands.  Optical surveys
at low redshift ($z < 0.05$) and low (Seyfert) luminosity
find obscured fractions of $\sim$0.65--0.75 \citep{1995ApJ...454...95M,
  1992ApJ...393...90H, 1982ApJ...256..410L}, implying there are 2--3
times more Type~2s than Type~1s in the local Universe.
High luminosity, radio-selected, and hence unbiased by orientation,  
samples with $z>0.3$ find an optical obscured fraction of $\sim$0.6, consistent
with a torus half-opening angle of $\sim$53\deg\ in Unification models
\citep{2000MNRAS.316..449W} and a luminosity dependence
\citep{2005MNRAS.359.1345G} consistent with the ``receding torus
model''. 
X-ray surveys, sensitive to gas rather than dust obscuration and
probing deeper into the obscured AGN population, find a wide range of
obscured fractions $\sim$0.1--0.8, decreasing with luminosity and
increasing with redshift \citep{2008A&A...490..905H,
  2006ApJ...652L..79T, 2005ApJ...635..864L, 2012ApJ...757..181S,
  2011ApJ...728...58B}, 
although \citet{2017Natur.549..488R}, using a local {\it Swift}/BAT
selected sample, showed that the dependence is primarily with
$L/L_{\rm Edd}$. 

The obscured fraction in the medium-$z$ 3CRR sample studied in this
paper is 0.68 when the optical classification (based on the presence
or absence of the broad emission lines in optical spectra) is used.
However, if the classification is based on X-rays, where
\nh~$=10^{22}$~cm$^{-2}$ is assumed to divide obscured from unobscured
sources, then four out of the five low-\nh\ NLRGs will qualify as
X-ray unobscured, and 3C~196 a quasar with \nh~$>10^{22}$~cm$^{-2}$ as
X-ray obscured, yielding an obscured fraction of 0.61.  The ratio of
X-ray unobscured (\nh~$<10^{22}$~cm$^{-2}$) to Compton-thin obscured
($10^{22} <$ \nh~$<1.5\times 10^{24}$) to CT (\nh~$ >
1.5\times10^{24}$) sources is then 1.9:2:1.

In the high-$z$ 3CRR sample the obscured fraction is lower. It is 0.42
if optical classification is used and 0.5 if X-ray classification is
used, the difference being due to two quasars with \nh~$>10^{22}$~cm$^{-2}$
classified as obscured in X-rays.  The ratio of X-ray unobscured to
Compton-thin obscured to CT sources is 2.5:1.4:1.

The difference between optical and X-ray obscured fractions comes from
four low-\nh, low $L/L_{\rm Edd}$ NLRGs in the medium-$z$ 3CRR sample
and two high-\nh, high $L/L_{\rm Edd}$ quasars in the high-$z$ sample.
In the former case the X-ray obscured fraction is lower in comparison
with the optical obscured fraction, while in the latter case (high
$L/L_{\rm Edd}$ sample) it is higher.

As shown above, the obscured fraction is an inaccurate tool for
measuring the level of obscuration in a sample. Not only does the
obscured fraction depend on the sample's wavelength selection,
luminosity, and redshift but also on whether optical or X-ray
classification is used. It also depends on the sample's $L/L_{\rm
  Edd}$ range, which defines the geometry of the obscuring material
(more puffed-up torus for lower $L/L_{\rm Edd}$, see 
Section~\ref{sec:geometry}) and number of sources with inconsistent
optical and X-ray type.

\subsection{Sources with inconsistent optical and  X-ray types} 
\label{sec:}

The obscured AGN fraction in the medium-$z$ and high-$z$ samples differs
slightly depending on whether the source classification is based on
optical spectra or X-ray data.  \citet{2014MNRAS.437.3550M} studied
AGN with a wide range of redshifts ($0.3<z<3.5$) in the {\it
  XMM}-COSMOS survey and found that setting the dividing line between
Type~1 and Type~2 at \nh~$=10^{21.5}$~cm$^{-2}$ rather than
10$^{22}$~cm$^{-2}$ gives a better correspondence between optical and
X-ray type. However, even then $\sim$30\% of AGN in their sample have
conflicting optical and X-ray classifications.
At dust extinctions $A_{\rm V}=5$--6~mag, the broad emission lines
H$\beta$ and H$\alpha$ are totally obscured. This corresponds to
column densities \nh~$=(0.9--1.1)\times 10^{22}$~cm$^{-2}$ for a
Galactic dust-to-gas ratio. A small (factor of a few) divergence from the
Galactic dust-to-gas ratio will result in inconsistent X-ray and
optical classifications around the dividing Type1/Type2 column density
of \nh~$=10^{22}$~cm$^{-2}$.

\citet{2014MNRAS.437.3550M} found that the AGN with conflicting
optical and X-ray type can be divided into two classes:
\begin{itemize}
\item optical Type~1 and X-ray Type~2 sources, which are
  high-luminosity broad-line AGN with X-rays absorbed by dust-free
  material lying at sub-parsec scales, and
\item optical Type~2 and X-ray Type~1 sources, which are
  low-luminosity, unobscured AGN where the broad lines are probably
  diluted by the host galaxy.
\end{itemize}

The radio-selected 3CRR sample can give further insight into the
nature of sources with inconsistent optical and X-ray classifications:

\begin{enumerate}
\item The high-$z$ sample has 2 quasars (optical Type~1) with high
  column density of \nh =10$^{22.7-23}$~cm$^{-2}$ and HR~$>0$ (X-ray
  Type~2).  These sources (3C~68.1, 325) have high $L/L_{\rm Edd} >
  0.3$ , intermediate viewing angles ($-3 < $~log \rcd~$< -2$), where
  our line of sight is skimming the edge of the accretion disk or
  torus.  In these high $L/L_{\rm Edd}$ sources, the X-rays are
  possibly obscured by gas in the strong, outflowing accretion disk
  wind \citep{2015ApJ...805..122L,2018MNRAS.480.5184N}, while the BLR
  is visible directly. %
  Because of strong UV radiation pressure (high $L/L_{\rm Edd}$), 
  the low-\nh\ gas and dust clouds are blown away.

\item The medium-$z$ 3CRR sample has 5 NLRGs (optical Type~2) with low
  column density \nh~$<10^{22}$~cm$^{-2}$ and quasar-like HR~$<0$
  (X-ray Type~1).  They have intermediate viewing angles, skimming the
  edge of the torus/accretion disk. These NLRGs have low $L/L_{\rm
    Edd}$, which allows low-\nh\ clouds to survive in the vicinity of
  the nucleus and results in a ``puffed-up'' torus (see
  Section~\ref{sec:expl_lowNH_NLRGS}), which can hide the broad-line
  region.
  
\end{enumerate}

In both the high- and medium-$z$ 3CRR samples, AGN with conflicting
optical/X-ray types have intermediate radio core fractions ($-3 <
$~log~\rcd~$< -2$), where viewing angles are skimming the edge of the
accretion disk or torus. In this regime, the torus and accretion disk
are most vulnerable to changes in the $L/L_{\rm Edd}$ ratio.
We find that sources classified as optical Type~1 and X-ray Type~2
(X-ray obscured quasars) have high $L/L_{\rm Edd}$ ratio, where the
strong accretion disk winds obscure the X-rays.  The optical Type~2
and X-ray Type~1 sources (unobscured NLRGs) are low $L/L_{\rm Edd}$
AGN, where the edge or atmosphere of the ``puffed-up'' dusty torus
provides obscuration for both the X-rays and the BLR.

\section{Summary}

A complete, flux-limited (10~Jy), low-frequency (178~MHz)
radio-selected, and so unbiased by the effects of orientation and
obscuration sample of $0.5 < z < 1$ 3CRR sources has now been observed
with \chandra .  The sample includes 14 quasars (no blazars), 29
NLRGs, and 1 LERG with similar (within $\sim$1.5~dex) 178~MHz
extended radio luminosities (i.e., similar intrinsic AGN luminosities).
All sources are radio luminous and of FR\,II type, meaning they all
harbor a powerful AGN in their nucleus.  The radio core fraction
\rcd\ provides an estimate of the viewing angle (with respect to the
radio jet) and so nuclear orientation. We study the dependence of
X-ray, mid-IR, and radio properties on orientation and obscuration and
other central engine parameters ($L/L_{\rm Edd}$), and compare our
results with the high-$z$ ($1< z < 2$) 3CRR sample
\citep{2013ApJ...773...15W} allowing investigation of redshift and
luminosity-dependent effects on obscuration relative to orientation.
We find: 

\begin{enumerate}

\item {\bf Modified AGN Unification.} Quasars in the medium-$z$ ($0.5
  < z < 1$ ) 3CRR sample have high observed X-ray luminosities
  \lx(0.5$-$8\,keV)~$\sim 10^{44.8}$--$10^{45.9}$~erg\,s$^{-1}$, soft
  hardness ratios (HR~$<0$), and high radio core fractions
  (log~\rcd~$>-2$), implying low obscuration (\nh~$<
  10^{22.5}$~cm$^{-2}$) and face-on orientation.  By contrast, NLRGs
  have 10--1000~times lower observed (uncorrected for obscuration)
  X-ray luminosities \lx(0.5$-$8\,keV)~$\sim
  10^{42.9}$--$10^{45.1}$~erg\,s$^{-1}$ despite having similar radio
  luminosities to quasars, a wide range of hardness ratios
  ($-0.6<HR<0.9$) and low radio core fractions (log~\rcd~$<-1.9$).
  This combination of properties implies a range of obscuration
  (\nh~$> 10^{21}$~cm$^{-2}$) and edge-on orientation. These
  properties together with the observed trend of increasing X-ray
  obscuration (expressed by \nh\ and decreasing \lx /$L$(178~MHz)
  with decreasing radio core fraction
  \rcd\ (Figure~\ref{fg:LxLr_HRvsR}), are consistent with the
  orientation-dependent obscuration of Unification models.  However,
  an additional variable, a range of $L/L_{\rm Edd}$, is needed to
  explain the large range of column densities (\nh~$=
  10^{21.5}$--$10^{23.5}$~cm$^{-2}$) found in NLRGs observed at
  intermediate viewing angles ($-3 <$~log~\rcd ~$< -2$) 
  and the sample's broad and smooth distributions of intrinsic column
  densities and intrinsic X-ray luminosities.

\item {\bf $L/L_{\rm Edd}$ dependence on redshift.} In the high-$z$
  3CRR sample \citep{2013ApJ...773...15W}, a simple Unification model
  was sufficient to explain the multiwavelength properties of the
  sample, suggesting a narrower range of $L/L_{\rm Edd}$ and
  orientation effects dominating the observed properties. We estimate
  that $L/L_{\rm Edd}$ is high $>$0.3, possibly due to higher gas
  supply in the denser galaxy environments at higher redshifts. The
  narrow range and higher values of $L/L_{\rm Edd}$ produce a bimodal
  distribution of \nh\ and a narrower distribution of intrinsic X-ray
  luminosities, peaking at higher \lx , in comparison with the
  medium-$z$ sample.
\item {\bf Low-\nh\ NLRGs.} Five NLRGs (3C~6.1, 175.1, 228, 263.1, 445) in
  the medium-$z$ sample show unusually low intrinsic column densities
  ($21 <$~log~\nh /cm$^{-2}$ $<22.1$). They have high, quasar-like
  \lx\ and \lxlr\ ratios, soft HR, low mid-IR emission, and  intermediate
  viewing angles. 
  Analysis of their properties suggest a low $L/L_{\rm Edd}$ 
  resulting in a puffed-up  dusty torus.  

\item {\bf Covering factor.} The medium-$z$ and high-$z$ samples
  have similar fractions of Compton-thick sources ($\sim$20\%), but
  there are relatively fewer quasars (32\% vs. 50\%) and
  more Compton-thin NLRGs in the medium-$z$
  sample (45\% vs. 29\%), implying a larger covering factor of the
  Compton-thin material or a ``puffed-up'' torus.  We interpret this
  as being due to $L/L_{\rm Edd}$ extending to lower values
  ($\sim$0.01) in the medium-$z$ 3CRR sample, allowing lower
  column density material to remain in the ``atmosphere'' of the
  torus.

\item {\bf Geometry of the obscuring material.} Assuming a random
  distribution of source orientation on the sky and a simple geometry
  in which the obscuring material lies in a disk or torus
  perpendicular to the radio jet, we conclude that Compton-thick
  obscuring material extends $\sim$12\deg\ above and below the 
  disk/torus midplane, additional Compton-thin obscuring material extends
  for another $\sim$31\deg\ with the density diminishing with viewing angle, 
  and the remaining $\sim$47\deg (torus opening angle) are
  unobscured. In the high-$z$ sample Compton thick material occupied
  12\deg\ below and above the midplane, Compton-thin material 18\deg,  
  and the torus opening angle was 60\deg.

\item {\bf Compton-thick sources.} Nine NLRGs (3C~55, 172, 184, 220.3, 
  225B, 277.2, 280, 427.1, 441) are likely Compton-thick based on
  several Compton-thick indicators: $L$\oiii/$L$(2--8~keV)~$\ge -0.25$,
  $L$(30\,$\mu$m)/$L$(2--8~keV)~$>1.8$, low radio core fraction
  (log~\rcd\ $\leq -3$) and/or strong 9.7$\mu$m silicate absorption.
  Comparison of different Compton-thick indicators shows that
  $L$(\oiii)/$L$(2--8~keV)~$\ge -0.25$ is most robust, available for
  both the radio-quiet and radio-loud AGN, and independent of
  $L/L_{\rm Edd}$. The $L$(30~$\mu$m)/$L$(2--8~keV) ratio is dependent
  on $L/L_{\rm Edd}$, and only Compton-thick sources with high,
  quasar-like $L/L_{\rm Edd}$ ratios have values $>$1.8.  The strength
  of the silicate absorption is affected by dust lying at host galaxy
  scales and dust related to mergers.

\item {\bf Obscured fractions.} The ratio of the unobscured (\nh~$\leq
  10^{22}$~cm$^{-2}$) to Compton-thin obscured to Compton-thick
  (\nh~$\geq 1.5\times10^{24}$~cm$^{-2}$) sources in the medium-$z$
  sample is 1.9:2:1, and the obscured fraction is $=0.61$.  In
  comparison, this ratio in the high-$z$ sample is 2.5:1.4:1, and the
  obscured fraction is 0.5, implying a larger torus opening angle
  (60\deg $\pm$8\deg\ vs. 47\deg $\pm$3\deg). If the sources in the
  medium-$z$ sample are divided according to optical spectral type, a
  slightly different ratio is found: quasars to Compton-thin NLRGs to
  Compton-thick NLRGs = 1.6:2.3:1, and the obscured fraction is 0.68.  
  The difference between the optical and X-ray derived obscured
  fractions is due to a few intermediate \nh\ sources with
  inconsistent optical and X-ray Type1/Type2 classifications.
  
\item {\bf Inconsistent optical and X-ray type sources.} Four
  low-\nh\ NLRGs from the medium-$z$ sample and two high-\nh\ quasars
  from the high-$z$ sample (3C~68.1, 325) have inconsistent optical
  and X-ray Type1/Type2 classifications.  These sources have
  intermediate inclination angles (i.e., lines of sight skimming the
  edge of the torus or accretion disk) and have
  \nh~$\sim$~10$^{22}$~cm$^{-2}$. For high $L/L_{\rm Edd} > 0.3$, we
  observe an optical Type~1, X-ray Type~2 source (obscured quasar),
  where the X-ray obscuration is due to a strong accretion disk wind,
  and for low $L/L_{\rm Edd} \sim 0.01$ an optical Type~2, X-ray
  Type~1 source (unobscured NLRG), where a puffed-up dusty torus
  provides obscuration and hides the broad-line region.
  
\end{enumerate}

\section*{Acknowledgements} 

We thank Prof. Gordon Richards for valuable comments that improved the
manuscript.  Support for this work was provided by the National
Aeronautics and Space Administration through \chandra\ Award Number:
GO3-14115X (JK), %
GO4-15102X (BJW, JK),
and by the \chandra\ X-ray Center (CXC), which is operated by the
Smithsonian Astrophysical Observatory for and on behalf of the
National Aeronautics Space Administration under contract NAS8-03060
(BJW, JK, MAz).
JB acknowledges support from the CONICYT-Chile grants Basal-CATA
PFB-06/2007 and Basal AFB-170002, FONDECYT Postdoctorados 3160439 and
the Ministry of Economy, Development, and Tourism's Millennium Science
Initiative through grant IC120009, awarded to The Millennium Institute
of Astrophysics, MAS.

The scientific results in this article are based to a significant
degree on observations made by the \chandra~X-ray Observatory (CXO).
This research has made use of data obtained from the \chandra\ Data
Archive, and software provided by the CXC in the application packages
CIAO \citep{2006SPIE.6270E..1VF} and {\it Sherpa}
\citep{2001SPIE.4477...76F}.

This research has made use of data provided by the National Radio
Astronomy Observatory, which is a facility of the National Science
Foundation operated under cooperative agreement by Associated
Universities, Inc. and data from the Sloan Digital Sky Survey (SDSS). 
Funding for the SDSS and SDSS-II has been provided by the Alfred
P. Sloan Foundation, the Participating Institutions, the National
Science Foundation, the U.S. Department of Energy, the National
Aeronautics and Space Administration, the Japanese Monbukagakusho, the
Max Planck Society, and the Higher Education Funding Council for
England. The SDSS Web Site is http://www.sdss.org/. 
The SDSS is managed by the Astrophysical Research Consortium for the
Participating Institutions. The Participating Institutions are the
American Museum of Natural History, Astrophysical Institute Potsdam,
University of Basel, University of Cambridge, Case Western Reserve
University, University of Chicago, Drexel University, Fermilab, the
Institute for Advanced Study, the Japan Participation Group, Johns
Hopkins University, the Joint Institute for Nuclear Astrophysics, the
Kavli Institute for Particle Astrophysics and Cosmology, the Korean
Scientist Group, the Chinese Academy of Sciences (LAMOST), Los Alamos
National Laboratory, the Max-Planck-Institute for Astronomy (MPIA),
the Max-Planck-Institute for Astrophysics (MPA), New Mexico State
University, Ohio State University, University of Pittsburgh,
University of Portsmouth, Princeton University, the United States
Naval Observatory, and the University of Washington.

This research is based on observations made by {\it Herschel}, which
is an ESA space observatory with science instruments provided by
European-led Principal Investigator consortia and with important
participation from NASA.

This work is based in part on observations made with the {\it Spitzer Space
Telescope}, which was operated by the Jet Propulsion Laboratory,
California Institute of Technology under a contract with NASA.

We acknowledge the use of Ned Wright's calculator
\citep{2006PASP..118.1711W} and NASA/IPAC Extragalactic Database
(NED), operated by the Jet Propulsion Laboratory, California Institute
of Technology, under contract with the National Aeronautics and Space
Administration.

\vspace{5mm}
\facilities{CXO, XMM, Herschel, Spitzer, Sloan}

\bibliographystyle{apj}
\bibliography{refs}



\clearpage

\begin{sidewaystable}[t]
\tablenum{1}
\begin{center}
\caption{X-ray Observations and Radio Data for the Medium Redshift 3CRR Sample}
\label{tb:obs} 
\end{center}
\vspace{-0.7cm}
\begin{minipage}{\textwidth}
\scriptsize{
\begin{tabular}{lrcrcccccccccccccc}
\hline
Name &  redshift & Source  & {\it Chandra} & Date Obs. &Exp. time &X-ray &RA & Dec. & Pos. &  Galactic &F$_{\nu}$(5GHz) & Ref. & F$_{\nu}$(5GHz)  & Ref. &
log L$_{\nu}$(5GHz)  & log \rcd &log L([OIII])~$^b$ \\
& &Type&OBSID& UT & ks \ \ \ \ \  &Ref. & J2000.0 & J2000 & Ref.  &  N$_{\rm H}$~$^{a}$ & {\it Core} &&{\it Total}&&{\it Total} &&[W]\\
& & &  & & & & && &$10^{20}$~cm$^{-2}$ & [mJy]  && [Jy] &&  [erg s$^{-1}$ Hz$^{-1}$] &&\\
\hline
3C 6.1   & 0.840 & G     &   3009 & 2002 Oct 15 &  36.49 & {\scriptsize 3,4,7,11,12}  &00:16:31.1 & +79:16:50 & M06 & 14.80& 4.4 &H94& 1.087 &H94& 34.58 & -2.39&35.37$^f$\\ 
3C 6.1   & 0.840 & G     &   4363 & 2002 Aug 26 &  19.90 & {\scriptsize 3,4,7,11,12}  &00:16:31.1 & +79:16:50 & M06 & 14.80& 4.4 &H94& 1.087 &H94& 34.58 & -2.39&35.37$^f$\\ 
3C 22    & 0.936 & G     &  14994 & 2013 Jun 05 &   9.35 &  &00:50:56.2 & +51:12:03 & M06 & 17.20 & 7.3 &F97& 0.76 & F93 & 34.54  & -2.01&36.44$^f$\\ 
3C 34    & 0.689 & G     &  16046 & 2014 Sep 25 &  11.92 &  15 &01:10:18.5 & +31:47:19 & M06 & 5.62 &1.03 &J95&0.381&J95 & 33.91 &-2.57 &36.44$^f$\\
3C 41    & 0.795 & G     &  16047 & 2014 Sep 03 &  11.89 & 15 &01:26:44.3 & +33:13:11 & M06 & 5.05 &0.64 & W99 & 1.45 & L80 & 34.65 &-3.35 &35.66$^f$\\
3C 49    & 0.621 & G/CSS &  14995 & 2013 Aug 31 &   9.45 &  &01:41:09.1 & +13:53:28 & D96 &  4.79 & 7.7 &L98&0.894 & H94 & 34.17  & -2.06&35.79\\ 
3C 55    & 0.735 & G/CT  &  16050 & 2014 Jun 15 &  11.92 & 15 &01:57:10.5 & +28:51:39 & M06& 5.48 &3.4&F93&0.88&F93& 34.35 &-2.40 &35.35$^f$\\
3C 138   & 0.759 & Q/CSS &  14996 & 2013 Mar 22 &   2.00 &  &05:21:09.8 & +16:38:22 & Z14 &  23.80& 485 &F89& 3.34 & F89 & 34.96  & -0.77 &36.46\\ 
3C 147   & 0.545 & Q/CSS &  14997 & 2013 Aug 26 &   2.00 &  &05:42:36.1 & +49:51:07 & M98 &  20.50& 2500&L98& 7.456& L98 & 34.96  & -0.30&36.79\\ 
3C 172   & 0.519 & G/CT  &  14998 & 2013 Sep 05 &   9.95 &  &07:02:08.3 & +25:13:53 & G04 &  7.98 & 0.5$^c$ &G04& 0.844 & L80 & 34.59  & -3.23&36.08$^f$\\ 
3C 175   & 0.770 & Q     &  14999 & 2013 Feb 21 &   2.00 &  &07:13:02.4 & +11:46:16 & M06 &  10.50& 23.5&B94& 0.687& G91 & 34.29  & -1.45&36.10\\ 
3C 175.1 & 0.920 & G     &  15000 & 2013 Feb 10 &   9.94 &  &07:14:04.6 & +14:36:22 & M06 &  8.99 & 1.1 &M06& 0.556& L80 & 34.39  & -2.70&35.96$^f$\\ 
3C 184   & 0.994 & G/CT  &   3226 & 2002 Sep 22 &  18.89 & {\scriptsize 3}  &07:39:24.4 & +70:23:10 & D96 & 3.45 & $<$0.2 & M06& 0.596 & L80 & 34.50  & $<$-3.47 &36.22$^f$\\ 
3C 184 $^d$  & 0.994 & G/CT  &  0028540601 & 2002 Mar 10 & 40.9 & {\scriptsize 3} & 07:39:24.4 & +70:23:10 & D96 & 3.45 & $<$0.2& M06& 0.596 & L80& 34.50  & $<$-3.47 &36.22$^f$\\ 
3C 184 $^d$  & 0.994 & G/CT  &  0028540201 & 2001 Sep 19 & 38.9 & {\scriptsize 3} & 07:39:24.4 & +70:23:10 & D96 &  3.45 & $<$0.2& M06& 0.596 & L80& 34.50  & $<$-3.47&36.22$^f$\\ 
3C 196   & 0.871 & Q     &  15001 & 2013 Mar 23 &  2.00 &  & 08:13:36.0 & +48:13:02 & M06 &  4.55 & 11.6&M06& 4.329& L80 & 35.22  & -2.57 &36.08\\ 
3C 207   & 0.681 & Q     &   2130 & 2000 Nov 04 & 37.54 & {\scriptsize 1,3,6,9,16,17,19} & 08:40:47.5 & +13:12:23 & M06 &  4.12 & 539&M06& 1.43&L80 & 34.47  & -0.22 &36.05\\ 
3C 216   & 0.670 & Q/CSS &  15002 & 2013 Feb 25 &  2.00 &  & 09:09:33.5 & +42:53:46 & M06 &  1.60 & 1050& H89 & 1.797& H89  & 34.57  & 0.15 &$<$35.46\\ 
3C 220.3 & 0.685 & G/CT  &  14992 & 2013 Jan 21 &  9.94 &  & 09:39:23.8 & +83:15:25 & H14 &  3.65 & $<$0.2& M06 & 0.636& L80  & 34.12  & $<$-3.50&36.00$^f$\\ 
3C 225B  & 0.580 & G/CT  &  16058 & 2014 Oct 18 & 11.92 & 15 & 09:42:15.4 & +13:45:50 & M18 & 3.50 &1.11&G04 &0.97& G04 & 34.14 &-2.94 &35.58$^f$\\
3C 226   & 0.818 & G     &  15003 & 2013 Oct 07 &  9.94 &  & 09:44:16.5 & +09:46:16 & M06 &  2.97 & 4.4$^c$& M06 & 0.636& L80  & 34.14   & -2.16&36.04$^f$\\ 
3C 228   & 0.552 & G     &   2095 & 2001 Jun 03 & 13.78 & {\scriptsize 3,12}  & 09:50:10.7 & +14:20:00 & G04 &  3.18 & 13.3 & G88& 1.132 & L80& 34.35  & -1.92  &35.36$^f$\\ 
3C 228   & 0.552 & G     &   2453 & 2001 Apr 23 & 10.61 & {\scriptsize 3,12}  & 09:50:10.7 & +14:20:00 & G04 &  3.18 & 13.3 & G88& 1.132 & L80& 34.35  & -1.92 &35.36$^f$ \\ 
3C 247   & 0.749 & G     &  16060 & 2014 Sep 26 & 11.64 & 15 & 10:58:59.0 & +43:01:24 &M97 & 1.06 & 3.5 &G88 & 0.95 & F14 & 34.40 &-2.43&35.92$^f$\\
3C 254   & 0.737 & Q     &   2209 & 2001 Mar 26 & 29.67 & {\scriptsize 8,5,3,13,18} & 11:14:38.7 & +40:37:20 & W12 &  1.90 & 19& H89 & 0.747 & L80 & 34.21 & -1.58 &36.71\\ 
3C 263   & 0.646 & Q     &   2126 & 2000 Oct 28 & 49.19 & {\scriptsize 3,5,7,10,13,18}  & 11:39:57.0 & +65:47:49 & M06 &  1.18 & 161& M06 & 1.033 & L80 & 34.32  & -0.73 &36.71\\ 
3C 263.1 & 0.824 & G     &  15004 & 2013 Mar 20 &  9.94 &  & 11:43:25.0 & +22:06:56 & M06 &  2.12 & 1.4& M06 & 0.775& H94  & 34.23  & -2.74 &36.31$^f$\\ 
3C 265   & 0.811 & G     &   2984 & 2002 Apr 25 & 58.92 & {\scriptsize 2,3}  & 11:45:28.9 & +31:33:46 & M06 &  1.90 & 2.65 & F93& 0.63 & F93& 34.13  & -2.37 &36.80\\ 
3C 268.1 & 0.970 & G     &  15005 & 2013 Jul 08 &  9.94 &  & 12:00:24.4 & +73:00:45 & M06 &  1.97 & 2.0 & G88& 2.602& L80  & 34.78   & -3.11&35.51$^f$\\ 
3C 275.1 & 0.557 & Q     &   2096 & 2001 Jun 02 & 24.76 & 15 & 12:43:57.7 & +16:22:53 &G04 & 1.98 & 207 & G04& 0.910 &F14& 34.07 & -0.53 & 35.62$^f$\\
3C 277.2 & 0.766 & G/CT  &  16063 & 2015 May 07 & 11.91 &15  & 12:53:33.3 & +15:42:31 & M06 &1.96 &0.48 & W99 &0.576& L80&34.21 &-3.08& 36.10$^f$\\
3C 280   & 0.996 & G/CT  &   2210 & 2001 Aug 27 & 63.52 & {\scriptsize 3,8,11,13,14} & 12:56:57.8 & +47:20:19 &Xray$^{e}$&  1.13 & $<$0.7 & M06& 1.519 & L80 & 34.55  & $<$-3.34&37.14$^f$ \\ 
3C 286   & 0.850 & Q/CSS &  15006 & 2013 Feb 26 &  2.00 &  & 13:31:08.2 & +30:30:32 & M98 &  1.15 & ...& ... & 7.584 & A95 & 35.49   & ...&35.99$^f$\\ 
3C 289   & 0.967 & G     &  15007 & 2013 Jul 28 &  9.70 &  & 13:45:26.2 & +49:46:32 & M06 &  1.15 & 0.8 & M06& 0.596 & L80& 34.38    & -2.87&35.46$^f$\\ 
3C 292   & 0.710 & G     &  17488 & 2014 Nov 21 &  7.97 & 15 & 13:50:41.8 & +64:29:35 & B06    &  2.12 & 1   & W99 & 0.702 & B91 &34.22 &-2.85&  36.33$^g$\\
3C 309.1 & 0.905 & Q/CSS &   3105 & 2002 Jan 28 & 16.95 & {\scriptsize 3} & 14:59:07.5 & +71:40:19 & M98 &  2.30 & 804& LM97 & 3.734 & L80 & 35.16  & -0.56 &36.70\\ 
3C 330   & 0.550 & G     &   2127 & 2001 Oct 16 & 44.18 & {\scriptsize 3} & 16:09:34.9 & +65:56:37 & G04 &  2.81 & 0.74 & F97& 2.35 & F93& 34.90    & -3.50&36.57$^f$\\ 
3C 334   & 0.555 & Q     &   2097 & 2001 Aug 22 & 32.47 & {\scriptsize 3} & 16:20:21.8 & +17:36:23 & G04 &  4.24 & 111 & B94 & 0.566 & L80 & 34.28 & -0.61 &36.37\\ 
3C 336   & 0.927 & Q     &  15008 & 2013 Mar 03 &  2.00 &  & 16:24:39.0 & +23:45:12 & M06 &  4.47 & 21.3 & M06& 0.685 & L80 & 34.43  & -1.49 &36.46\\ 
3C 337   & 0.635 & G     &  15009 & 2013 Oct 05 &  9.95 &  & 16:28:52.5 & +44:19:06 & M06 &  1.05 & 0.3 & M06& 0.904 & L80& 34.50   & -3.48 &34.76$^f$\\ 
3C 340   & 0.775 & G     &  15010 & 2013 Oct 20 &  9.95 &  & 16:29:36.5 & +23:20:12 & M06 &  4.04 & 1.2 & M06& 0.685 & L80 & 34.40   & -2.76&35.96$^f$\\ 
3C 343   & 0.988 & G/CSS &15011 & 2013 Apr 28 &  9.94 &  & 16:34:33.7 & +62:45:35 & K81 &  2.67 & $<$300 & P81& 1.48 & L80 & 34.78  & $<$-0.59&35.68 \\ 
3C 343.1 & 0.750 & G/CSS &  15012 & 2013 Feb 25 &  9.94 &  & 16:38:28.1 & +62:34:44 & P11 &  2.70 & $<$200 & G88& 1.192 & L80 & 34.64  & $<$-0.70 &35.71\\ 
3C 352   & 0.807 & G     &  15013 & 2013 Oct 10 &  9.95 &  & 17:10:44.1 & +46:01:28 & M06 &  2.41 & 3.4& M06 & 0.467 & L80& 34.24 & -2.13 &36.66$^f$\\ 
3C 380   & 0.692 & Q/CSS &   3124 & 2002 May 20 &  5.33 & {\scriptsize 3} & 18:29:31.7 & +48:44:46 & P06 &  5.67 & 4500 & LM97& 7.447 & L80 & 35.42  & 0.18 &36.76\\ 
3C 427.1 & 0.572 & LINER/CT& 2194 & 2002 Jan 27 & 39.45 & {\scriptsize 3} & 21:04:06.9 & +76:33:10 & G04 &  10.90& 1.0$^c$ & G04& 0.953& L80  & 34.51  & -2.98 &36.07$^g$\\ 
3C 441   & 0.708 & G/CT  &  15656 & 2013 Jun 26 &  6.98 &  & 22:06:04.9 & +29:29:19 & D96 &  8.32 & 3.5 & F97& 1.005& H94  & 34.56   & -2.46&35.68$^f$\\ 
3C 455   & 0.543 & G/CSS &  15014 & 2013 Aug 13 &  9.95 &  & 22:55:03.8 & +13:13:34 & B94 &  4.99 & 1.6$^c$ & B94& 0.923 & L80 & 34.49   & -2.76&36.07\\ 
\hline
\end{tabular}

$^a$~Galactic equivalent hydrogen column density from \citet{1990ARA&A..28..215D}.  
$^b$~The \oiii $\lambda 5007$ luminosity from \citep{2004MNRAS.349..503G}.
$^c$~The 5~GHz data are unavailable, so an 8~GHz flux was used to
calculate the 5~GHz flux assuming a radio spectral index of $\alpha = 0.7$
for the radio-lobes and 
$\alpha = 0.3$ for the radio-core ($F_{\nu} \propto \nu^{-\alpha}$). 
$^{d}~$XMM data.  
$^e$~No radio core position
available - position from \chandra\ image. 
$^f$~L(\oiii) determined from \oii $\lambda$3727 emission 
using the L(\oii) vs. L(\oiii) relation
from \cite{2004MNRAS.349..503G}. 
$^g$~L(\oiii) determined from  radio luminosity
L(151MHz) \citep{2004MNRAS.349..503G}. \\
X-ray references: 
 1:  \citet{2002A&A...381..795B},  
 2:  \citet{2004MNRAS.354L..43B},  
 3:  \citet{2006MNRAS.366..339B},  
 4:  \citet{2007MNRAS.381.1109B},  
 5:  \citet{2003MNRAS.339.1163C},  
 6:  \citet{2004PhDT........11C},  
 7:  \citet{2005ApJ...626..733C},  
 8:  \citet{2003ApJ...584..643D},  
 9:  \citet{2003A&A...401..505G},  
10:  \citet{2002ApJ...581..948H},  
11:  \citet{2004ApJ...612..729H},  
12:  \citet{2009MNRAS.396.1929H},  
13:  \citet{2010ApJ...723.1447H},  
14:  \citet{2011ApJS..197...24M},  
15:  \citet{2018ApJS..234....7M},  
16:  \citet{2004ApJ...608..698S},  
17:  \citet{2004ApJ...605L.105S},  
18:  \citet{2011ApJS..196....2S},  
19:  \citet{2005ApJ...630..721T},  
Radio references: 
A95 -     \citet{1995A&AS..112..235A},  
B94 -     \citet{1994AJ....108..766B},  
B06 -     \citet{2006MNRAS.366..339B},
F89 -     \citet{1989A&A...217...44F},  
F93 -     \citet{1993AJ....105.1690F},  
F97 -     \citet{1997AJ....114.2292F},  
G88 -     \citet{1988A&A...199...73G},  
G91 -     \citet{1991ApJS...75.1011G},  
G04 -     \citet{2004MNRAS.351..845G},  
H89 -     \citet{1989AJ.....98.1208H},  
H94 -     \citet{1994AJ....107..471H},  
L80 -     \citet{1980MNRAS.190..903L},  
L98 -     \citet{1998MNRAS.299..467L},  
LM97 -    \citet{1997A&AS..122..235L},  
M97 -     \citet{1997ApJS..112..415M},
M06 -     \citet{2006MNRAS.372..113M},  
P81 -     \citet{1981ApJ...248...61P},  
P88 -     \citet{1988ApJ...328..114P}.  
Positions from radio data:
B94 -    \citet{1994A&AS..105...91B}, 
D96 -     \citet{1996AJ....111.1945D}, 
H14 -     \citet{2014ApJ...790...46H}, 
K81 -     \citet{1981A&AS...45..367K}, 
M98 -     \citet{1998AJ....116..516M}, 
P06 -     \citet{2006AJ....131.1872P}, 
P11 -     \citet{2011AJ....142...89P},
W99 -     \citet{1999MNRAS.309.1017W},
Positions from optical and radio:
Z14 -     \citet{2014AJ....147...95Z},  
Optical position from SDSS:
W12 -     \citet{2012AJ....144...49W} 
}
\end{minipage}
\end{sidewaystable}


\clearpage

\begin{sidewaystable}[t]
\tablenum{2}
\setlength{\tabcolsep}{4pt}
\begin{center}
\caption{X-ray Source Parameters}
\label{tb:flux}
\end{center}
\vspace{-0.7cm}
\begin{minipage}{\textwidth}
  \scriptsize{
  \begin{tabular}{lrclrccrrcrrcr}
\hline
 Name & {\it Chandra} & Net Cts & Bkgrd. Cts & F(0.5$-$8 keV) & log
 L(0.5$-$8\,keV) & $\Gamma$ & \nh\ \ \ \ \ \ \ \ \ & f(1 keV) \ \ \ \ &
 Reduced & F(0.5$-$8 keV) &F(0.5$-$8 keV) & log L(0.5$-$8 keV)
  & HR\ \ \ \ \ \\
 & OBSID &(0.5-8 keV) & (0.5-8 keV) & Srcflux \ \ \ \ \ \ \ & Srcflux &
 &$10^{22}$cm$^{-2}$& $10^{-6}$\ \ \ \ \ \ & $\chi^2$ & Observed\ \ \ \ \ \ &
 Intrinsic\ \ \ \ \ \ &  &  \\
\ \ (1)&(2) \ \ &(3)& \ \ \ \ \ \ (4)&\ (5) \ \ \ \ \ \ \ \ \ \ &(6)&(7)&(8) \ \ \ \ \ \ \ \ \
&(9)\ \ \ \ \ \ \ \ &(10)&(11) \ \ \ \ \ \ \ &(12) \ \ \ \ \ \ &(13)&(14)\ \ \ \ \ \\
\hline
3C 6.1  & 3009 & $1718.8\pm 41.5$ & $1.22\pm 0.08$ & $33.60^{+0.80}_{-0.90}$ & $45.04^{+0.01}_{-0.01}$ & 1.76$^{+0.10}_{-0.09}$ & $0.32^{+0.13}_{-0.12}$ & $14.52^{+1.44}_{-1.28}$ & 0.7 & $43.59^{+5.03}_{-5.05}$ & $47.33^{+7.20}_{-5.81}$ & $45.19^{+0.06}_{-0.06}$ & $-0.29^{+0.02}_{-0.02}$ \\
 3C 6.1  & 4363 & $811.1\pm 28.5$ & $0.87\pm 0.07$ & $28.90^{+1.10}_{-1.00}$ & $44.98^{+0.02}_{-0.02}$ & 1.62$^{+0.12}_{-0.12}$ & $0.26^{+0.14}_{-0.12}$ & $11.26^{+1.39}_{-1.23}$ & 0.7 & $39.97^{+7.38}_{-5.55}$ & $42.91^{+8.76}_{-7.35}$ & $45.15^{+0.08}_{-0.08}$ & $-0.29^{+0.03}_{-0.03}$ \\
 3C 22  & 14994 & $68.8\pm 8.3$ & $0.22\pm 0.03$ & $4.38^{+0.72}_{-0.65}$ & $44.27^{+0.07}_{-0.07}$ & 1.9 & $21.24^{+16.02}_{-7.36}$ & $11.31^{+6.21}_{-3.64}$ & 0.5 & $12.53^{+5.62}_{-5.03}$ & $31.14^{+15.71}_{-10.35}$ & $45.13^{+0.18}_{-0.18}$ & $0.59^{+0.11}_{-0.08}$ \\
 3C 34 & 16046 & $73.7\pm 8.6$ & $0.32\pm 0.04$ & $6.18^{+0.79}_{-0.79}$ & $ 44.12^{+0.06}_{-0.05}$ & 1.9 &$12.32^{+3.09}_{-3.09}$ & $56.31^{+13.44}_{-13.44}$ & 0.21 & $5.70^{+1.52}_{-1.52}$ & $25.37^{+0.33}_{-0.33}$ &$44.74^{+0.01}_{-0.01}$  & $0.67_{-0.08}^{+0.09}$\\
 3C 41 & 16047 & $37.5\pm 6.1$ & $0.47\pm 0.05$ & $2.40^{+0.46}_{-0.52}$ & $ 43.87^{+0.09}_{-0.09}$ & 1.9 & $29.45^{+13.11}_{-13.11}$ & $48.37^{+21.75}_{-21.75}$ & 0.21& $2.86^{+1.42}_{-1.42}$  & $16.81^{+0.30}_{-0.30}$ &$44.71^{+0.01}_{-0.01}$ & $0.78_{-0.08}^{+0.12}$\\
 3C 49  & 14995 & $161.8\pm 12.7$ & $0.24\pm 0.03$ & $12.30^{+1.00}_{-1.00}$ & $44.29^{+0.03}_{-0.04}$ & 1.9 & $6.38^{+1.11}_{-0.91}$ & $17.64^{+2.63}_{-2.49}$ & 0.2 & $25.16^{+3.46}_{-3.22}$ & $52.81^{+6.33}_{-8.09}$ & $44.92^{+0.05}_{-0.07}$ & $0.39^{+0.09}_{-0.06}$ \\
 3C 55 & 16050 & $15.6\pm 4.0$ & $0.35\pm 0.04$ & $17.70^{+4.20}_{-4.90}$ & $ 44.65^{+0.12}_{-0.11}$ & 1.9 & ... & $2.27^{+0.98}_{-0.98}$& 0.4& $1.08^{+0.40}_{-0.40}$ &... &... & $-0.12_{-0.27}^{+0.22}$\\
 3C 138  & 14996 & $388.9\pm 19.7$ & $0.10\pm 0.02$ & $193.00\pm 10.00$ & $45.70^{+0.02}_{-0.02}$ & 1.9 & $<0.96$ & $71.07^{+7.51}_{-6.74}$ & 1.2 & $192.48^{+24.94}_{-29.60}$ & $211.64^{+17.16}_{-18.91}$ & $45.74^{+0.03}_{-0.04}$ & $-0.15^{+0.05}_{-0.05}$ \\
 3C 147  & 14997 & $151.0\pm 12.3$ & $0.04\pm 0.01$ & $81.30^{+6.70}_{-6.60}$ & $44.98^{+0.03}_{-0.04}$ & 1.9 & $<0.95$ & $30.61^{+4.48}_{-4.21}$ & 0.3 & $79.41^{+12.55}_{-10.78}$ & $91.97^{+11.84}_{-17.82}$ & $45.03^{+0.05}_{-0.09}$ & $-0.31^{+0.07}_{-0.08}$ \\
 3C 172  & 14998 & $31.7\pm 5.7$ & $0.30\pm 0.05$ & $1.84^{+0.44}_{-0.37}$ & $43.28^{+0.09}_{-0.10}$ & 1.9 & $82.97^{+75.16}_{-42.29}$ & $31.92^{+82.59}_{-21.31}$ & 0.5 & $6.88^{+13.59}_{-5.21}$ & $135.73^{+70.93}_{-87.35}$ & $45.15^{+0.18}_{-0.45}$ & $0.73^{+0.13}_{-0.10}$ \\
 3C 175  & 14999 & $354.9\pm 18.8$ & $0.07\pm 0.03$ & $155.00^{+8.00}_{-9.00}$ & $45.62^{+0.02}_{-0.03}$ & 1.9 & $<0.56$ & $54.81^{+5.41}_{-5.01}$ & 0.7 & $152.51^{+13.15}_{-17.48}$ & $162.99^{+14.80}_{-16.42}$ & $45.64^{+0.04}_{-0.05}$ & $-0.32^{+0.05}_{-0.05}$ \\
 3C 175.1  & 15000 & $88.7\pm 9.4$ & $0.31\pm 0.04$ & $6.72^{+0.76}_{-0.75}$ & $44.44^{+0.05}_{-0.05}$ & 1.9 & $1.23^{+0.79}_{-0.56}$ & $3.73^{+0.80}_{-0.72}$ & 0.2 & $8.62^{+2.00}_{-1.77}$ & $11.40^{+2.00}_{-2.03}$ & $44.67^{+0.07}_{-0.08}$ & $-0.21^{+0.10}_{-0.10}$ \\
 3C 184  & 3226 & $47.5\pm 6.9$ & $0.47\pm 0.05$ & $1.17^{+0.20}_{-0.17}$ & $43.76^{+0.07}_{-0.07}$ & 1.9 & $<2.23$ & $0.24^{+0.13}_{-0.09}$ & 0.3 & $2.07^{+0.21}_{-0.27}$ & $0.72^{+0.35}_{-0.34}$ & $43.55^{+0.17}_{-0.28}$ & $0.30^{+0.15}_{-0.12}$ \\
3C 184 $^a$ & XMM &  $776 \pm 65$   & ... & $1.75^{+0.27}_{-0.24}$ & $43.94^{+0.06}_{-0.06}$  & $1.4^{+0.3}_{-0.2}$ & $48.7^{+22.0}_{-12.1}$ & $24^{+11}_{-10}$ & 39.3&   $17^{+7}_{-6}$ & ... & $44.8^{+0.1}_{-0.2}$ & ... \\
 3C 196  & 15001 & $89.9\pm 9.5$ & $0.07\pm 0.02$ & $32.60^{+3.60}_{-3.60}$ & $45.07^{+0.05}_{-0.05}$ & 1.9 & $2.68^{+1.17}_{-0.85}$ & $24.85^{+5.60}_{-5.06}$ & 0.1 & $48.60^{+11.40}_{-11.55}$ & $70.97^{+14.98}_{-13.87}$ & $45.41^{+0.08}_{-0.09}$ & $-0.07^{+0.10}_{-0.10}$ \\
 3C 207  & 2130 & $6462.7\pm 80.4$ & $8.34\pm 0.21$ & $85.40^{+1.10}_{-1.10}$ & $45.23^{+0.01}_{-0.01}$ & 2.15$^{+0.07}_{-0.06}$ & $0.29^{+0.05}_{-0.04}$ & $66.05^{+3.76}_{-3.53}$ & 0.9 & $141.60^{+12.41}_{-8.78}$ & $165.25^{+13.72}_{-12.74}$ & $45.52^{+0.03}_{-0.03}$ & $-0.29^{+0.01}_{-0.01}$ \\
 3C 216  & 15002 & $247.9\pm 15.7$ & $0.07\pm 0.02$ & $89.90\pm 5.80$ & $45.23^{+0.03}_{-0.03}$ & 1.9 & $0.43^{+0.18}_{-0.15}$ & $39.73^{+4.53}_{-4.29}$ & 0.6 & $102.39^{+9.91}_{-14.21}$ & $119.58^{+11.94}_{-15.37}$ & $45.36^{+0.04}_{-0.06}$ & $-0.34^{+0.05}_{-0.06}$ \\
3C 220.3  & 14992 & $5.7\pm 2.4$ & $0.28\pm 0.04$ & $0.46^{+0.22}_{-0.17}$ & $42.97^{+0.17}_{-0.20}$ & 1.9 & ... & $<0.45$ & 0.1 & ... & ... & ... & $-0.33^{+0.29}_{-0.46}$ \\
 3C 225B & 16058 & $12.6\pm 3.6$ & $0.40\pm 0.04$ & $ 0.84^{+0.23}_{-0.27}$ & $ 43.07^{+0.14}_{-0.12}$ & 1.9 & ... & $2.04^{+0.93}_{-0.93}$ &0.2 & $0.98^{+0.35}_{-0.35}$ & ... & ... & $-0.23_{-0.30}^{+0.23}$\\
 3C 226  & 15003 & $58.8\pm 7.7$ & $0.21\pm 0.04$ & $3.67^{+0.57}_{-0.51}$ & $44.05^{+0.06}_{-0.06}$ & 1.9 & $16.23^{+7.57}_{-4.78}$ & $7.90^{+2.89}_{-2.42}$ & 0.5 & $8.99^{+3.30}_{-3.38}$ & $22.40^{+9.14}_{-6.90}$ & $44.84^{+0.15}_{-0.16}$ & $0.56^{+0.13}_{-0.08}$ \\
 3C 228  & 2095 & $341.6\pm 18.5$ & $0.45\pm 0.07$ & $14.20^{+0.80}_{-0.70}$ & $44.23^{+0.02}_{-0.02}$ & 1.9 & $0.11^{+0.07}_{-0.06}$ & $5.49^{+0.53}_{-0.51}$ & 0.5 & $15.41^{+1.87}_{-1.57}$ & $16.65^{+1.26}_{-1.76}$ & $44.30^{+0.03}_{-0.05}$ & $-0.57^{+0.04}_{-0.04}$ \\
 3C 228  & 2453 & $251.6\pm 15.9$ & $0.43\pm 0.07$ & $13.20^{+0.80}_{-0.80}$ & $44.20^{+0.03}_{-0.03}$ & 1.9 & $<0.24$ & $4.56^{+0.53}_{-0.39}$ & 0.6 & $13.40^{+1.70}_{-1.82}$ & $13.66^{+1.06}_{-1.86}$ & $44.21^{+0.03}_{-0.06}$ & $-0.58^{+0.05}_{-0.05}$ \\
 3C 247 & 16060 & $42.7\pm 6.6$ & $0.33\pm 0.04$ & $ 2.78^{+0.44}_{-0.51}$ & $ 43.87^{+0.07}_{-0.07}$ & 1.9 & $7.56^{+2.87}_{-2.87}$  & $23.53^{+7.42}_{-7.42}$ & 0.4 & $3.99^{+0.95}_{-0.95}$ & $10.48^{+0.37}_{-0.37}$ &$44.44^{+0.02}_{-0.02}$ & $ 0.43_{-0.13}^{+0.15}$\\
 3C 254  & 2209 & $5087.4\pm 71.3$ & $1.64\pm 0.11$ & $88.60^{+1.20}_{-1.20}$ & $45.33^{+0.01}_{-0.01}$ & 1.99$^{+0.06}_{-0.06}$ & $0.08^{+0.04}_{-0.04}$ & $47.56^{+2.60}_{-2.48}$ & 0.6 & $128.68^{+7.87}_{-10.20}$ & $132.12^{+6.28}_{-6.77}$ & $45.50^{+0.02}_{-0.02}$ & $-0.43^{+0.01}_{-0.01}$ \\
 3C 263  & 2126 & $9061.1\pm 95.2$ & $2.89\pm 0.18$ & $88.60^{+0.90}_{-0.90}$ & $45.19^{+0.00}_{-0.00}$ & 1.89$^{+0.04}_{-0.03}$ & $<0.06$ & $48.84^{+1.62}_{-0.93}$ & 0.9 & $145.38^{+4.85}_{-4.89}$ & $145.95^{+4.01}_{-4.42}$ & $45.41^{+0.01}_{-0.01}$ & $-0.32^{+0.01}_{-0.01}$ \\
 3C 263.1  & 15004 & $423.8\pm 20.6$ & $0.21\pm 0.03$ & $30.70\pm 1.50$ & $44.98^{+0.02}_{-0.02}$ & 1.9 & $0.21^{+0.12}_{-0.10}$ & $11.84^{+0.98}_{-0.95}$ & 0.8 & $32.01^{+3.87}_{-2.27}$ & $35.47^{+2.84}_{-3.13}$ & $45.05^{+0.03}_{-0.04}$ & $-0.39^{+0.04}_{-0.04}$ \\
 3C 265  & 2984 & $362.3\pm 19.1$ & $2.68\pm 0.17$ & $2.53\pm 0.15$ & $43.88^{+0.03}_{-0.03}$ & 1.9 & $35.68^{+11.71}_{-7.49}$ & $11.35^{+3.42}_{-2.23}$ & 0.5 & $10.28^{+2.57}_{-2.51}$ & $33.74^{+7.04}_{-9.33}$ & $45.01^{+0.08}_{-0.14}$ & $0.45^{+0.04}_{-0.05}$ \\
 3C 268.1  & 15005 & $48.8\pm 7.0$ & $0.22\pm 0.05$ & $2.44^{+0.45}_{-0.40}$ & $44.06^{+0.07}_{-0.08}$ & 1.9 & $24.00^{+11.21}_{-6.83}$ & $8.70^{+3.52}_{-2.73}$ & 0.2 & $9.72^{+2.97}_{-3.83}$ & $27.11^{+8.61}_{-11.64}$ & $45.10^{+0.12}_{-0.24}$ & $0.82^{+0.09}_{-0.06}$ \\
 3C 275.1 & 2096 & $4085.1\pm 63.9$ & $0.94\pm 0.07$ & $95.20^{+1.40}_{-1.50}$ & $ 45.09^{+0.01}_{-0.01}$ & $1.85^{+0.05}_{-0.05}$ & $0.02^{+0.01}_{-0.01}$ & $242.0_{-11.4}^{+11.4}$ & 1.3 & $99.79^{+4.55}_{-4.55}$ &$112.5^{+2.44}_{-2.44}$ &$45.16^{+0.01}_{-0.01}$ & $-0.46_{-0.01}^{+0.01}$\\
 3C 277.2 & 16063 & $10.7\pm 3.3$ &  0$.27\pm 0.04$ & $ 0.47^{+0.14}_{-0.17}$ & $ 43.12^{+0.15}_{-0.13}$ & 1.9 & ... & $1.45^{+0.83}_{-0.83}$  & 0.4& $0.69_{-0.38}^{+0.38}$ &... &... & $-0.44_{-0.29}^{+0.21}$\\
 3C 280  & 2210 & $116.6\pm 11.0$ & $5.39\pm 0.21$ & $0.70^{+0.07}_{-0.07}$ & $43.54^{+0.04}_{-0.05}$ & 1.9 & $21.83^{+13.77}_{-6.99}$ & $1.97^{+0.88}_{-0.54}$ & 0.6 & $2.43^{+0.83}_{-0.82}$ & $6.10^{+1.49}_{-2.14}$ & $44.48^{+0.09}_{-0.19}$ & $0.10^{+0.08}_{-0.10}$ \\
 3C 286  & 15006 & $118.9\pm 10.9$ & $0.10\pm 0.02$ & $45.90^{+4.10}_{-4.00}$ & $45.19^{+0.04}_{-0.04}$ & 1.9 & $<0.30$ & $14.31^{+2.46}_{-1.43}$ & 0.5 & $41.47^{+4.83}_{-7.34}$ & $42.40^{+5.27}_{-6.71}$ & $45.16^{+0.05}_{-0.07}$ & $-0.61^{+0.06}_{-0.08}$ \\
 3C 289  & 15007 & $55.7\pm 7.5$ & $0.32\pm 0.04$ & $2.38^{+0.45}_{-0.40}$ & $44.04^{+0.08}_{-0.08}$ & 1.9 & $16.52^{+10.83}_{-5.22}$ & $6.88^{+3.15}_{-2.13}$ & 1.0 & $7.92^{+4.26}_{-3.24}$ & $21.84^{+5.98}_{-8.63}$ & $45.01^{+0.11}_{-0.22}$ & $0.70^{+0.11}_{-0.08}$ \\
 3C 292 & 17488 & $59.6\pm 7.7$ & $0.43\pm 0.05$ & $ 9.24^{+1.26}_{-1.36}$ & $ 44.33^{+0.06}_{-0.06}$ & 1.9 & $20.03^{+6.41}_{-4.60}$ & $98.63^{+25.28}_{-29.29}$ & 0.4& $7.68^{+2.74}_{-2.74}$ &$44.29^{+0.51}_{-0.51}$&$45.01^{+0.01}_{-0.01}$  & $ 0.86_{-0.05}^{+0.08}$ \\
 3C 309.1  & 3105 & $5306.3\pm 72.8$ & $0.69\pm 0.09$ & $163.00^{+3.00}_{-2.00}$ & $45.81^{+0.01}_{-0.01}$ & 1.62$^{+0.05}_{-0.03}$ & $<0.12$ & $61.91^{+2.48}_{-1.23}$ & 0.7 & $231.20^{+10.08}_{-8.39}$ & $232.37^{+9.02}_{-11.96}$ & $45.96^{+0.02}_{-0.02}$ & $-0.49^{+0.01}_{-0.01}$ \\
 3C 330  & 2127 & $128.0\pm 11.4$ & $1.02\pm 0.11$ & $1.61^{+0.14}_{-0.14}$ & $43.28^{+0.04}_{-0.04}$ & 1.9 & $22.88^{+20.29}_{-10.49}$ &$3.38^{+2.33}_{-1.25}$ & 0.5 & $4.11^{+1.31}_{-1.48}$ & $11.91^{+4.13}_{-4.15}$ & $44.59^{+0.13}_{-0.19}$ & $-0.08^{+0.08}_{-0.09}$ \\
 3C 334  & 2097 & $7223.5\pm 85.0$ & $8.52\pm 0.31$ & $133.00^{+1.00}_{-2.00}$ & $45.21^{+0.00}_{-0.01}$ & 1.90$^{+0.03}_{-0.02}$ & $<0.03$ & $49.88^{+1.10}_{-0.65}$ & 0.7 & $149.45^{+2.45}_{-4.78}$ & $148.32^{+3.75}_{-5.13}$ & $45.26^{+0.01}_{-0.02}$ & $-0.57^{+0.01}_{-0.01}$ \\
 3C 336  & 15008 & $193.9\pm 13.9$ & $0.06\pm 0.02$ & $72.20^{+5.30}_{-5.20}$ & $45.48^{+0.03}_{-0.03}$ & 1.9 & $<0.84$ & $29.87^{+3.69}_{-3.50}$ & 0.1 & $82.93^{+12.74}_{-9.17}$ & $89.72^{+9.38}_{-12.74}$ & $45.57^{+0.04}_{-0.07}$ & $-0.47^{+0.06}_{-0.07}$ \\
 3C 337  & 15009 & $9.8\pm 3.2$ & $0.25\pm 0.05$ & $0.52^{+0.23}_{-0.18}$ & $42.94^{+0.16}_{-0.18}$ & 1.9 & ... & $0.12^{+0.10}_{-0.10}$ & 0.5 & $1.52^{+1.03}_{-1.13}$ &... & ... & $0.38^{+0.34}_{-0.22}$ \\
 3C 340  & 15010 & $87.8\pm 9.4$ & $0.20\pm 0.05$ & $5.82^{+0.69}_{-0.68}$ & $44.20^{+0.05}_{-0.05}$ & 1.9 & $6.12^{+2.20}_{-1.58}$ & $7.44^{+1.86}_{-1.63}$ & 0.5 & $11.15^{+2.79}_{-2.67}$ & $21.58^{+4.52}_{-3.17}$ & $44.77^{+0.08}_{-0.07}$ & $0.31^{+0.10}_{-0.10}$ \\
 3C 343  & 15011 & $17.7\pm 4.2$ & $0.26\pm 0.04$ & $1.48^{+0.36}_{-0.31}$ & $43.86^{+0.09}_{-0.10}$ & 1.9 & ... & $0.57^{+0.20}_{-0.18}$ & 0.2 & $1.60^{+0.74}_{-0.83}$ & ... & ... & $-0.44^{+0.17}_{-0.24}$ \\
 3C 343.1  & 15012 & $47.7\pm 6.9$ & $0.28\pm 0.04$ & $3.46^{+0.54}_{-0.50}$ & $43.94^{+0.06}_{-0.07}$ & 1.9 & $1.94^{+1.77}_{-1.14}$ & $2.28^{+0.92}_{-0.79}$ & 0.4 & $4.35^{+2.08}_{-1.38}$ & $6.93^{+2.35}_{-2.83}$ & $44.24^{+0.13}_{-0.23}$ & $-0.25^{+0.12}_{-0.15}$ \\
 3C 352  & 15013 & $135.7\pm 11.7$ & $0.26\pm 0.04$ & $9.30^{+0.80}_{-0.85}$ & $44.44^{+0.04}_{-0.04}$ & 1.9 & $3.44^{+0.92}_{-0.74}$ & $8.57^{+1.43}_{-1.34}$ & 0.5 & $14.92^{+2.12}_{-2.18}$ & $25.04^{+4.89}_{-3.79}$ & $44.87^{+0.08}_{-0.07}$ & $0.07^{+0.08}_{-0.09}$ \\
 3C 380  & 3124 & $2642.8\pm 51.4$ & $0.23\pm 0.03$ & $324.00^{+7.00}_{-6.00}$ & $45.82^{+0.01}_{-0.01}$ & 1.91$^{+0.08}_{-0.08}$ & $0.05^{+0.06}_{-nan}$ & $148.97^{+11.22}_{-10.45}$ & 0.7 & $423.44^{+41.57}_{-41.64}$ & $445.74^{+32.53}_{-34.26}$ & $45.96^{+0.03}_{-0.03}$ & $-0.34^{+0.02}_{-0.02}$ \\
 3C 427.1  & 2194 & $31.2\pm 5.7$ & $1.84\pm 0.10$ & $0.52^{+0.11}_{-0.09}$ & $42.83^{+0.08}_{-0.08}$ & 1.9 & $26.35^{+25.91}_{10.09}$ & $0.11^{+0.05}_{-0.04}$ & 0.1 & $0.86^{+0.14}_{-0.11}$ & $0.39^{+0.16}_{-0.19}$ & $43.10^{+0.15}_{-0.30}$ & $0.07^{+0.17}_{-0.19}$ \\
 3C 441  & 15656 & $1.9\pm 1.4$ & $0.14\pm 0.04$ &
 $0.24^{+0.23}_{-0.14}$ & $42.71^{+0.30}_{-0.40}$ & 1.9 & ... &
 $<0.45$ & 0.1 & $0.89^{+0.40}_{-0.68}$ &  ... & ... & $0.64^{+0.36}_{-0.06}$ \\
 3C 455  & 15014 & $151.7\pm 12.3$ & $0.26\pm 0.04$ & $13.00\pm 1.00$ & $44.18^{+0.03}_{-0.03}$ & 1.9 & $<0.81$ & $4.83^{+0.81}_{-0.71}$ & 0.6 & $12.81^{+1.90}_{-2.56}$ & $14.71^{+1.69}_{-2.06}$ & $44.23^{+0.05}_{-0.07}$ & $-0.38^{+0.07}_{-0.08}$ \\
\hline
\end{tabular}
}

Notes: Column (1) source name, (2) {\it Chandra} OBSID,
(3) and (4) {\it Chandra} source and background counts,
(5) and (6) rest frame (K-corrected) fluxes (in
$10^{-14}$~erg\,cm$^{-2}$\,s$^{-1}$) and luminosities (in
erg\,s$^{-1}$) determined from {\it Srcflux} and corrected for Galactic
\nh\ (both quoted with 1$\sigma$ errors), (7), (8), and (9)
the X-ray power-law slope $\Gamma$, intrinsic \nh\ 
(in units of 10$^{22}$\,cm$^{-2}$), and normalization of the power-law at 1~keV
($10^{-6}$~photons\,cm$^{-2}$\,s$^{-1}$\,keV$^{-1}$) (all with
1$\sigma$ errors) from {\it Sherpa} spectral fitting.  Sources with $<$30
counts were modeled with a power-law ($\Gamma=1.9$) and Galactic
\nh\ (Table~\ref{tb:obs}). Spectral fits for sources with 30$-$700~counts also included
intrinsic absorption. A few complex spectra also included a soft
excess and/or a 6.4~keV fluorescence Fe K$\alpha$ line (see
Table~\ref{tb:complex} for details). Fluxes and luminosities quoted in
column (11), (12), and (13) include these features. For sources with 
$>$700 counts, $\Gamma$ was freed after an initial
fit with $\Gamma=1.9$. 
Column (11) rest frame  (``Observed'') fluxes  (in $10^{-14}$~erg\,cm$^{-2}$\,s$^{-1}$) corrected
for Galactic \nh\ only. Column (12) rest frame ``intrinsic'' fluxes corrected for
both the Galactic and intrinsic \nh , column (13) rest frame luminosities (in erg\,s$^{-1}$)
all derived from the best-fit {\it Sherpa} model.  The
fluxes of piled-up sources are corrected for pile-up (see Section~\ref{sec:complex}).  
Column (14) hardness ratios calculated using
the Bayesian Estimation of Hardness Ratios BEHR (\citealp{2006ApJ...652..610P}; Section~\ref{sec:dat_proc}). \\  
$^a$~XMM data from \cite{2006MNRAS.366..339B}. Note that the net
counts, fluxes and luminosities are quoted in the 2--10keV band.  
\end{minipage}
\end{sidewaystable}



\clearpage

\begin{deluxetable}{llcccllcccc}
\tablenum{3}
\label{tb:complex}
\tablewidth{0pt}
\tablecaption{X-ray Parameters of Sources with Complex Spectra and/or
  Pile-up}
\tablehead{
  \colhead{Name} & \colhead{\it Chandra} & \colhead{Type $^a$} &
  \colhead{$\chi^2$} & 
  \colhead{\fe\ } &
  \colhead{\fe\ } & \colhead{\fe\ } & \colhead{Soft excess} &
  \colhead{Soft excess} & \colhead{Pileup} \\
  \colhead{} & \colhead{OBSID} & \colhead{} & \colhead{}  & 
  \colhead{FWHM $^b$} & \colhead{Pos. $^c$} & \colhead{Ampl. $^d$} &
  \colhead{Ampl. $^d$} & \colhead{$\Gamma$} & \colhead{fraction} 
}
\startdata
3C 006.1 & 3009 & G & 0.7 & ... & ... & ... & ... & ... & 0.076 \\ 
3C 006.1 & 4363 & G & 0.7 &  ... & ... & ... & ... & ... & 0.065 \\ 
3C 184 & 3226 & G & 0.4 &  $1.31^{+0.67}_{-0.55}$ & ... & $1.91^{+1.10}_{-0.86}$ & ... & ... & ... \\ 
3C 207 & 2130 & Q & 0.9 &  ... & ... & ... & ... & ... & 0.350 \\ 
3C 254 & 2209 & Q & 0.7 &  ... & ... & ... & ... & ... & 0.296 \\ 
3C 263 & 2126 & Q & 1.2 &  ... & ... & ... & ... & ... & 0.403 \\ 
3C 265 & 2984 & G & 0.6 &  $0.21^{+0.13}_{-0.13}$ & $6.49^{+0.14}_{-0.11}$ & $5.56^{+3.33}_{-3.33}$ & $1.49^{+0.37}_{-0.35}$ & $2.57^{+0.51}_{-0.47}$ & ... \\ 
3C 280 & 2210 & G & 0.7 & ... & ... & ... & $0.08^{+0.04}_{-0.04}$ & $3.25^{+0.93}_{-0.83}$ & ... \\ 
3C 309.1 & 3105 & Q & 0.7 & ... & ... & ... & ... & ... & 0.067 \\ 
3C 330 & 2127 & G & 0.4 &  $1.21^{+2.29}_{-0.58}$ & ... & $1.08^{+0.63}_{-0.63}$ & $2.06^{+0.44}_{-0.35}$ & $2.50^{+0.35}_{-0.56}$ & ... \\ 
3C 334 & 2097 & Q & 0.8 & ... & ... & ... & ... & ... & 0.085 \\ 
3C 380 & 3124 & Q & 0.7 &  ... & ... & ... & ... & ... & 0.230 \\ 
3C 427.1 & 2194 & G & 0.1 &  $<0.005$ & ... & $179.8^{+297.0}_{-297.0}$ & $0.62^{+0.29}_{-0.29}$ & $2.40^{+0.92}_{-0.99}$& ... \\ 
\enddata
\tablecomments{Sources with complex spectra are fitted in {\it
    Sherpa} with an 
  absorbed power-law (see Table~\ref{tb:flux} for the best-fit parameters) and
  a 6.4~keV fluorescence \fe\ line, and/or a soft excess.  A
  ``...'' indicates that the relevant fit was not needed for that source.
  If the best-fit parameter value = 0, a 1$\sigma$ upper limit is quoted. 
$^a$~Source Type: Q = quasar, G = Narrow Line Radio Galaxy
(NLRG). 
  $^b$~\fe\ line full width at half maximum (FWHM) in keV. 
  $^c$~\fe\ line position in keV.  
  $^d$~\fe\ line and soft excess amplitudes are in units of 10$^{-6}$ photons\,cm$^{-2}$\,s$^{-1}$\,keV$^{-1}$. 
\\
}
\end{deluxetable}


 
\clearpage
\begin{deluxetable}{lcccccc}
\tabletypesize{\footnotesize}
\tablenum{4}
\tablecaption{Source properties from Bayesian X-ray Analysis (BXA) X-ray spectral fit.}
\tablehead{
  \colhead{Name} & \colhead{$\log$ L(2-10keV) $^a$} & \colhead{$f_\mathrm{scat}$ $^b$} & \colhead{log \nh\ $^c$} & \colhead{P(\nh $<10^{22}$)} & \colhead{P($10^{22}<$\nh $<10^{24}$)} & \colhead{P(\nh $>10^{24}$)} \\
  & \colhead{erg\,s$^{-1}$} &  & \colhead{cm$^{-2}$} &  &  & 
}
\startdata
3C006.1* & $45.08^{+0.03}_{-0.02}$ & $7.17^{+1.74}_{-3.44} \%$ & $21.99^{+0.08}_{-0.10}$ &  56\% &  44\% &   0\% \\
   3C022 & $45.32^{+0.25}_{-0.17}$ & $1.24^{+1.01}_{-0.71} \%$ & $23.55^{+0.20}_{-0.17}$ &   0\% &  93\% &   6\% \\
   3C034 & $44.78^{+0.13}_{-0.11}$ & $0.20^{+0.65}_{-0.17} \%$ & $23.11^{+0.09}_{-0.10}$ &   0\% & 100\% &   0\% \\
   3C041 & $44.98^{+0.21}_{-0.18}$ & $0.32^{+0.55}_{-0.26} \%$ & $23.61^{+0.17}_{-0.16}$ &   0\% &  95\% &   4\% \\
   3C049 & $44.93^{+0.07}_{-0.08}$ & $0.07^{+0.30}_{-0.05} \%$ & $22.79 \pm 0.06$ &   0\% & 100\% &   0\% \\
   3C055 & $45.34^{+0.22}_{-0.36}$ & $0.63^{+1.00}_{-0.36} \%$ & $25.16^{+0.56}_{-0.65}$ &   1\% &   1\% &  97\% \\
   3C138 & $45.66^{+0.04}_{-0.03}$ & $1.27^{+5.14}_{-1.21} \%$ & $21.84^{+0.12}_{-0.13}$ &  89\% &  10\% &   0\% \\
   3C147 & $45.03^{+0.08}_{-0.06}$ & $2.54^{+4.56}_{-2.45} \%$ & $21.96 \pm 0.17$ &  59\% &  41\% &   0\% \\
   3C172 & $45.27^{+0.23}_{-0.26}$ & $0.05^{+0.14}_{-0.03} \%$ & $24.01 \pm 0.13$ &   0\% &  47\% &  52\% \\
   3C175 & $45.58 \pm 0.03$ & $0.46^{+4.02}_{-0.43} \%$ & $21.39^{+0.15}_{-0.18}$ &  99\% &   0\% &   0\% \\
 3C175.1 & $44.63 \pm 0.07$ & $0.25^{+2.12}_{-0.22} \%$ & $22.19 \pm 0.14$ &   9\% &  91\% &   0\% \\
   3C184 & $44.99^{+1.10}_{-0.27}$ & $2.02^{+2.27}_{-1.92} \%$ & $23.74^{+1.71}_{-0.23}$ &   0\% &  62\% &  37\% \\
   3C196 & $45.41 \pm 0.08$ & $0.13^{+1.30}_{-0.10} \%$ & $22.46 \pm 0.11$ &   0\% & 100\% &   0\% \\
  3C207* & $45.34 \pm 0.01$ & $0.18^{+0.79}_{-0.15} \%$ & $21.25 \pm 0.02$ & 100\% &   0\% &   0\% \\
   3C216 & $45.29^{+0.04}_{-0.05}$ & $0.51^{+3.72}_{-0.48} \%$ & $21.28^{+0.21}_{-0.40}$ &  98\% &   0\% &   1\% \\
 3C220.3 & $43.04^{+1.32}_{-0.27}$ & $1.06^{+4.03}_{-1.00} \%$ & $21.17^{+3.85}_{-0.80}$ &  64\% &   2\% &  32\% \\
  3C225B & $43.33^{+1.41}_{-0.29}$ & $1.42^{+3.99}_{-1.34} \%$ & $21.83^{+3.49}_{-1.23}$ &  50\% &   2\% &  46\% \\
   3C226 & $44.93 \pm 0.13$ & $1.36^{+1.24}_{-0.92} \%$ & $23.24 \pm 0.13$ &   0\% & 100\% &   0\% \\
   3C228 & $44.24 \pm 0.03$ & $0.42^{+3.17}_{-0.38} \%$ & $20.53^{+0.30}_{-0.34}$ &  99\% &   0\% &   0\% \\
   3C247 & $44.45 \pm 0.14$ & $2.05^{+3.20}_{-1.76} \%$ & $22.94^{+0.15}_{-0.18}$ &   0\% & 100\% &   0\% \\
  3C254* & $45.37 \pm 0.01$ & $0.37^{+2.65}_{-0.32} \%$ & $20.04^{+0.06}_{-0.03}$ & 100\% &   0\% &   0\% \\
  3C263* & $45.86 \pm 0.01$ & $9.97^{+0.03}_{-0.04} \%$ & $22.80 \pm 0.01$ &   0\% & 100\% &   0\% \\
 3C263.1 & $44.99^{+0.04}_{-0.02}$ & $0.32^{+2.69}_{-0.29} \%$ & $20.85^{+0.38}_{-0.48}$ &  97\% &   0\% &   2\% \\
   3C265 & $45.03^{+0.10}_{-0.08}$ & $2.59^{+0.74}_{-0.65} \%$ & $23.52^{+0.07}_{-0.05}$ &   0\% & 100\% &   0\% \\
 3C268.1 & $45.13^{+0.12}_{-0.13}$ & $0.09^{+0.39}_{-0.07} \%$ & $23.42 \pm 0.11$ &   0\% & 100\% &   0\% \\
 3C275.1 & $45.09 \pm 0.01$ & $0.24^{+2.09}_{-0.21} \%$ & $20.55^{+0.12}_{-0.14}$ & 100\% &   0\% &   0\% \\
 3C277.2 & $44.57^{+0.53}_{-1.27}$ & $1.77^{+4.01}_{-1.38} \%$ & $24.61^{+0.97}_{-3.72}$ &  29\% &   3\% &  67\% \\
   3C280 & $45.85^{+0.15}_{-1.11}$ & $0.22^{+2.91}_{-0.09} \%$ & $25.10^{+0.64}_{-1.46}$ &   0\% &  19\% &  80\% \\
   3C286 & $45.21^{+1.06}_{-0.05}$ & $1.08^{+5.87}_{-1.03} \%$ & $20.47^{+4.46}_{-0.34}$ &  75\% &   0\% &  24\% \\
   3C289 & $45.07^{+0.16}_{-0.11}$ & $0.09^{+0.33}_{-0.07} \%$ & $23.29 \pm 0.11$ &   0\% &  99\% &   0\% \\
   3C292 & $45.07^{+0.12}_{-0.11}$ & $0.15^{+0.38}_{-0.13} \%$ & $23.33 \pm 0.09$ &   0\% & 100\% &   0\% \\
3C309.1* & $45.82 \pm 0.01$ & $0.22^{+1.95}_{-0.20} \%$ & $20.14^{+0.17}_{-0.11}$ & 100\% &   0\% &   0\% \\
   3C330 & $44.29^{+0.20}_{-0.13}$ & $5.31 \pm 2.09 \%$ & $23.48^{+0.22}_{-0.15}$ &   0\% &  91\% &   9\% \\
  3C334* & $45.19 \pm 0.01$ & $0.20^{+1.60}_{-0.18} \%$ & $20.19^{+0.18}_{-0.12}$ & 100\% &   0\% &   0\% \\
   3C336 & $45.51 \pm 0.05$ & $0.31^{+3.06}_{-0.27} \%$ & $21.48^{+0.20}_{-0.35}$ &  99\% &   0\% &   0\% \\
   3C337 & $43.73^{+0.47}_{-0.25}$ & $0.21^{+1.88}_{-0.18} \%$ & $22.88^{+0.56}_{-0.36}$ &   2\% &  84\% &  13\% \\
   3C340 & $44.88^{+0.09}_{-0.11}$ & $1.93^{+1.58}_{-1.12} \%$ & $22.94 \pm 0.10$ &   0\% & 100\% &   0\% \\
   3C343 & $43.85^{+0.22}_{-0.14}$ & $0.57^{+4.08}_{-0.54} \%$ & $21.43^{+0.78}_{-0.87}$ &  80\% &   4\% &  14\% \\
 3C343.1 & $44.23 \pm 0.11$ & $0.57^{+3.50}_{-0.53} \%$ & $22.25 \pm 0.16$ &   7\% &  92\% &   0\% \\
   3C352 & $44.89 \pm 0.08$ & $0.32^{+1.83}_{-0.29} \%$ & $22.55 \pm 0.08$ &   0\% & 100\% &   0\% \\
  3C380* & $45.82 \pm 0.01$ & $0.19^{+1.72}_{-0.17} \%$ & $20.33^{+0.29}_{-0.21}$ & 100\% &   0\% &   0\% \\
 3C427.1 & $44.42^{+0.70}_{-0.80}$ & $1.09^{+5.15}_{-0.94} \%$ & $24.05^{+1.42}_{-0.73}$ &   0\% &  49\% &  50\% \\
   3C441 & $43.22^{+1.06}_{-0.76}$ & $0.23^{+2.43}_{-0.21} \%$ & $22.98^{+1.76}_{-1.14}$ &  20\% &  54\% &  24\% \\
   3C455 & $44.16^{+0.05}_{-0.04}$ & $0.47^{+4.08}_{-0.43} \%$ & $20.98^{+0.38}_{-0.54}$ &  98\% &   0\% &   1\% \\
\enddata
\tablecomments{``*'' indicates a piled-up source. $^a$~Intrinsic 2$-$10~keV
  luminosity corrected for absorption.  $^b$~Strength of the
  scattered power-law relative to the intrinsic continuum power-law
  (3$\sigma$ range). $^c$~Intrinsic column density. Last three columns
  show the HBM probability that the intrinsic column density is
  unobscured (\nh~$<10^{22}$cm$^{-2}$), obscured Compton-thin
  (\nh~$= 10^{22}-10^{24}$cm$^{-2}$), obscured Compton-thick
  (\nh~$>10^{24}$cm$^{-2}$) }
\label{tb:BXApost}
\end{deluxetable}



\begin{deluxetable}{lcccccc}
\tablenum{5}
\label{tb:CT_ind}
\tablecaption{Compton-thick and borderline CT candidates}
%
%
\tablehead{
 \colhead{Name} & 
 \colhead{log L(\oiii)/L(2-8keV)} & 
 \colhead{log L(30$\mu$m)/L(2-8keV)} &
 \colhead{log L(0.5-8keV)/L(178MHz)} &
 \colhead{log \rcd $< -3$} &
\colhead{9.7~$\mu$m absorption\,$^a$}&
\colhead{HBM$^b$} \\
 \colhead{} & 
 \colhead{$>-0.25$} & 
 \colhead{$>1.8$} &
 \colhead{$<0$} &
 \colhead{} & 
 \colhead{} &
 \colhead{P\%} 
}
\startdata
\multicolumn{7}{c}{Compton-thick sources:} \\
3C\,55 & x & \checkmark &\checkmark  & x & strong & 97\\
3C\,172 & x & $>0.8$   &  \checkmark & \checkmark & ... &52 \\
3C\,184 & \checkmark &  \checkmark   &\checkmark  &  \checkmark& strong&37 \\
3C\,220.3 & \checkmark & \checkmark & \checkmark & \checkmark & ... &32 \\
3C\,225B  & \checkmark & ... &\checkmark  & \checkmark & ...& 46\\
3C\,277.2 & \checkmark & ... & \checkmark &\checkmark  & ...& 67\\
3C\,280   & \checkmark & \checkmark & \checkmark & \checkmark&no&80 \\
3C\,427.1$^c$ & \checkmark & $>1.2$ & \checkmark & \checkmark & ... &50 \\
3C\,441   & \checkmark & \checkmark & \checkmark & x &strong&24 \\
\hline
\multicolumn{7}{c}{Highly obscured/borderline CT candidates:}\\
3C\,330 &  x& $>1.2$    &\checkmark  &  \checkmark &moderate&9\\
3C\,337 & x & $>1.2$  &\checkmark  &  \checkmark & ... &13\\
3C\,343 $^d$   & x     & \checkmark & \checkmark & ... &strong&14 \\
\enddata
\tablecomments{ 
``...'' means no available data. $^a$~9.7\,$\mu$m silicate absorption. Strong silicate absorption is defined as having optical depth $\tau > 0.3$. 
$^b$~Probability of the source being CT from the Hierarchical Bayesian Model (based on X-ray data alone).
$^c$~3C\,427.1 as a LERG has a low mid-IR (30$\mu$m) luminosity. 
$^d$~The radio core fraction \rcd\ is
not available for this CSS source. 
}
\end{deluxetable}



\begin{deluxetable}{lccccc}
\tablenum{6}
\tablewidth{0pt}
\label{tb:geom}
\tablecaption{Geometry of the torus in the medium-$z$ and high-$z$ 3CRR samples}
\tablehead{
 \colhead{} &
   \colhead{} &
   \multicolumn{2}{c}{Medium-$z$} &
   \multicolumn{2}{c}{High-$z$}\\
     \colhead{Type} &  \colhead{log N$_{\rm H}$/cm$^{-2}$} &  \colhead{No. of sources} &  \colhead{cone angle$^a$} &  \colhead{No. of sources} &  \colhead{cone angle$^a$} 
   }
\startdata
Quasar & $<$ 21.5          & 14/44=32\% & 47\deg $\pm$3\deg  & 19/38=50\% & 60\deg $\pm$8\deg \\
Intermediate & 21.5--22.5  &  6/44=14\% & 10\deg $\pm$4\deg  &  3/38=8\%  &  5\deg $\pm$2\deg \\
NLRG & 22.5--24            &  15/44=34\% & 21\deg $\pm$2\deg  &  8/38=21\% & 13\deg $\pm$2\deg \\
CT NLRG & $>$ 24           &  9/44=20\% & 12\deg $\pm$3\deg  &  8/38=21\% & 12\deg $\pm$4\deg \\
\enddata
\tablecomments{$^a$ cone angle below/above the equatorial plane}
\end{deluxetable}




\newpage

\clearpage  

\clearpage

\begin{figure}
\epsscale{0.7}
\plotone{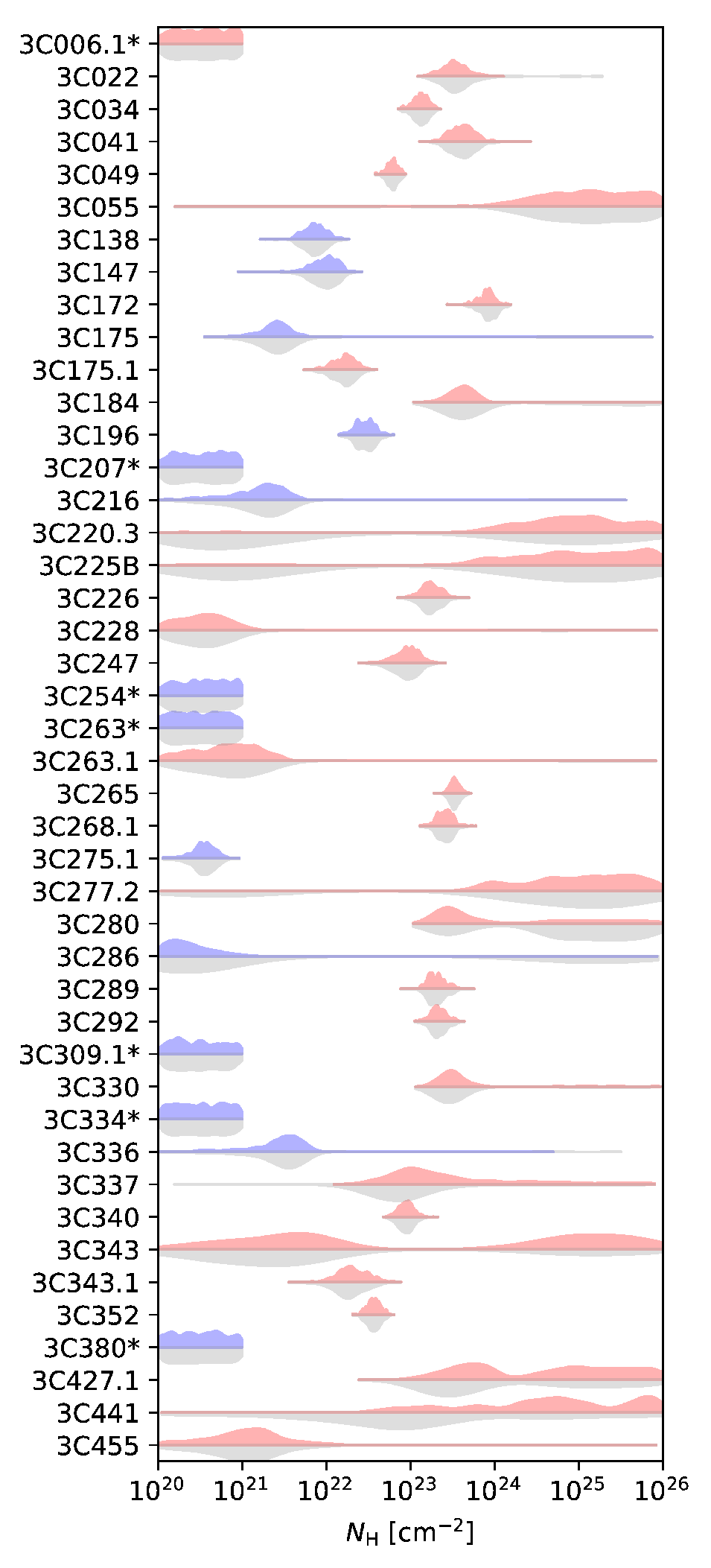}
\caption{Constraints on the intrinsic column density \nh\ for each
  source in the sample from the Hierarchical Bayesian Modeling (HBM).
  Gray distributions show probabilities when assuming (as a first step
  of the HBM) flat, uninformative priors.  Results with posterior
  probabilities (after incorporating information from the whole
  sample) are shown in blue for quasars and in red for NLRGs.  Thicker
  vertical dimension of contours implies higher probability. Sources
  with substantial pileup, marked with an asterisk, were manually
  given an unobscured (\nh $<10^{21}$~cm$^{-2}$) solution.}
\label{fg:HBM_NH}
\end{figure}

\begin{figure}
\epsscale{0.7}
\plotone{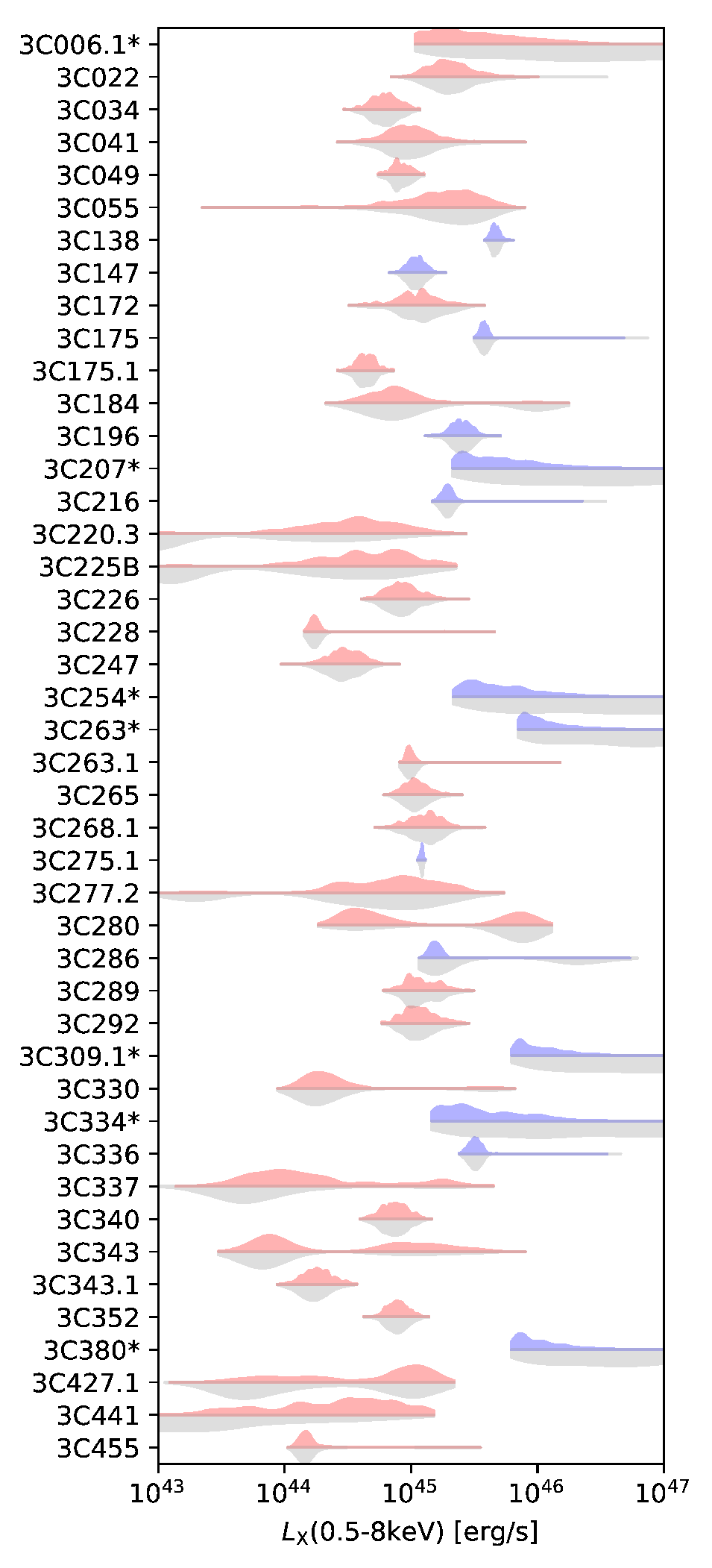}
\caption{Constraints on the intrinsic 0.5$-$8~keV X-ray luminosity for
  each source in the sample from the Hierarchical Bayesian Modeling.
  Gray distributions show probabilities when assuming (as a first step
  of the HBM) flat, uninformative priors.  Results with posterior
  probabilities (after incorporating information from the whole
  sample) are shown in blue for quasars and in red for NLRGs.  Thicker
  vertical dimension of contours implies higher probability.  Sources
  with substantial pileup are marked with an asterisk.}
\label{fg:HBM_Lx}
\end{figure}

\begin{figure}
  \epsscale{0.9}
  \plotone{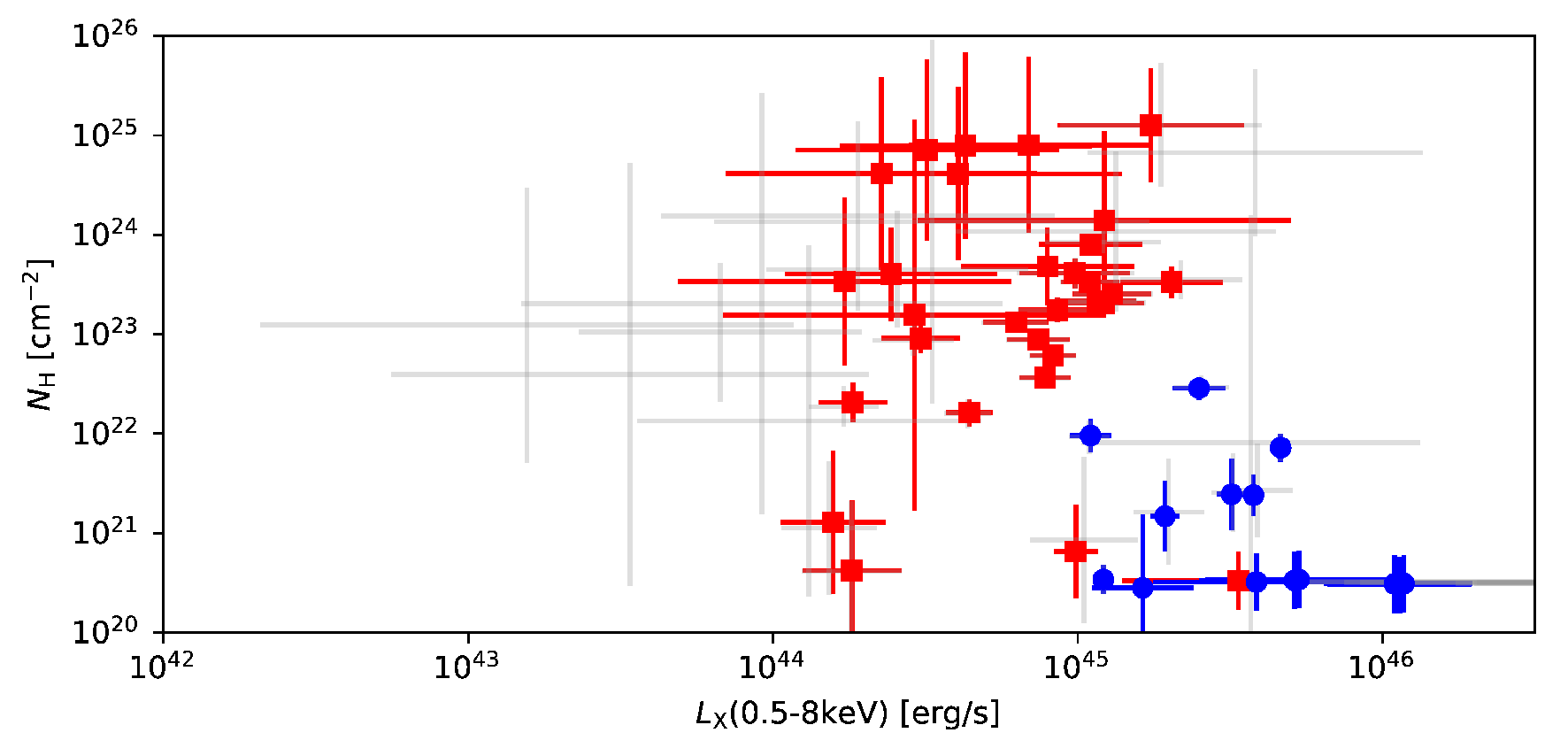}
\caption{Intrinsic X-ray luminosity $L(0.5-8~{\rm keV})$ versus
  intrinsic column density \nh\ from the Hierarchical Bayesian
  Modeling (HBM). Results of the first HBM run which assumed flat,
  uninformative priors are plotted in gray. Results with posterior
  probabilities (after incorporating information from the whole
  sample) are plotted in blue for quasars and red in NLRGs.  The HBM
  is able to tighten the constraints on several low-information
  sources. The few quasars with substantial pileup, which were
  manually set to be unobscured with \nh~$<10^{21}$~cm$^{-2}$ have
  luminosities $>10^{45.5}$~erg~s$^{-1}$.}
\label{fg:HBM_NH_L}
\end{figure}

\begin{figure}
  \epsscale{0.6}
  \plotone{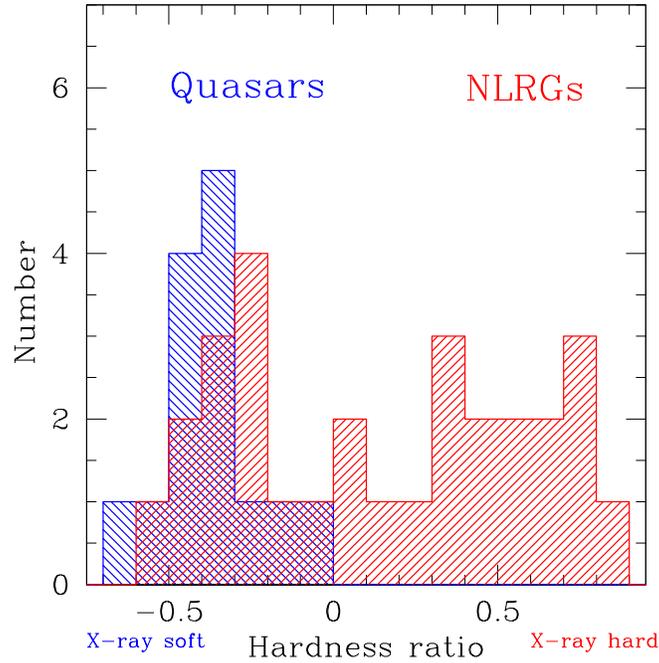}
  \vspace{-0.3in}
\caption{Histograms of the X-ray hardness ratios for quasars (blue)
  and NLRGs (red). The NLRGs with quasar-like HR~$<0$
  are: 3C~6.1, 175.1, 220.3, 228, 263.1, 330, 343, 343.1, 455.}
\label{fg:HRdist}
\end{figure}

\clearpage
\begin{figure}
\epsscale{0.9}
\plotone{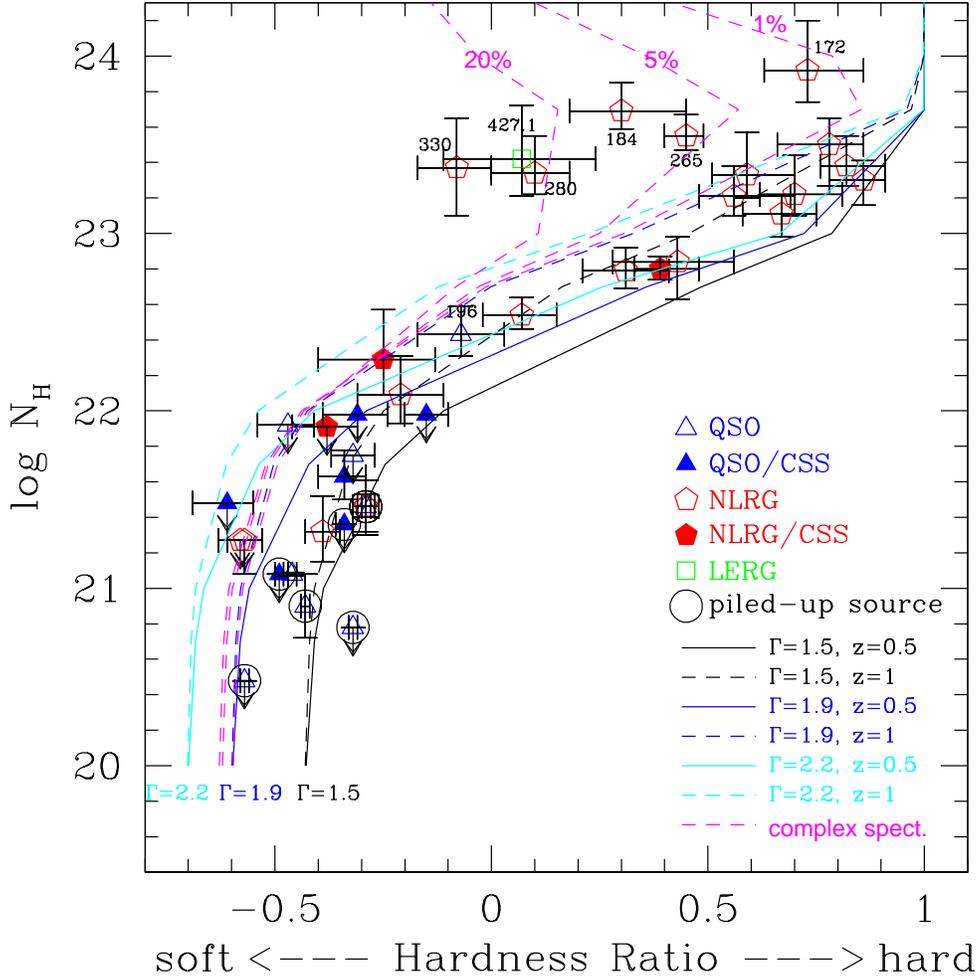}
\vspace{-0.5in}
\caption{The intrinsic column density \nh\ fitted with {\it Sherpa}
  (available for sources with $> $30~cts) as a function of the
  observed X-ray hardness ratio (Section~\ref{sec:result_Lx}). Lines
  show the relation between \nh\ and hardness ratio for an absorbed
  power-law with $\Gamma$ = 1.5 (blue), 1.9 (black), or 2.2 (green) at
  two values of redshift $z=0.5$ (solid lines) and $z=1$ (dotted
  lines) spanning the sample's redshift range. The 
  \nh\ ranges from $10^{20}$ to $10^{25}$~cm$^{-2}$.  Red dashed lines
  show the absorbed power-law model with $\Gamma = 1.9$ at $z=1$ to
  which a scattered component was added with a 1\%, 5\%, and 20\%
  normalization, relative to the intrinsic power-law. This additional
  component is needed to explain softer hardness ratios in comparison
  with the model predictions in 3C~172, 184, 265, 280, 330, 427.1 The
  different symbols indicate the class of source as shown in the
  legend.
}
\label{fg:NHvsHR}
\end{figure}

\clearpage
\begin{figure}
\epsscale{0.8}
\plotone{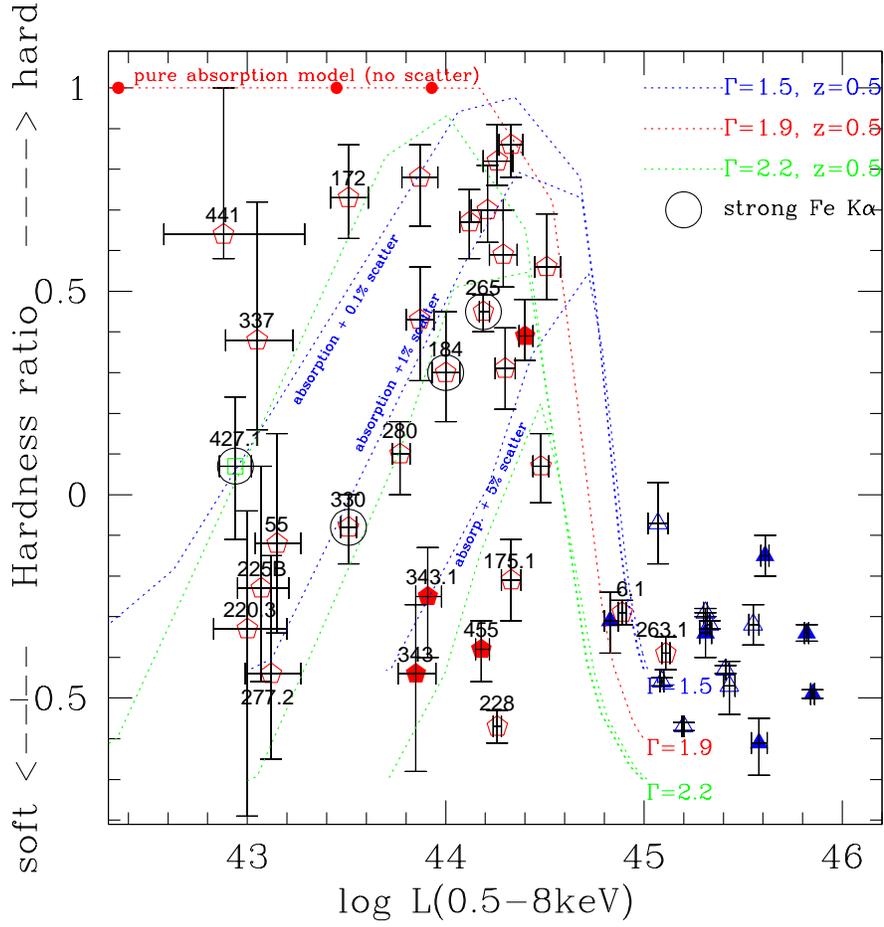}
\vspace{-0.5in}
\caption{X-ray hardness ratio as a function of 0.5--8~keV X-ray
  luminosity not corrected for intrinsic absorption. The different symbols
  indicate the source type and are the same as in
  Figure~\ref{fg:NHvsHR}. Sources with strong iron K$\alpha$ are
  circled. The red dotted curve shows a power-law model with
  $\Gamma=1.9$ absorbed by intrinsic column density ranging from \nh =
  $1 \times 10^{20}$~cm$^{-2}$ (lower right corner) to $5 \times
  10^{24}$~cm$^{-2}$ (upper left corner), where large red dots (from right
  to left) indicate \nh\ = (1,2,5)$\times 10^{24}$~cm$^{-2}$. Other
  dotted curves show absorbed (\nh~between $1\times 10^{20} - 1\times
  10^{25}$~cm$^{-2}$) power-law models with $\Gamma$ = 1.5 (blue) and
  $\Gamma$=2.2 (green) with an added scattered light component
  normalized to 0.1\%,1\%, and 5\% of the intrinsic AGN continuum. All
  curves are for $z=0.5$.  Most NLRGs require an additional, scattered
  light component.} 
\label{fg:LxvsHR}
\end{figure}

\clearpage
\begin{figure}
\epsscale{0.47}
\plotone{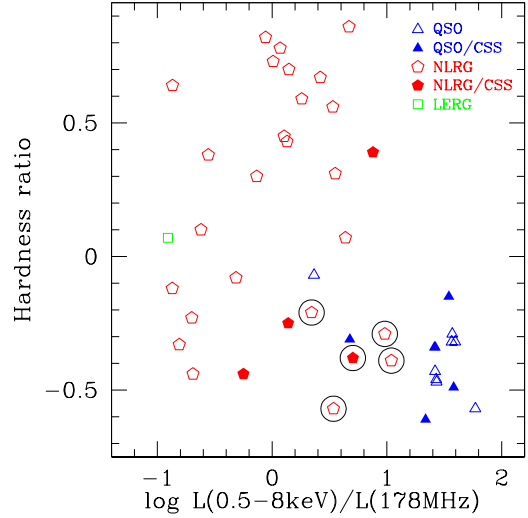}
\vspace{-0.5in}
\caption{X-ray hardness ratio as a function of  the ratio of 0.5$-$8~keV
  luminosity (uncorrected for intrinsic \nh ) to the total 178~MHz
  radio luminosity. Symbol shapes indicate object class as in
  Figure~\ref{fg:NHvsHR}.   The low-\nh\ NLRGs are circled.}
\label{fg:HRvsLxLr}
\end{figure}

\vspace{-0.8in}
\begin{figure}
\epsscale{1}
\plottwo{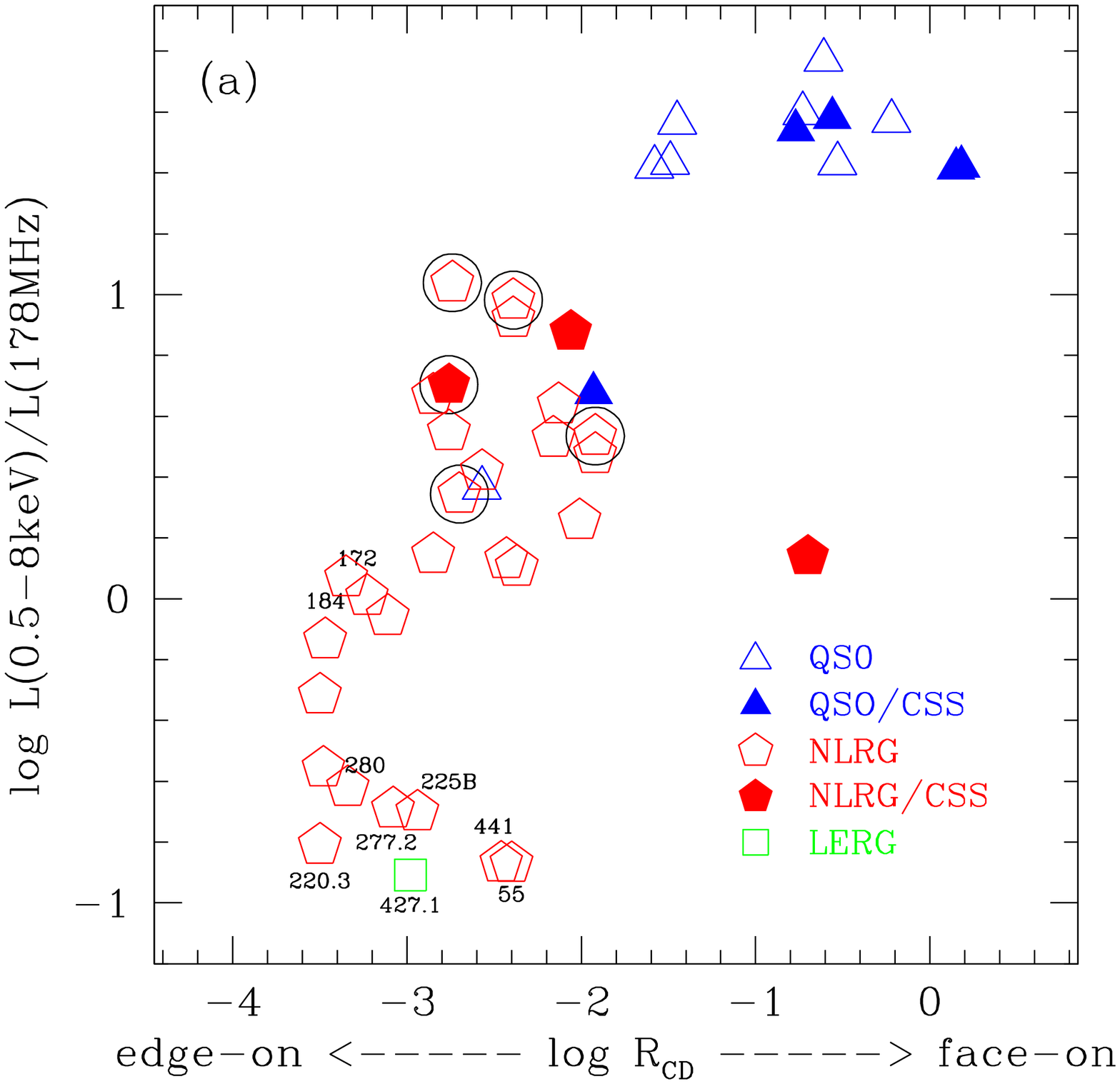}{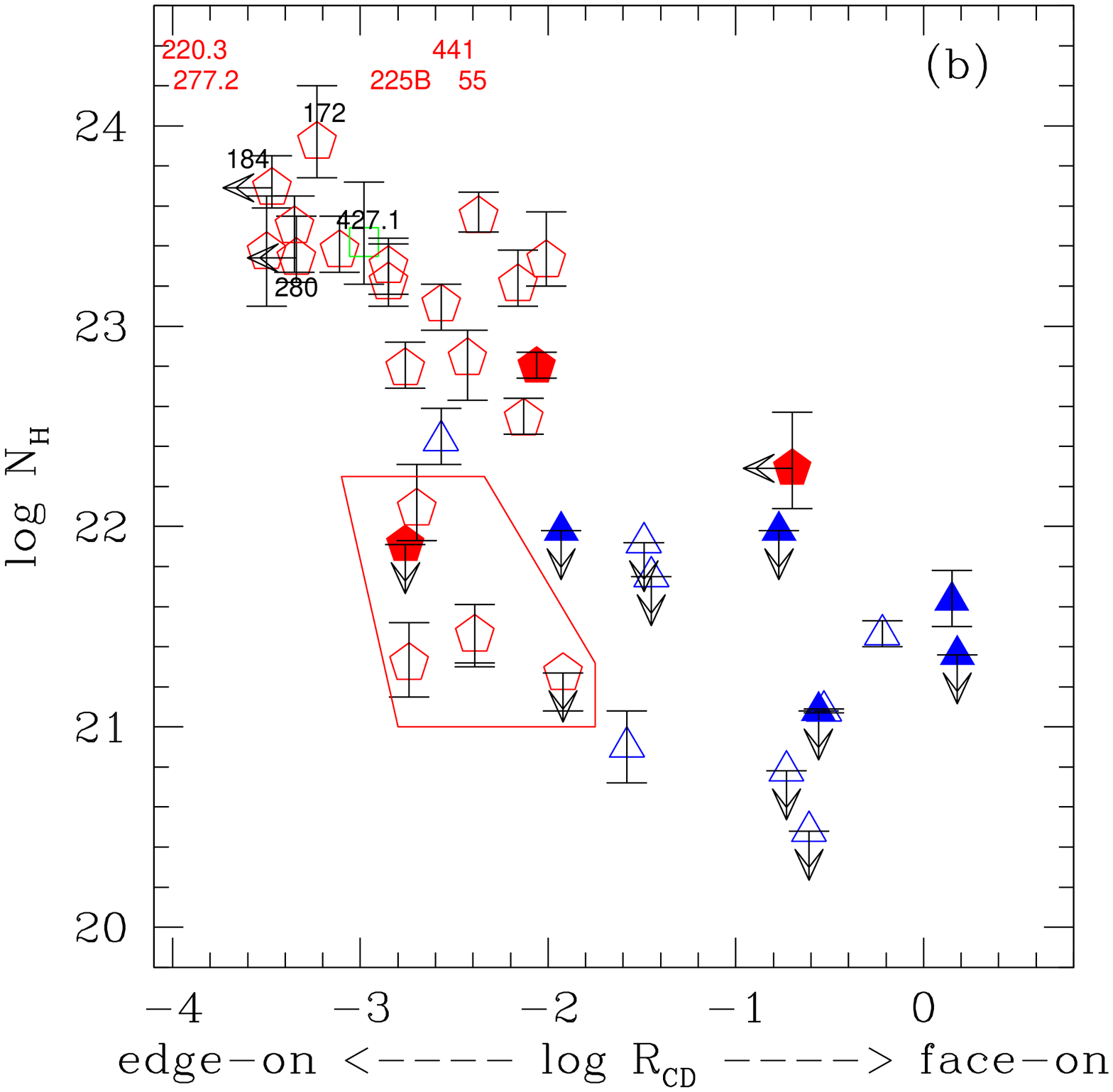}
\vspace{-0.5in}
\caption{The ratio of 0.5--8~keV luminosity, uncorrected for \nh , to
  the total 178~MHz luminosity ({\it left}) and intrinsic equivalent
  hydrogen column density (\nh, {\it right}) estimated from spectral
  fits as a function of the radio core fraction \rcd .  A strong trend
  with \rcd\ for both parameters is consistent with the
  orientation-dependent obscuration of Unification
  models. Compton-thick sources are named in both figures (however in
  figure on the right 3C~55, 220.3, 225B, 277.2, 441 with no
  \nh\ estimate due to low $S/N$ have only their names indicated at
  their log \rcd\ values and log~\nh /cm$^{-2} >$24).  The low
  \nh\ NLRGs are circled in figure on the left and enclosed in a red
  contour in figure on the right. Symbols in both figures indicate
  source type as in Figure~\ref{fg:NHvsHR}.}
\label{fg:LxLr_HRvsR}
\end{figure}

\clearpage
\begin{figure}
\epsscale{0.95}
  \plottwo{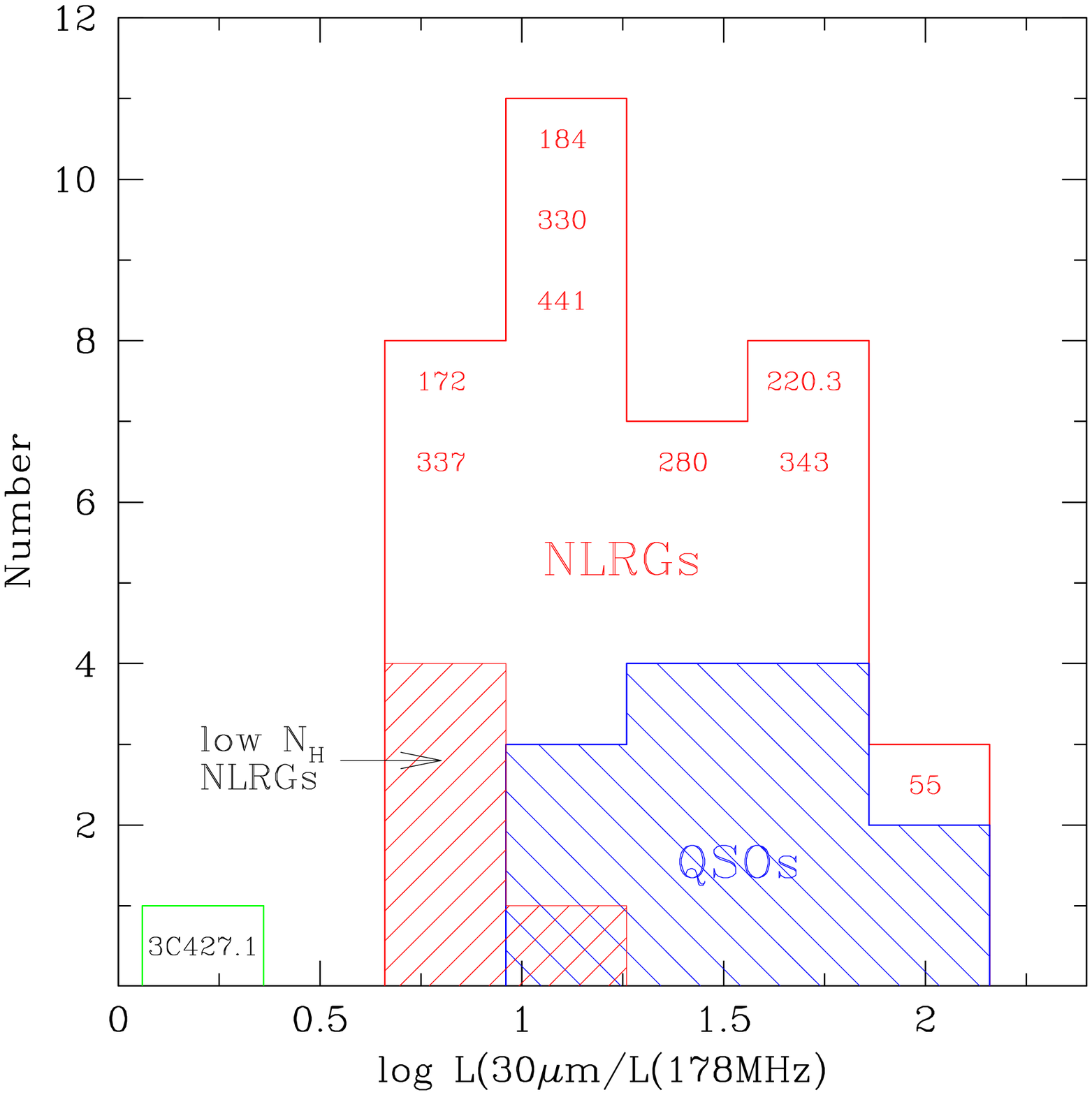}{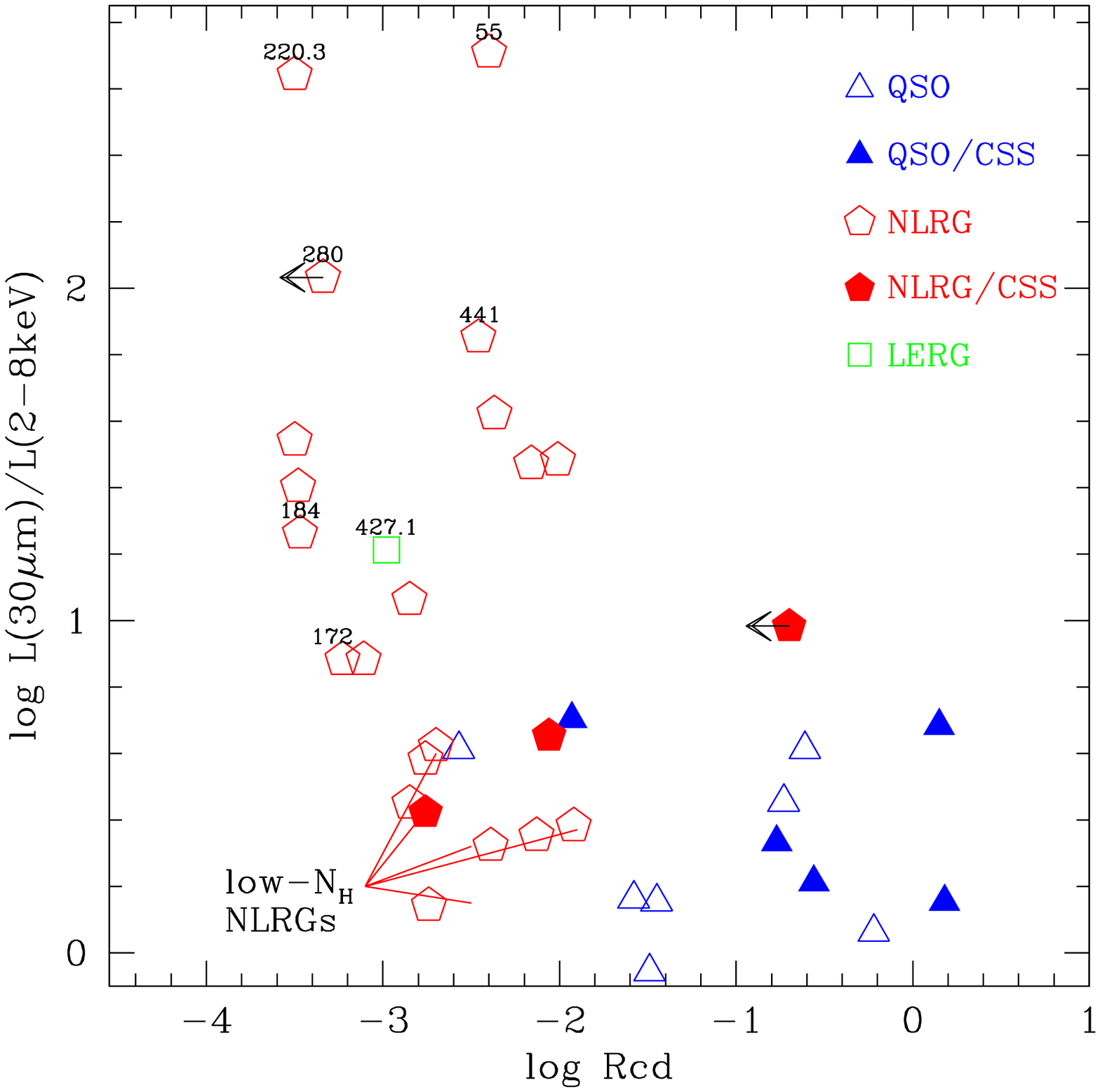}
  \vspace{-0.5in}
\caption{{\it (Left:)} Histogram of the 30~$\mu$m to 178~MHz
  luminosity ratios. Quasars are plotted in blue and NLRGs in red.
  The low-\nh\ NLRGs are shown by the red hatched histogram. 
  3C~427.1, plotted in green, has low
  MIR emission, as expected for LERGs
  \citep{2016AJ....151..120W}. Compton-thick and borderline
  Compton-thick sources are indicated by their 3C identification. {\it
    (Right:)} The ratio of 30~$\mu$m to 2-8~keV luminosity (not
  corrected for \nh ) as a function of the radio core
  fraction \rcd . Different symbols indicate source type as in
  Figure~\ref{fg:NHvsHR}. The Compton-thick sources are named and
  NLRGs with low \nh\ are indicated in both figures.}
\label{fg:MIRfigs}
\end{figure}

%
%
%
%
%

\vspace{-1in}

\begin{figure}
\epsscale{0.5}
\plotone{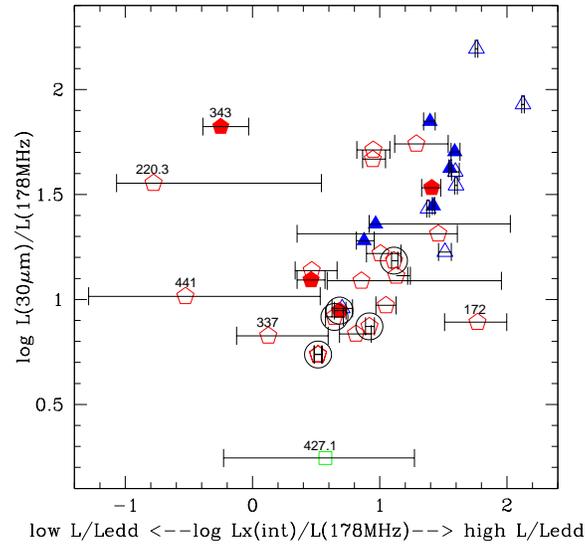}
\vspace{-0.5in}
\caption{The dependence of the 30~$\mu$m on the intrinsic X-ray
  emission (from HBM modelling) both normalized to 178~MHz luminosity.
  The intrinsic X-ray luminosity depends on $L/L_{\rm Edd}$. Symbols
  indicate source type as in Figure~\ref{fg:NHvsHR}. Most quasars and
  NLRGs follow a strong correlation where the mid-IR increases with
  intrinsic $L_{\rm X}$ i.e., $L/L_{\rm Edd}$.  The outliers are:
  3C~220.3 which lenses a background submm galaxy resulting in a
  higher than expected 30~$\mu$m luminosity, 3C~343 possibly another
  lens candidate, 3C~172, 441 IR-weak, Compton-thick sources,
  3C~427.1 a LERG expected to have low IR emission, and 3C~337 a
  highly obscured source with low $L/L_{\rm Edd}$
  (Sections~\ref{sec:expl_lowNH_NLRGS},~\ref{sec:otherCT}).  
  Low-\nh\ NLRGs are circled in black and show preferably lower mid-IR
  emission and $L/L_{\rm Edd}$ ratios.}
\label{fg:LMIR_Ledd_fig}
\end{figure}

\clearpage
\begin{figure}
\plotone{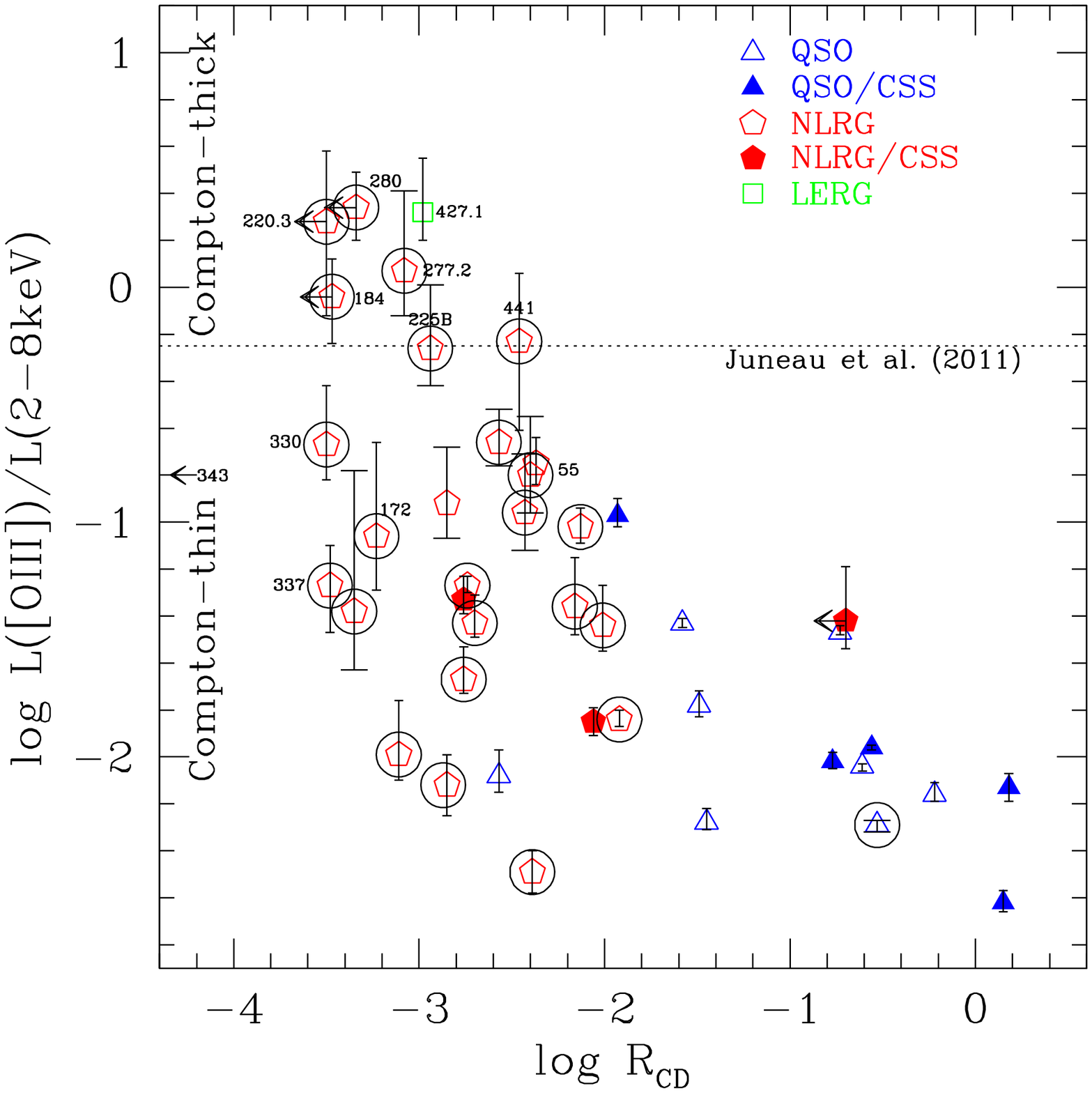}
\vspace{-0.5in}
\caption{The ratio of L(\oiii) to the 2$-$8~keV X-ray
  luminosity (not corrected for obscuration) as a function of radio
  core fraction \rcd .  Symbols indicate source type as in Figure~\ref{fg:NHvsHR}.
  For sources lacking \oiii\ measurements, values were estimated from
  \oii\ measurements following \citet{2004MNRAS.349..503G} and are
  circled.  3C~427.1 and 3C~292 have \oiii\ estimated from 151~MHz
  radio luminosity.  The dotted line is the dividing line between Compton-thin and
  Compton-thick sources reported by \citep{2011ApJ...736..104J}. 
  Compton-thick and borderline CT sources are indicated by their 3C identification.}
\label{fg:O3LxvsR}
\end{figure}

\clearpage
\begin{figure}
\plottwo{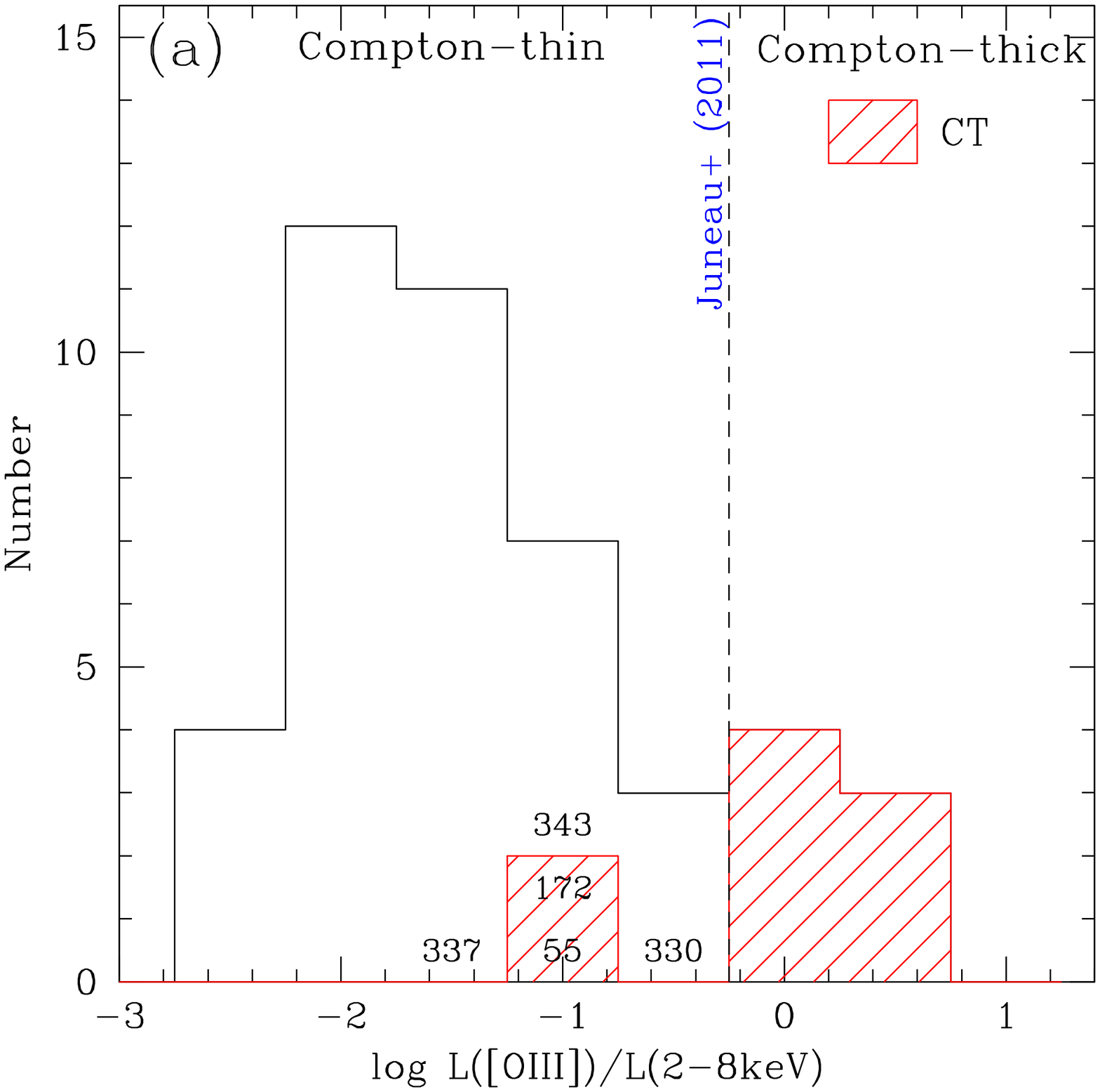}{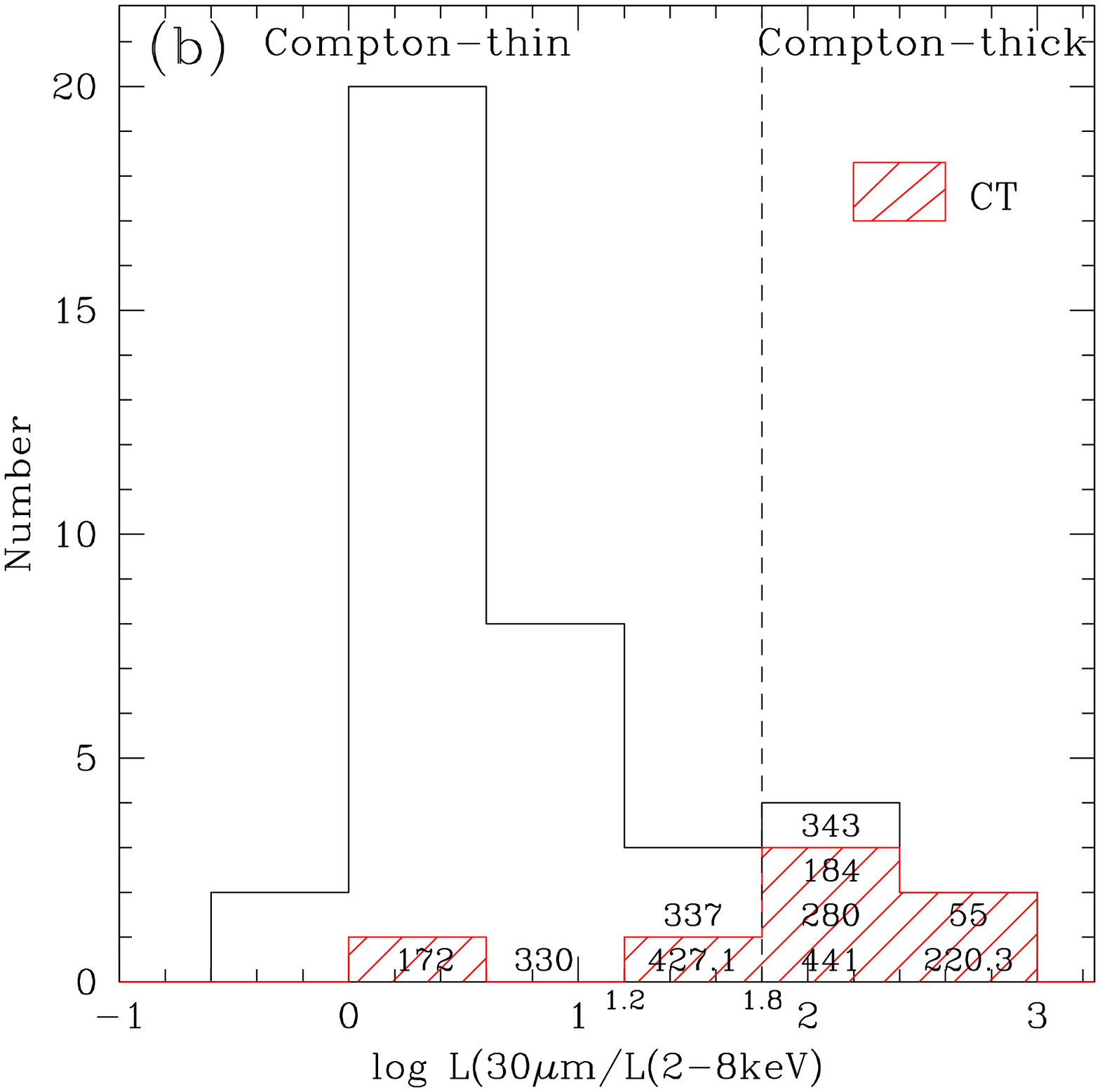}
\vspace{-0.5in}
\caption{Histograms of high obscuration indicators: the ratio of
  \oiii$\lambda$5007 to the observed 2$-$8~keV X-ray luminosity ({\it
    left}) and 30~$\mu$m to the observed 2--8~keV luminosity ({\it
    right}). The 2$-$8~keV X-ray luminosities are uncorrected
  for intrinsic \nh.  Compton-thick sources are denoted by the red
  hatched regions. The L(\oiii)/L(2--8~keV)$\ge -0.25$ value from
  \cite{2011ApJ...736..104J} (dashed line) finds exclusively CT
  sources, but misses 3C~55 and 3C~172. 
  The $L$(30~$\mu$m)/$L$(2--8~keV)$>1.8$ ratio finds most Compton-thick
  sources except for 3C~172 and 3C~427.1.
}
\label{fg:o3_Lx_MIR_Lx_fig}
\end{figure}

\vspace{-1.1in}

\begin{figure}
\epsscale{0.55}
\plotone{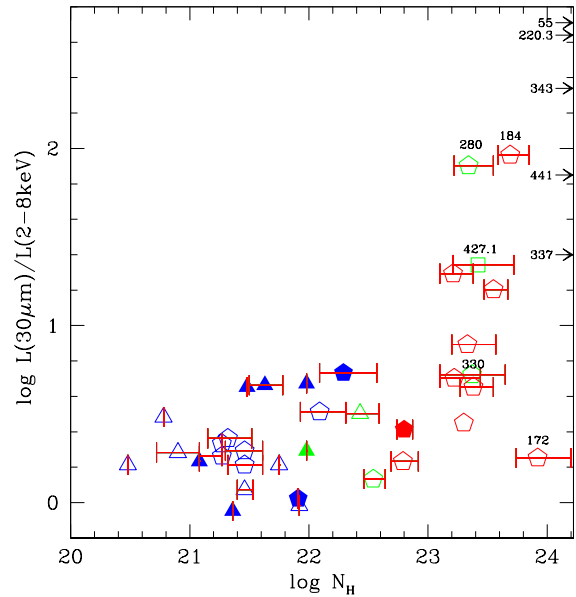}
\vspace{-0.6in}
\caption{Dependence of the 30~$\mu$m to 2$-$8~keV X-ray (uncorrected
  for \nh ) luminosity ratio on intrinsic \nh . For many sources with
  \nh~$>10^{23}$\,cm$^{-2}$ the L(30\,$\mu$m)/L(2$-$8\,keV) ratio
  substantially increases above 1. Symbols denote source type as in
  Figure~\ref{fg:NHvsHR}.  Compton-thick and borderline Compton-thick
  sources are labeled. Low $S/N$ Compton-thick sources without an
  \nh\ estimate (due to low $S/N$) are indicated along the ordinate
  only by their 3C identification.}
\label{fg:MIR_NH_fig}
\end{figure}

\clearpage
\begin{figure}
  \plotone{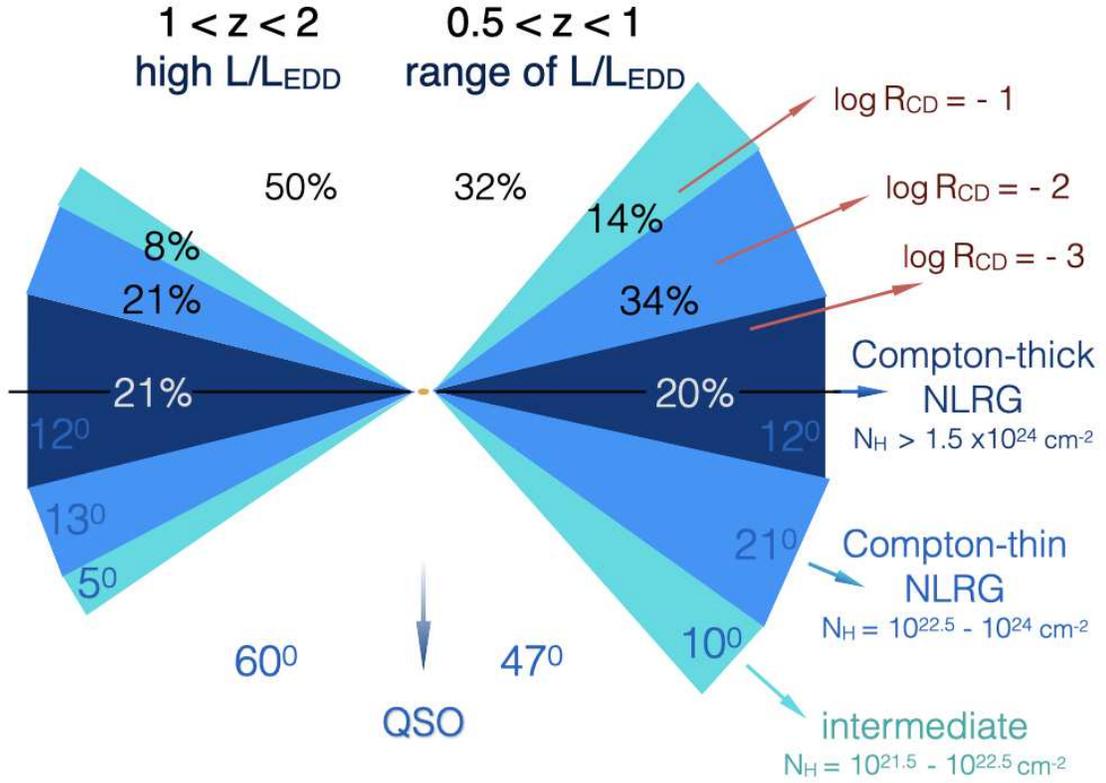}
\caption{Schematic of geometry of the circumnuclear obscuring dusty
  region inferred from the number of sources as a function of
  \nh\ in the high-$z$ sample (\citealp{2013ApJ...773...15W};
  represented by the left side of the diagram) and the medium-$z$ 3CRR
  sample (represented by the right side of the diagram; see also
  Table~\ref{tb:geom} for details). Percentages show how many sources
  are in each category: QSO (\nh~ $<10^{21.5}$\,cm$^{-2}$),
  intermediate sources (\nh $=10^{21.5-22.5}$\,cm$^{-2}$; light blue),
  Compton-thin NLRGs (\nh $=10^{22.5-24}$\,cm$^{-2}$; blue), and
  Compton-thick NLRGs (\nh~$>1.5 \times 10^{24}$\,cm$^{-2}$, dark
  blue).  Red arrows show lines of sight for which the radio core
  fraction is $\log R_{\rm CD}=-1, -2, -3$.  The torus in the high-$z$
  sample is more compact due to high $L/L_{\rm Edd}$, while in the
  medium-$z$ sample the torus is ``puffed-up'', which we interpret as
  due to a larger range of $L/L_{\rm Edd}$ ratios extending to lower
  values in comparison with the high-$z$ sample (see
  Section~\ref{sec:geometry}).}
\label{fg:geometry} 
\end{figure}
 
\vspace{-0.5in}

\clearpage
\begin{figure}
\epsscale{1}
\plottwo{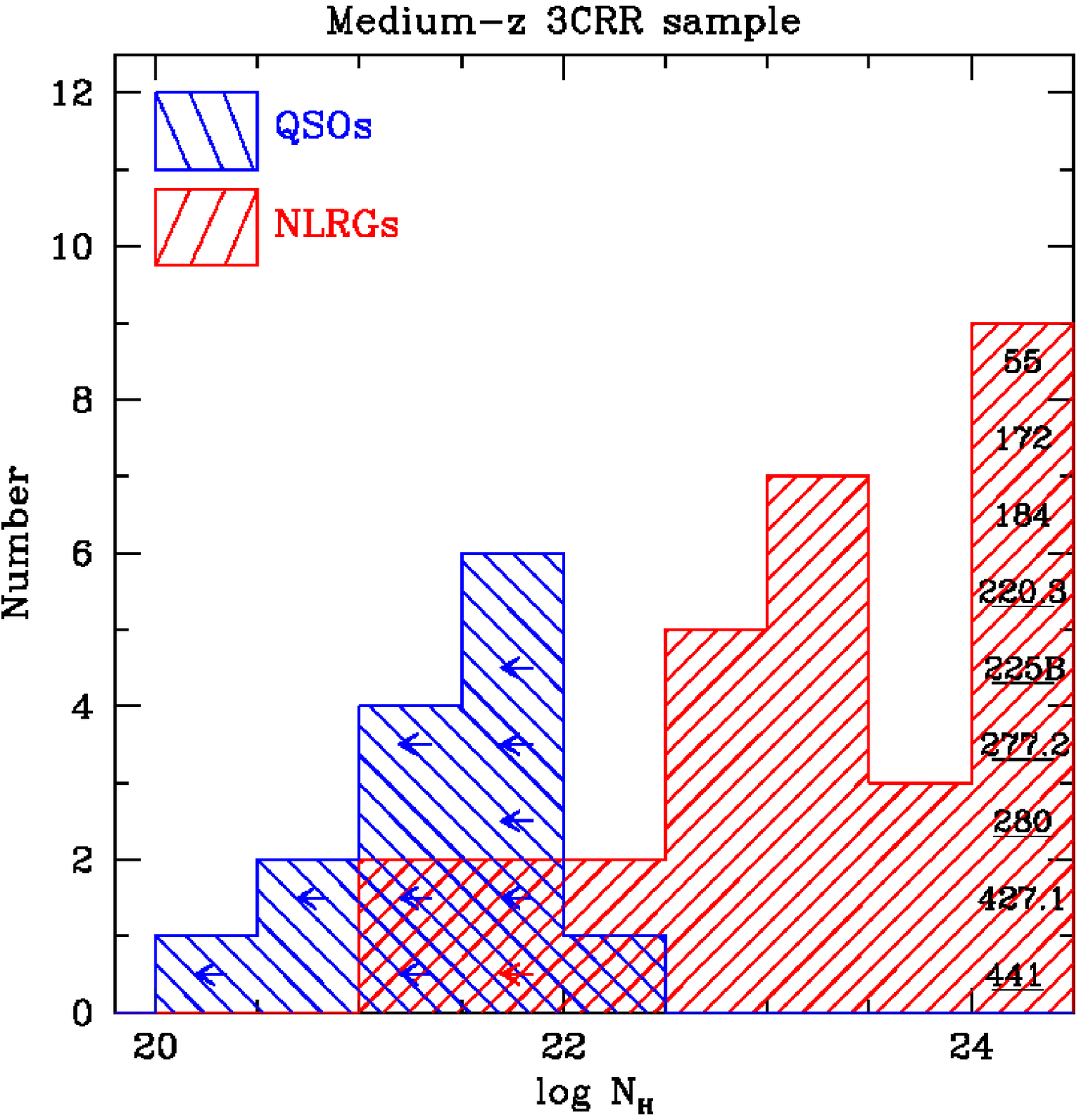}{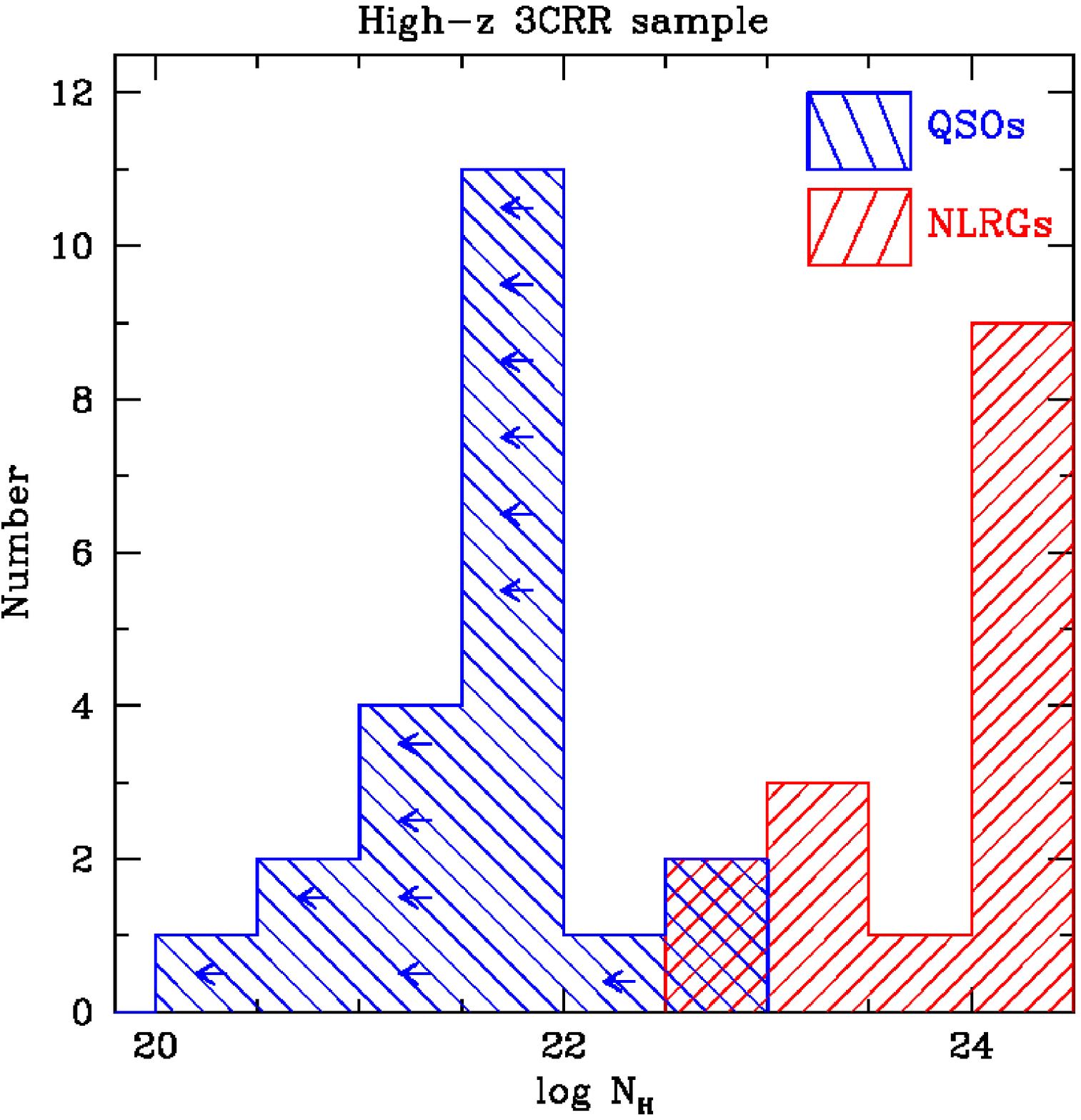}
%
\caption{Histograms of the X-ray equivalent intrinsic hydrogen column
  density \nh\ for the medium-$z$ 3CRR sample {\it (left)} and the
  high-$z$ 3CRR sample \citep{2013ApJ...773...15W}
  {\it(right)}. Quasars are shown in blue and NLRGs in red. Upper
  limits, mostly for quasars with no evidence for intrinsic
  absorption, are indicated by arrows. Compton-thick NLRGs in the
  medium-$z$ sample with no measurement of \nh\ due to low $S/N$ are
  indicated by underlined 3C identification.  Note the low \nh\ ($\le
  10^{22}$~cm$^{-2}$) NLRGs in the medium-$z$ sample, not present in 
  the high-$z$ sample.}
\label{fg:NHdist}
\end{figure}


\begin{figure}
  \epsscale{0.65}
\plotone{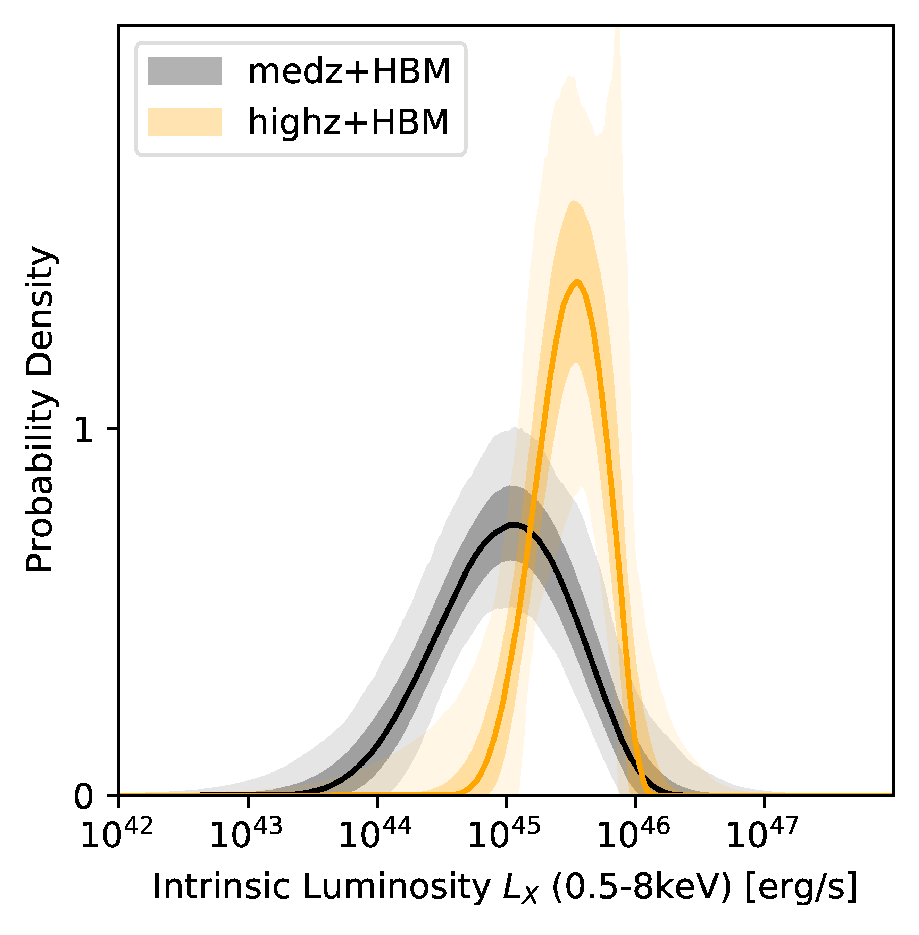}
\caption{Intrinsic 0.5--8~keV luminosity distributions of the
  medium-redshift (gray) and high-redshift (yellow) population derived
  from the HBM modeling of the two samples.  The high-luminosity tail
  ($L_{\rm X} > 10^{46}$~erg~s$^{-1}$) of the medium-$z$ sample is due
  to a simplistic treatment of a few piled-up quasars for which
  \nh\ $<10^{21}$~cm$^{-2}$ was assumed.  Shaded areas give 68\% and
  99\% confidence intervals. 
}
  \label{fig:pop-lum-dist}
\end{figure}
 
\clearpage
\begin{figure}
  \epsscale{0.7}
  \plotone{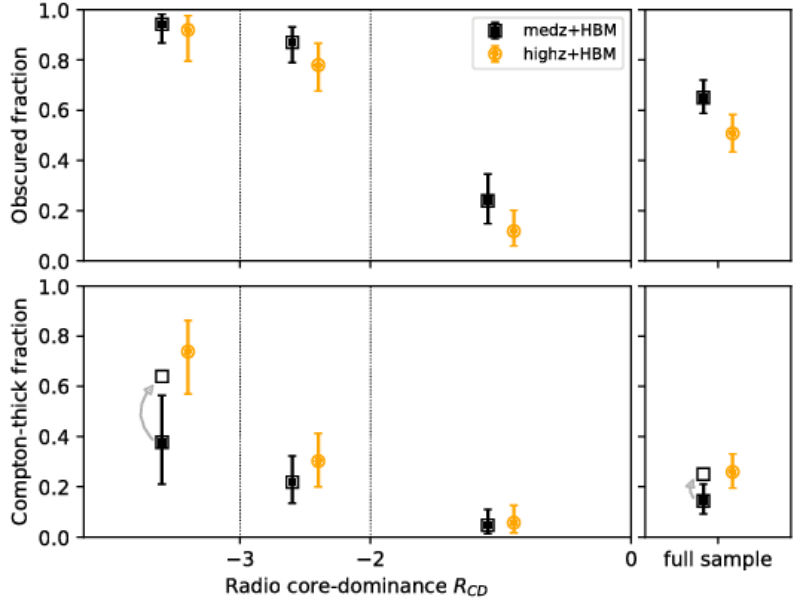}
\caption{The total obscured (\nh~$>10^{22}$~cm$^{-2}$) fraction
  ({\em{top}}) and Compton-thick (\nh~$>10^{24}$~cm$^{-2}$) fraction
  ({\em{bottom}}) from the Hierarchical Bayesian modeling.  Fractions
  are shown in three bins of log~\rcd\ (three left-most pairs of 
  panels) showing an increase with inclination. Fractions for the full
  medium-$z$ and high-$z$ samples are shown in the right-most pair 
  of panels. Open squares indicate Compton-thick fractions estimated from
  multiwavelength data (Section~\ref{sec:CT_fractions}). }
 \label{fig:Obscuration-fractions}
\end{figure}

\section{Appendix}
\subsection{Hierarchical Bayesian Model}

Inference on the obscured fraction in this sample is challenging
because of substantial uncertainties of measurements. For several
observations, short exposures give large uncertainties on the
line-of-sight obscuration, which can additionally be degenerate with
the intrinsic AGN luminosity. 
We want to incorporate these uncertainties to produce realistic
estimates of the obscured fraction. A self-consistent framework to do
this is a Hierarchical Bayesian Model (HBM). We begin by writing down
Bayes theorem for an individual object:

\[
p(\theta|D)=\frac{p(D|\theta)\times p(\theta)}{p(D)}
\]
The posterior probability distribution $p(\theta|D)$ of the parameters
$\theta=(L_{\rm{X}},\NH)$ is primarily shaped by
the likelihood function $p(D|\theta)$, given by the Poisson
count probability \citep{Cash1979} comparing the detected counts
$c_{i}$ to the assumed X-ray spectral model $m_{i}$ propagated
through the detector response:  

\[
p(D|\theta)=\prod_{i}\rm{Poisson}(c_{i};m_{i}(\theta))
\]
The second ingredient is $p(\theta)$, which normally is the prior of
the Bayesian computation, describing the prior knowledge of the
parameters $\theta$. In a Hierarchical Bayesian Model, we estimate
$p(\theta)$ simultaneously from the observations themselves.

To this end, we define only the shape of $p(\theta)$ assuming
population distributions with hyper-parameters (parameters of the
prior distribution) $H$:

\[
p(\theta,H)=p(L_{\rm{X}},H)\times p(\NH,H)
\]

As an example, the population distribution could be described by Gaussian 
distributions whose parameters would be the hyper-parameters H.

For the column density $\NH$ we assume a log-uniform distribution
within three bins (``unobscured'' $20-22$, ``Compton-thin obscured'',
$22-24$ and ``Compton-thick'', $24-26$). The relative ratio is defined
by the obscured fraction, $f_{\rm{obsc}}$ and the fraction of
obscured AGN that are Compton-thick, $f_{\rm{CT}}$:

\[
p(\NH,f_{\rm{obsc}},f_{\rm{CT}})=\left\{ \begin{array}{ll}
1-f_{\rm{obsc}} & \rm{if\,}\NH<10^{22}\rm{cm}^{-2}\\
f_{\rm{obsc}}\times(1-f_{\rm{CT}}) & \rm{if\,}\NH=10^{22-24}\rm{cm}^{-2}\\
f_{\rm{CT}} & \rm{if\,}\NH>10^{24}\rm{cm}^{-2}
\end{array}\right.
\]

For the luminosity distribution we adopt the flexible Beta distribution
(adopting a Gaussian or Student-t distribution instead does not change
the results significantly):

\[
p(L_{\rm{X}},\mu,\sigma,a,b)=\rm{Beta}(\log L_{X};\mu,\sigma,a,b)\times p(\mu,\sigma,a,b).
\]

For the population hyper-parameters, we adopt a uniform prior on
the mean logarithmic luminosity $\mu$, a log-uniform prior on the
population dispersion scale $\sigma$ and uniform priors on the shape
parameters $a$ and $b$. 

\begin{figure}

  \epsscale{0.6}
  \plotone{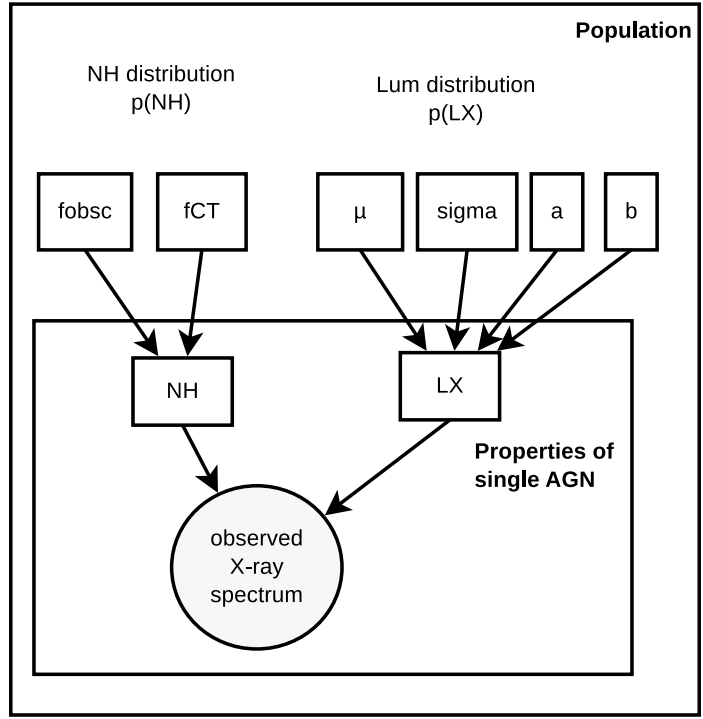}
\caption{\label{fig:Hierarchical-Bayesian-model.}Hierarchical Bayesian model.
The circle indicates observations and rectangles indicate parameters.
In the top half, a population distribution with some hyper-parameters
$(\mu,\sigma)$ generated AGN with some properties ($L_{\rm{X}}$,
$\NH$). In the bottom half, these in turn generated the observed
X-ray spectral data. The Hierarchical Bayesian model quantifies the
probability of this multi-level process as a function of its parameters.}
\end{figure}

We can then estimate the parameters $\theta$ for all sources (their
$L_{\rm{X}}$, $\NH$) simultaneously with the hyper-parameters
$H=(f_{\rm{obsc}},f_{\rm{CT}},\mu,\sigma,a,b)$. That means,
we explore a $(6+N\times2)$-dimensional parameter space:

\[
p(H,\theta_{0},\theta_{1}...,\theta_{N})=\prod_{i}p(D|\theta_{i})\times p(\theta_{i},H)
\]
and as usual in Bayesian inference, derive marginalized probability
distributions on the physical parameters (e.g., $\NH$), but also the
population distribution in $\NH$. 

In this analysis, the source parameters influence the population
distributions. 
At the same time, if the population distribution is well-constrained
by the majority of sources, a source with poor observational
constraints can benefit from the population distribution, as it gives
a prior where the parameters are most likely. Thus, the Hierarchical
Bayesian Model strengthens weak observations (the population informs
inference of individual objects down the hierarchy) and allows
inference of the population distribution (individual objects inform
the population distribution up the hierarchy). The model is
illustrated in Figure~\ref{fig:Hierarchical-Bayesian-model.}.

In practice, we compute the hierarchical model in two steps. First, we
compute $p(\theta_{i}|D)$ for each object under uninformative (flat)
priors. This is a simple X-ray spectral analysis with the BXA
(Bayesian X-ray Analysis; \citealp{Buchner2014}) module for Sherpa
\citep{2006SPIE.6270E..1VF}, assuming an AGN with obscuration
(\texttt{BNTORUS} model; \citealp{2011MNRAS.413.1206B}) with a warm,
mirror power law added (similar to the scattered AGN light component in
Section~\ref{sec:complex}).  All normalizations have wide log-uniform  
(uninformative) priors, the intrinsic photon index is assigned a Gaussian prior
centered at $1.95$ with standard deviation $0.15$. The warm mirror
normalization can reach up to $10\%$ of the intrinsic AGN powerlaw
component.  This standard setup is described e.g. in
\citet{Buchner2014}. The X-ray spectral analysis produces posterior
distributions in $p(\theta_{i}|D)$ that are described by equally
probable posterior samples $\theta_{ij}$ (as may be familiar from
Markov chain Monte Carlo analyses). In our case, we select $M=1000$
posterior samples for each source.  These samples cluster where the
posterior is most probable, and thus can be used as weight points in
Monte Carlo integrations.

To constrain the population parameters, we then evaluate the
population distribution at the object posterior sample locations:

\[
p(H)=\prod_{i}\frac{1}{M}\sum_{j}p(\theta_{ij},H)
\]
and only need to explore a 6-dimensional parameter space. This is
well-defined because the $\theta_{ij}$ samples indicate where the
population distributions have most weight. If the samples from
one object are clustered distant from another object's samples, then
the population distribution must spread to cover both. 
On the other hand, because the population distribution is a 
probability distribution normalised to unity, extremely wide distributions
give low probabilities at any specific location. Therefore the population
distribution will prefer to cover the samples. 
If uncertainties (cluster widths) become large, both narrow and broad 
population distributions are similarly probable.
Thereby, this formalism self-consistently carries forward the uncertainties
from each source analysis into the uncertainties on population parameters.

\subsection{Object constraints with flat population priors}
\label{sec:flat_priors}

%
%

We first analyze the spectra of sources independently with BXA and
report the $\NH$ and $L_{\rm{X}}$ constraints under flat priors. The
probability distributions of $\NH$ and $L_{\rm{X}}$ 
for each source are presented in Figure~\ref{fg:HBM_NH} and
\ref{fg:HBM_Lx} respectively, as gray contours.

We correct for significantly piled-up ($>20\%$) sources, where naive
spectral analysis may be biased by assuming these are unobscured
luminous AGN, in the following way. We assume a log-uniform column
density probability $\log\NH=20-21$. For the luminosity distribution,
we take lower limit of the luminosity derived from spectral analysis
as a lower limit on the true luminosity, and assume a log-uniform
distribution extending to very high luminosities ($\log L_{\rm X}=\log
L_{\rm X,min}-47$). The population model will truncate the
high-luminosity end based on other sources (see Figure~\ref{fg:HBM_NH}
and \ref{fg:HBM_Lx}, where gray shapes show probability distributions
from spectral analysis, and red shapes show updated probability
distributions after reweighing by the HBM analysis). This suppresses
the extremely high luminosities of the piled-up sources (marked with
an asterisk), and some Compton-thick, high-luminosity secondary
solutions (e.g. in 3C184 and 3C280) based on the Compton-thick
fractions of the well-constrained (by HBM) sources.

Figure~\ref{fg:HBM_NH_L} 
shows the constraints on both the luminosity and column density for all
sources in the medium-$z$ 3CRR sample.

\subsection{HBM constraints}


We now use the HBM to estimate the intrinsic luminosity distribution
and the obscured fractions of the population. The luminosity
distribution for the medium-$z$ 3CRR sample is shown in red in
Figure~\ref{fig:pop-lum-dist}. It is centered at $\mu=45.1\pm1.2$
and
$\sigma=4.5\pm1.0$ wide. The shape of the distribution is described by
$a=7.7\pm1.8$ and $b=3.6\pm2.6$, indicating a right-skewed, steeply
falling distribution.

We perform the same analysis for the high-$z$ 3CRR sample analysed in
\cite{2013ApJ...773...15W} and find consistent results in the spectral
analysis and reported obscured and Compton-thick fractions.  The X-ray
luminosity distribution for the high-$z$ sample is shown in black in
Figure~\ref{fig:pop-lum-dist}.


Finally, to investigate the dependence of obscured and Compton-thick
fractions on orientation (i.e. \rcd ), we modify the HBM to allow
three different groups (log \rcd$<-3$ , $-3 <$~log \rcd~$<-2$ ,
$-2<$~log \rcd$<0$) to have different obscured fractions, while still
enforcing the same luminosity distribution for all groups. The results
are presented in Figure~\ref{fig:Obscuration-fractions}, 
with the total obscured fraction in the top panels, and the
Compton-thick fraction in the bottom panels. The obscured fraction
increases to $>70\%$ for more intermediate and edge-on viewing angles
(log $R_{CD}<-2$). It is remarkably low ($\lesssim20\%$) for face-on
sources.  The Compton-thick fraction is small until the lowest
log~$R_{CD}<-3$ values are reached; in those edge-on sources it
reaches $\tilde 60\%$.  Within the uncertainties, the obscuration
results from the two samples are consistent with each other.


The obscured fractions are shown in
Figure~\ref{fig:Obscuration-fractions}.  We obtain an upper limit on
the Compton-thick fraction of all AGN in both medium- and high-$z$
samples of 20\% and an obscured fraction of $55\pm10\%$. The low
Compton-thick fraction (found from the HBM) is due to the few secure
Compton-thick candidates (see Figure~\ref{fg:HBM_NH_L}). 
However when information from multiwavelength data is included in
estimating the number of CT sources (Section~\ref{sec:CT_fractions}) -
the CT fractions for the medium-$z$ sample and the most edge-on
inclination sources (log~\rcd~$<-3$) increases and approaches those
found for the high-$z$ sample. All obscured and CT fractions, derived
with the HBM, are listed in Table~\ref{tab:hbmobscfrac}. 

\tablenum{7}
\begin{table}
\begin{center}
\label{tab:hbmobscfrac}
\caption{Obscured fractions derived with the HBM.}
\begin{tabular}{ccc}
 & Obscured fraction & Compton-thick fraction \\
\hline
& Medium-z Sample: &  \\
full sample &$0.57^{+0.09}_{+0.08}$ & $0.07^{+0.07}_{+0.04}$ \\
$-2 <$~log \rcd~$< 0$ & $0.19^{+0.11}_{+0.08}$ & $0.05^{+0.07}_{+0.04}$ \\
$-3 <$~log \rcd~$< -2$ & $0.81^{+0.08}_{+0.11}$ & $0.11^{+0.12}_{+0.07}$ \\
log \rcd~$<-3$ &  $0.92^{+0.05}_{+0.10}$ & $0.34^{+0.21}_{+0.17}$ \\

&&\\
& High-z Sample: & \\
full sample & $0.51^{+0.08}_{+0.07}$ & $0.26^{+0.07}_{+0.06}$ \\
$-2 <$~log \rcd~$< 0$ & $0.12^{+0.08}_{+0.06}$ & $0.06^{+0.07}_{+0.04}$ \\
$-3 <$~log \rcd~$< -2$ & $0.78^{+0.09}_{+0.10}$ & $0.30^{+0.11}_{+0.10}$ \\
log \rcd$<-3$ & $0.92^{+0.06}_{+0.12}$ & $0.74^{+0.12}_{+0.17}$  \\
\hline
\end{tabular}
\end{center}
\end{table}

\end{document}